\documentclass[
    preprint,
    titlepage,
    preprintnumbers,
    superscriptaddress,
    nofootinbib,
    aps,
    prd,
    floatfix,
    tightenlines
]{revtex4-2}

\usepackage[T1]{fontenc}

\usepackage{xcolor}
\usepackage{amsmath,amssymb}
\usepackage{bbm}
\usepackage{subfiles}
\usepackage{cancel}
\usepackage{slashed}
\usepackage{multirow}
\usepackage{myfuncs}
\usepackage{setspace}
\usepackage{soul}
\usepackage{forest}

\definecolor{H_D_He_naDM}{HTML}{E69F00}
\definecolor{D_He_naDM}{HTML}{D55E00}
\definecolor{H_D_naDM}{HTML}{56B4E9}
\definecolor{D_naDM}{HTML}{0072B2}
\definecolor{H_naDM}{HTML}{CC79A7}
\definecolor{N_SIDM}{HTML}{009E73}

\usepackage[
    backref=page,
    colorlinks=true,
    linkcolor=red,
    citecolor=blue,
    urlcolor=violet,
    filecolor=purple
]{hyperref}

\makeatletter
\AtBeginDocument{%
  \let\BR@origNAT@sort@cites\NAT@sort@cites
  \def\NAT@sort@cites#1{%
    \BR@origNAT@sort@cites{#1}%
    \@for\BR@citekey:=\NAT@cite@list\do{%
      \edef\BR@citekey{%
        \expandafter\@firstofone\BR@citekey\@empty
      }%
      \expandafter\Hy@backout\expandafter{\BR@citekey}%
    }%
  }%
  \def\hyper@natlinkstart#1{%
    \hyper@linkstart{cite}{cite.#1}%
    \def\hyper@nat@current{#1}%
  }%
}
\makeatother

\backrefparscanfalse
\makeatletter
\def\BibitemShut#1{%
    \csname bibitem#1\endcsname
    \backrefprint
}
\makeatother

\usepackage[capitalize]{cleveref}
\crefname{section}{Sec.}{Secs.}
\Crefname{section}{Sec.}{Secs.}
\crefname{appendix}{App.}{Apps.}
\Crefname{appendix}{App.}{Apps.}
\Crefname{figure}{Fig.}{Figs.}
\crefname{figure}{Fig.}{Figs.}

\definecolor{DarkRed}{rgb}{0.6,0,0}
\definecolor{DarkGreen}{rgb}{0,0.6,0}
\definecolor{DarkOrange}{rgb}{1,0.35,0}
\definecolor{DarkBlue}{rgb}{0,0,0.6}

\makeatletter
\let\l@subsubsection\@gobbletwo
\makeatother

\AtBeginDocument{%
  \global\count\footins=1000\relax
}

\begin{document}

\title{The Dark Side of the Periodic Table: \\ Theory and Signatures of Nuclear Atomic Dark Matter}

\author{Pouya Asadi}
\email{pasadi@ucsc.edu}
\affiliation{Department of Physics and Santa Cruz Institute for Particle Physics, University of California Santa Cruz, Santa Cruz, CA 95064, USA}

\author{Zackaria Chacko}
\email{zchacko@umd.edu}
\affiliation{Maryland Center for Fundamental Physics, Department of Physics, University of Maryland, College Park, MD 20742, USA}

\author{David Curtin}
\email{d.curtin@utoronto.ca}
\affiliation{Department of Physics, University of Toronto, Toronto, ON M5S 1A7, Canada}

\author{Micah Mellors}
\email{m.mellors@mail.utoronto.ca}
\affiliation{Department of Physics, University of Toronto, Toronto, ON M5S 1A7, Canada}

\author{Ethan T.~Neil}
\email{ethan.neil@colorado.edu}
\affiliation{Department of Physics, University of Colorado, Boulder, CO 80309, USA}

\begin{abstract}
We study the theory and signatures of a Mirror Twin Higgs (MTH) scenario in which stable twin neutrons constitute most of the dark matter (DM), while a subdominant component of twin elements survives as nuclear atomic dark matter (naDM). 
This framework offers a solution to the \textit{matter coincidence problem} ($\Omega_{\rm DM}\simeq5\Omega_b$),
as the twin $\mathbb{Z}_2$ symmetry can correlate the visible and DM mass scales and their number densities.
We analyze twin nucleon freeze-out and the onset of twin Big Bang nucleosynthesis (BBN) to estimate the total naDM abundance, and determine which elements in the dark periodic table could be populated.
An appreciable but subdominant naDM component generally requires a fractional nucleon mass splitting near the percent level, with twin-BBN creating heavy elements up to twin carbon. 
Strong twin neutron--proton scattering allows nearly all of the dark matter, not only the atomic sub-component, to participate in \textit{dark acoustic oscillations} (DAO) at early times, giving rise to a characteristic two-stage DAO signature that may be detected with future cosmological probes.
Interactions between the twin and visible sectors, including the irreducible Higgs portal, make this scenario a well-motivated target for future directional direct-detection searches, and the naDM sub-component is expected to exhibit highly exotic astrophysics.
We outline a supersymmetric twin Froggatt--Nielsen realization in which spontaneous $\mathbb{Z}_2$ breaking in the flavon sector can generate the required  fermion spectrum and Higgs potential.
\end{abstract}

\maketitle

\pagebreak

\hypersetup{linkcolor=black}

\begingroup
  \fontsize{11}{12.7}\selectfont
  \setlength{\parskip}{0pt}
  \tableofcontents
\endgroup

\clearpage
\hypersetup{linkcolor=red}

\begin{spacing}{1.1}


\newpage

\section{Introduction}\label{sec: intro}

Despite comprising most of the matter in the universe, the microscopic nature of dark matter (DM) remains unknown, see Refs.~\cite{Cirelli:2024ssz,Bozorgnia:2024pwk} for recent reviews.
Although single-component paradigms such as the standard weakly interacting massive particle (WIMP) are attractive for their simplicity, given the multitude of fields and interactions in the Standard Model (SM), it is quite plausible that the dark sector exhibits a comparable degree of complexity.
From a phenomenological perspective, however, the enormous space of possible complex dark sectors is daunting and, absent further empirical motivation, runs counter to the economy of single-component candidates.

There is, however, an important empirical clue, which both suggests a connection between the dark and visible sectors, and favours non-minimal dark sectors: the \textit{matter coincidence problem}.
The present-day energy densities of baryonic SM matter and DM are strikingly similar, at roughly 5\% and 25\% respectively of the total energy density of the universe.
If the dark sector parameters are completely independent of the SM, this similarity is highly atypical.
The SM matter abundance depends on the QCD confinement scale, which---through dimensional transmutation---is exponentially sensitive to the value of the strong gauge coupling in the deep ultraviolet (UV).
Likewise, the DM mass and relic abundance could naturally span many orders of magnitude; in particular, the relic abundance generally has a steep dependence on other DM parameters, making such coincidences unlikely. 
That the two sectors nevertheless end up with energy densities within a factor of a few could therefore be an important clue pointing towards a common origin of visible and dark matter abundances. 
Explaining this similarity requires introducing ingredients beyond a minimal dark sector.
Models that can explain the observed ratio of abundances merit a more detailed study and should be at the forefront of our experimental program. 

Several different approaches have been proposed to address the matter coincidence problem, see e.g.
Refs.~\cite{Hodges:1993yb,Mohapatra:2000rk,Berezhiani:2000gw,Foot:2003jt,Berezhiani:2003xm,Foot:2004pq,Bai:2013xga,Newstead:2014jva,Farina:2015uea,Ibe:2019ena,Beauchesne:2020mih,Murgui:2021eqf,Ritter:2022opo,Rosa:2022sym,Brzeminski:2023wza,Banerjee:2024xhn,Chung:2024nnj,Cox:2025wxk,Girmohanta:2026eqg,Bai:2026ufp,Asadi:2026zbc}. 
For non-relativistic relics, solving the problem amounts to dynamically explaining the observation that
\begin{equation}
    m_b\,n_b \sim m_{\rm dark}\,n_{\rm dark}\, ,
    \label{eq:coinc_def}
\end{equation}
where $m_b$ ($m_{\rm dark}$) represents the SM baryon (DM) mass and $n_b$ ($n_{\rm dark}$) denotes the asymmetric SM (DM) number density.
The simplest---though by no means unique---way to account for Eq.~\eqref{eq:coinc_def} is to explain the similarities of the masses and number densities separately,
\begin{equation}
    m_b \sim m_{\rm dark},
    \qquad
    n_b \sim n_{\rm dark}\, .
    \label{eq:coinc_sol_separate}
\end{equation}
We focus on this possibility in the present work.
Since the SM nucleon mass is set predominantly by the QCD confinement scale, a particularly economical way to obtain
$m_{\rm dark}\sim m_b$
is for the DM mass to likewise arise from a confining gauge theory whose confinement scale is correlated with, and close to, the QCD confinement scale $\Lambda_{\rm QCD}$. This is most naturally enforced by a symmetry between the two sectors, which could  correlate the number densities as well.

The simplest realization of such a symmetry between the two sectors is a \Ztwo\@ that duplicates the SM matter fields and gauge forces in a dark mirror sector~\cite{Hodges:1993yb,Mohapatra:2000rk,Berezhiani:2000gw,Berezhiani:2003xm,Foot:2003jt,Foot:2004pq,An:2009vq,Higaki:2013vuv,Addazi:2015cua,Lonsdale:2018xwd,Ritter:2021hgu,Bodas:2024idn,Mohapatra:2025bdl} (for reviews with many more references, see Refs.~\cite{Berezhiani:2005ek,Ciarcelluti:2010zz,Foot:2014mia}). However, an exact \Ztwo\  is phenomenologically disfavoured by the cosmological constraints on $\DNeff$ and, somewhat more model-dependently, by the collider bounds on invisible Higgs decays. 
In addition, since the dark proton is the lightest baryonic state, this framework is also disfavoured by the cosmological bounds on DM self-interactions and on the fraction of DM that is acoustic during the cosmic microwave background (CMB) epoch. 

This then motivates a small explicit breaking of the \Ztwo\ symmetry that relates the two sectors, so that the dark neutron is the lightest baryonic state in the dark sector and constitutes the primary component of DM. Furthermore, this explicit breaking would relax the cosmological constraints from $\DNeff$~\cite{Berezhiani:1995am,Berezhiani:1995yi} and also allow the collider bounds on this scenario to be satisfied. However, because of the hierarchy problem of the SM, any such explicit breaking would, in general, lead to the SM Higgs and its \Ztwo\ counterpart having very different vevs, so that the masses of the baryons in the two sectors would be expected to be completely different. Therefore, any realization of this scenario must be tied to a solution of the hierarchy problem of the SM~\cite{Bodas:2024idn}.

These considerations motivate the realization of this scenario within the framework of the Mirror Twin Higgs (MTH) solution to the little hierarchy problem~\cite{Chacko:2005pe,Barbieri:2005ri,Chacko:2005vw}. In MTH models, the Higgs mass is protected against radiative corrections up to scales of order 5-10 TeV by the combination of a global symmetry and a discrete \Ztwo\ symmetry that relates the SM to a mirror copy of itself. 
The top partners are part of the mirror sector and are therefore neutral under the SM gauge group. They therefore evade conventional LHC searches for coloured top partners \cite{ATLAS:2020dsf,Gravili:2021zdr,ATLAS:2021kxv,CMS:2021eha,CMS:2023mny,CMS:2021beq}. See Refs.~\cite{Batell:2022tif, Asadi:2026mip} for recent reviews. 

The \Ztwo\ symmetry correlates the twin and SM QCD confinement scales, yielding twin baryon masses, and in particular a twin neutron mass, comparable to the SM proton mass.
The mild explicit breaking of the \Ztwo\ symmetry will not affect the protection of the Higgs mass against radiative corrections. When complemented by a mechanism that co-generates the twin and SM baryon asymmetries~\cite{Farina:2016ndq,Feng:2020urb,Kilic:2021zqu,Bittar:2024ryj,Alonso-Alvarez:2023bat}, this framework has all the essential ingredients to address both the little hierarchy problem and the matter coincidence problem~\cite{Barbieri:2005ri,Farina:2015uea,Farina:2016ndq,Barbieri:2017opf,Beauchesne:2020mih,Feng:2020urb,Kilic:2021zqu,Bittar:2024ryj,Alonso-Alvarez:2023bat}.\footnote{Alternative DM candidates in the MTH framework have been considered in Refs.~\cite{GarciaGarcia:2015fol,GarciaGarcia:2015pnn,Hochberg:2018vdo,Terning:2019hgj}.} In this scenario, a source of asymmetric reheating below the decoupling temperature at which the visible and twin sector decouple from each other must also be introduced~\cite{Chacko:2016hvu,Craig:2016lyx} to reduce the twin photon and twin neutrino contributions to $\DNeff \lesssim 0.1$ ~\cite{Goldstein:2026iuu}. However, this does not affect the relative abundances of DM and SM matter, and so does not affect the resolution of the matter coincidence problem. 
The \Ztwo\ breaking must render the twin neutron the lightest twin baryon, ensuring a viable DM candidate.

These considerations lead us to focus on \Ztwo\ breaking through non-identical effective Yukawa couplings in the visible and twin sectors.
Such a breaking can modify the twin quark masses sufficiently to reverse the twin neutron--proton mass ordering, making the twin neutron lighter than the twin proton.
As mentioned, some degree of \Ztwo\ breaking is already required phenomenologically in MTH models, e.g. since an exactly mirror-symmetric spectrum would generically lead to unacceptably large Higgs decays into the twin sector \cite{Burdman:2014zta,ParticleDataGroup:2024cfk}.
Flavour constructions in which 
$\mathbb{Z}_2$-symmetric dynamics are responsible for generating the fermion mass hierarchies, but in which the $\mathbb{Z}_2$ is spontaneously broken, provide a natural way to alter the relative twin fermion masses and thereby realize a spectrum in which the twin neutron is the lightest stable baryon. (This general possibility was first proposed in the context of Twin Higgs models in Ref.~\cite{Barbieri:2017opf}.)
We discuss below the possibility of generating such a structure from Froggatt--Nielsen (FN)~\cite{Froggatt:1978nt,Leurer:1992wg,Leurer:1993gy} models of flavour as a proof of principle, though this is by no means a unique solution and our analysis will not hinge on its details. 
In such a framework, the low-energy twin spectrum can differ substantially from that of the SM even for modest changes in the UV parameters.

We therefore explore a scenario in which stable \emph{twin neutron dark matter} (TNDM) with a mass of $m_{n,B} \approx 5~\mathrm{GeV}$ makes up the majority of DM in our universe, assuming that an unspecified baryogenesis mechanism respects the $\mathbb{Z}_2$ symmetry, co-generating equal amounts of visible- and twin-baryon number (though we explore $\mathcal{O}(1)$ deviations from that benchmark assumption). This setup offers a true solution to the matter coincidence problem, with the twin $\mathbb{Z}_2$ symmetry correlating both masses and abundances in the visible and dark sectors.\footnote{This is in marked contrast to \textit{co-genesis} models where visible and dark baryon number may be co-generated, but no fundamental mechanism generically and simultaneously enforces $m_b \sim m_\mathrm{dark}$ along with $n_b \sim n_\mathrm{dark}$.} Our study is released alongside a short companion letter~\cite{Asadi:2026oae}, which introduces the main ideas and focuses on the most distinctive new signatures.

Although the TNDM is uncharged under the long-range twin electromagnetic force, it retains residual twin strong and electroweak interactions and therefore can be an example of
self-interacting DM (SIDM), with self-interaction cross-sections in the $\sigma/m \sim \mathcal{O}(0.1 - 1)~\mathrm{cm}^2/\mathrm{g}$ range \cite{Tulin:2017ara,Cline:2022leq}. 
We assume that twin electromagnetism (EM) remains unbroken, a feature that gives rise to several distinctive phenomenological signatures.
As such, although the DM abundance is expected to be dominated by twin neutrons, a subdominant population of twin atoms can also be present and realize a version of atomic DM (aDM)~\cite{Kaplan:2009de} with ``dark nuclear physics.'' 
We refer to these atomic relics, and to the corresponding phenomenology more broadly, as \textit{nuclear atomic DM} (naDM).

The main distinguishing feature of our analysis from earlier work is a detailed treatment of the twin nuclear-atomic component while simultaneously using the twin $\mathbb{Z}_2$ and a UV solution to the flavour problem to explicitly solve the coincidence problem with $\sim 5~\mathrm{GeV}$ twin neutron dark matter and $n_b \approx n_\mathrm{dark}$.
While we focus on our MTH-FN realization, the presence of a naDM subcomponent is a generic consequence of this class of mirror-symmetric solutions to the matter coincidence problem, and could emerge from other UV completions~\cite{Lonsdale:2018xwd,Beauchesne:2020mih,Bodas:2024idn}.
The main goal is to determine how the twin-sector parameters---in particular the masses of the twin nucleons and pions, the twin electron mass, and the twin Fermi constant---control the existence and relic abundance of naDM and explore its phenomenology.

In \cref{sec: model definition}, we identify the regions of parameter space in which different naDM states can exist. We use a combination of non-relativistic modelling and lattice results to estimate the low-energy twin QCD parameters, such as masses, nuclear binding energies, and scattering lengths.
The twin proton is unstable unless the twin electron is much heavier than in the standard MTH. Since it would then need to be bound in nuclei to generate a naDM sub-component, the dominant uncertainty on the naDM \emph{abundance}  derives from the uncertainty in the twin deuteron binding energy $B_{D, B}$, which controls the onset of twin big bang nucleosynthesis (BBN).
We therefore utilize the available lattice results to bracket the uncertainty by considering limiting assumptions for the deuteron and helium binding energies.

An interesting aspect of the model's phase structure is that \textit{nuclear} and \textit{atomic} stability need not coincide. After recombination, a bound twin electron can be captured by a twin proton inside the nucleus, \(p_B^++e_B^-\to n_B+\nu_B\), transmuting the atom into a state with lower atomic number \(\Znum\) and, if the daughter is unbound, causing it to fragment into lighter nuclei and free twin neutrons. 
This process corresponds to (twin) electron capture.
When kinematically allowed, these decays can be rapid on cosmological timescales, complicating the late-time abundance and composition of naDM. In particular, twin hydrogen is always unstable in our parameter space, even where the free twin proton is stable against \(\beta^+\) decay.

In \cref{sec: cosmo} we then solve for the twin proton abundance at the onset of twin BBN and use that to estimate the naDM relic population. We find that an appreciable naDM component typically requires a twin neutron--proton mass splitting at the percent level together with sufficiently light twin electrons. 
For smaller mass splittings, too large a fraction of the DM can remain in naDM, in tension with constraints from structure formation, whereas larger splittings lead to efficient twin proton decay into twin neutrons and a pure-TNDM scenario.

We also find that the twin-deuterium bottleneck can result in very unfamiliar outcomes from twin BBN.  
In the SM, the low deuteron binding energy of $\sim 2~\mathrm{MeV}$ suppresses the formation of deuterons and hence heavier nuclei until temperatures below an MeV. 
At that temperature, the Coulomb barrier greatly suppresses fusion of twin elements with atomic mass higher than five or six, meaning neutrons rapidly get caught up in light nuclei, with only trace abundances of heavier elements produced.
In the twin sector, the large range of possible deuteron binding energies (both due to its dependence on the twin nucleon/pion masses and its theoretical uncertainty) means that if deuterons are bound and once the twin deuteron bottleneck resolves, the twin sector temperature may still be high enough to overcome the Coulomb barrier for the fusion of atomic nuclei with atomic numbers $\Znum$ up to some maximum value $\Znum_{\max}$, 
with the majority of nucleons ending up as the heaviest possible state~\cite{Krnjaic:2014xza, Hardy:2014mqa}.
We find $\Znum_{\max} \lesssim 6$.
Therefore, assuming that no other aspects of twin nuclear physics (such as unstable intermediate states) limit the formation of heavy twin elements, the naDM parameter space can be divided into distinct phases, where the majority of twin atoms take the form of twin deuterium, helium, lithium, beryllium, boron, or carbon. 
This can produce some fascinatingly exotic variations on late-time atomic dark matter phenomenology, though theoretical uncertainties on the twin nuclear binding energies and interaction rates make the locations of these phase boundaries, and hence the \emph{composition} of naDM, highly uncertain.

The main phenomenological signals and constraints associated with this framework are discussed in \cref{sec: observables,sec: terrestrial observables}. In \cref{sec: observables} we study dark acoustic oscillations (DAOs), TNDM self-interaction constraints from Bullet Cluster observations, 
and the ionization and distribution of naDM in the Milky Way. 
We find that reasonable estimates of TNDM self-interactions are safe from Bullet Cluster constraints. However, the interactions between the TNDM and the naDM are strong enough to keep them in equilibrium until fairly late times, giving rise to a unique two-scale DAO signature due to early (small scale today) decoupling of the dominant TNDM component from the oscillating charged twin nuclei, and late (large scale today) decoupling of twin nuclei from dark radiation. 
The signatures of the first DAO stage involving the full twin baryon DM population are on the cusp of our current constraints at galactic or smaller scales, motivating future dedicated studies and probes of DAOs in this regime. 
Other astrophysical and cosmological signatures can be highly exotic, like primordial twin dust and high-metallicity mirror stars. We comment on them briefly but leave detailed investigation for future studies.
Together, these complementary observables can test both the dominant self-interacting TNDM component, and a subdominant naDM population. 

In \cref{sec: terrestrial observables} we pay particular attention to the direct detection signals of our model.
In particular, the irreducible Higgs portal makes these models well-motivated targets for upcoming directional direct detection searches in the $\mathcal{O}(\mathrm{GeV})$ mass range. 
A photon kinetic-mixing portal, present in some UV completions, can further generate additional nuclear- and electron-recoil signals accessible to next-generation experiments. 
Finally, current and future colliders can probe both the general MTH-type signatures of this setup, and possibly detect the modifications to the twin-fermion spectrum that are unique to the naDM solution of the coincidence problem.

Our proof-of-principle UV-realization of this naDM MTH scenario is sketched out in \cref{sec: uv realization}. We show how a supersymmetric twin Froggatt--Nielsen construction with generic flavour charges can realize spontaneous $\mathbb{Z}_2$ breaking in the flavon sector, which generates the different effective Yukawas in the visible and twin sector and, in turn, implements our naDM scenario with stable TNDM. This preferentially populates certain regions of our more general naDM infrared (IR) parameter space, which may be particularly motivated for detailed investigation, though we emphasize that other UV completions could provide different theory priors. 
Intriguingly, the same spontaneous $\mathbb{Z}_2$ breaking in the flavon sector radiatively feeds into the Higgs sector, generating the $\mathbb{Z}_2$-breaking soft mass required for generating an asymmetry between the visible and the twin Higgs vacuum expectation value (vev) of $v_B/v_A \sim 3-9$. 
Such a setup might therefore provide a common explanation for all the $\mathbb{Z}_2$ breaking in the naDM MTH framework, though this needs to be confirmed in a more detailed model-building study that builds on our rough sketch.

Overall, as we conclude in \cref{sec: conclusion}, our analysis points to the MTH scenario as an attractive and arguably one of the most fully-developed solutions to the matter coincidence problem to date, while also motivating the rich phenomenology of naDM. This opens up the possibility that correlated collider, astrophysical, and cosmological measurements may 
not only point us towards the solutions of the little hierarchy problem and the particle nature of dark matter, but also explain the intriguing similarity in abundance of dark and visible matter in our universe today.

We provide additional technical details in several appendices. \Cref{app: auxiliary} clarifies our conventions for the auxiliary material accompanying the source file, while \cref{app: lattice fits} describes the lattice fits used in our calculations. \Cref{app: cosmo BG,app: np freeze-out} review the relevant cosmological background, including twin neutron--proton freeze-out. Finally, \cref{app: direct detection BG} presents details of the direct-detection calculation, including the DM velocity distribution, electron-screening effects, and our precise definition of the neutrino floor, while \cref{app: effective baryon yukawas} derives the effective couplings of twin baryons to the Higgs relevant for Higgs-portal direct detection.


\section{Model Definition}\label{sec: model definition}

In this section, we define a Twin Higgs model that provides a motivated framework for addressing the matter coincidence problem and study the properties of its low-energy spectrum. 
This relies on some source of soft \Ztwo\ breaking generating differences in the effective Yukawas between the two sectors, such that the masses of the twin fermions can be treated as free parameters.  (In order to avoid spoiling the solution to the little hierarchy problem, we assume the twin top Yukawa remains approximately equal to the SM top Yukawa.)

We focus on the scenario where the twin up--down quark mass ordering is reversed, thus realizing the TNDM possibility of stable twin neutrons as the dominant DM component. A possible  sub-component of twin-nuclear bound states realizes a version of aDM~\cite{Kaplan:2009de} with dark nuclear physics. We use ``nuclear atomic Dark Matter'' (naDM) to refer to both the atomic sub-component in particular, and this IR model setup in general.
Our goal is to set the stage for the subsequent investigation of its cosmological history and phenomenology by demonstrating how to systematically map between the twin-quark-level parameters and twin-nuclear parameters, in order to determine the possible twin-nuclear spectra that can arise from these scenarios.

This naDM MTH model solves the matter coincidence problem  because the underlying $\mathbb{Z}_2$ symmetry naturally aligns the twin and SM QCD scales and corresponding baryon masses at the $\mathcal{O}(1)$ level. Simultaneously, the same $\mathbb{Z}_2$ can enforce production of equal (or $\mathcal{O}(1)$ similar) amounts of twin and SM baryon number, either via some mechanism which unifies the two baryon numbers in a $\mathbb{Z}_2$-symmetric fashion or via some high-scale thermal mechanism that is insensitive to the small (and different) Higgs vevs and fermion masses in the two sectors. 
Our benchmark scenario is one where the twin neutron has a mass of $m_{n,B} \approx 5~\mathrm{GeV}$ and both sectors have equal asymmetric baryon number densities, but we also explore $m_{n,B} \sim (1 - 10)~\mathrm{GeV}$ to understand how the phenomenology depends on twin neutron mass, and to cover the possibility of $\mathcal{O}(1)$ differences in the twin and visible baryon numbers.
Two additional ingredients complete the definition of our benchmark:
\begin{itemize}

\item We assume that twin electromagnetism remains unbroken. Although models with a broken twin electromagnetic symmetry are also viable~\cite{Batell:2019ptb}, an unbroken twin $\mathrm{U}(1)$ gives rise to a much richer cosmological and astrophysical phenomenology through the existence of twin atoms and long-range interactions. Furthermore, breaking twin electromagnetism would introduce additional sources of \Ztwo\ breaking beyond what is needed to solve the coincidence problem, and is therefore less minimal.

\item We assume that the twin sector is colder than the visible sector in order to satisfy the stringent constraints from $\DNeff$, the effective number of light degrees of freedom (DOF) at recombination. We remain agnostic about the origin of this temperature asymmetry and simply take it as an input to our phenomenological analysis (see Refs.~\cite{Chacko:2016hvu, Craig:2016lyx,Adshead:2016xxj,Alonso-Alvarez:2023bat} for representative mechanisms for asymmetric reheating after the two sectors decouple).

\end{itemize}

In summary, our IR model uses an MTH framework with \Ztwo\ breaking in the non-top Yukawas, a stable twin neutron, unbroken twin electromagnetism, and a temperature asymmetry between the two sectors. To study the phenomenology of this framework, we first determine its low-energy spectrum. In \cref{subsec: dark quark lagrangian}, we introduce the relevant UV and IR parameters and identify a convenient set of low-energy quantities to label the parameter space of the model. 
We then investigate the properties of the twin hadrons (\cref{subsec: dark hadron masses}) and twin nuclei (\cref{subsec: dark binding energies}). We also carefully examine the stability of twin nuclei and twin atoms (\cref{subsec:stability}), before mapping the distinct regions of the model's parameter space based on the possibility and nature of the dominant naDM sub-components (\cref{subsec: phase diagram}). In particular, we identify the regions of parameter space where a naDM population can potentially exist, giving rise to a variety of distinctive cosmological and experimental signatures.

\subsection{Dark Quark Level Lagrangian}\label{subsec: dark quark lagrangian}

We start with the standard Mirror Twin Higgs model~\cite{Chacko:2005pe}, with low-energy Lagrangian
\begin{equation}\label{eq: twin Lagrangian}
    \mathcal{L} = \mathcal{L}_A + \mathcal{L}_B + V(H_A,H_B),
\end{equation}
where sector $A$ corresponds to the SM and sector $B$ to the dark (twin) sector.
The Higgs potential is given by ~\cite{Chacko:2005pe, Craig:2015pha}:
\begin{eqnarray}
    V = &  m_H^2&(|H_A|^2 + |H_B|^2) + \lambda_H(|H_A|^2 + |H_B|^2)^2 + \delta (|H_A|^4 + |H_B|^4) +  \Delta m^2 |H_A|^2 \ .
\end{eqnarray}
Here, $m_H$ and $\lambda_H$  are the SU(4) invariant mass and quartic, $\delta$ is the SU(4)-breaking but $\mathbb{Z}_2$ preserving quartic, and $\Delta m^2$ is the soft $\mathbb{Z}_2$-breaking Higgs mass.
We define $f^2 = 2(\langle H_A^2 \rangle + \langle H_B^2 \rangle)$, while $v^2 \equiv v_A^2 = 2 \langle H_A^2 \rangle = (246~\mathrm{GeV})^2$ is the measured  electroweak-symmetry-breaking Higgs vev in the visible sector.
The 125 GeV visible Higgs is identified with the pseudo Nambu-Goldstone boson of the approximate SU(4) symmetry. 
While the above potential is not SU(4)-symmetric, it is exact at the level of one-loop quadratic mass corrections as long as the $\mathbb{Z}_2$ is only broken softly. This constitutes the \emph{Twin Higgs Mechanism}, which stabilizes the Higgs mass up to scales of order 5-10 TeV without requiring new coloured states at LHC-accessible energies. 
Above 5-10 TeV, a supersymmetric or composite UV completion of the Twin Higgs is needed for a complete solution to the hierarchy problem~\cite{Chang:2006ra, Batra:2008jy,Craig:2013fga,Geller:2014kta,Barbieri:2015lqa}. Given this approximate low-energy protection, introducing additional $\mathbb{Z}_2$ breaking effects in the non-top Yukawa couplings does not ruin the MTH solution to the little hierarchy problem~\cite{Craig:2015pha}. 

The visible Higgs boson is given by $h \approx {\rm Re} (H_A) + \frac{1}{\sqrt{2}} \frac{v}{f} {\rm Re} (H_B) + \mathcal{O}(v^2/f^2)$, realizing a Higgs portal coupling to the dark sector and giving rise to invisible Higgs decays with branching ratio $\sim v^2/f^2$. This necessitates $f/v \gtrsim 3$, e.g. see Refs.~\cite{Craig:2015pha,Bittar:2024ryj} and references therein, but perturbing the theory away from the $\mathbb{Z}_2$-symmetric limit of $v/f = 1/\sqrt{2}$ imposes a tree-level tuning of severity $\sim f^2/v^2$ in the Barbieri-Giudice measure of tuning \cite{Barbieri:1987fn}. For natural Twin Higgs solutions to the little hierarchy problem, we therefore typically assume $3 \lesssim f/v \lesssim 10$, corresponding to $\Delta m^2 \sim \mathcal{O}(100~\mathrm{GeV})$ and $\delta \sim \mathcal{O}(0.05)$. 
See Refs.~\cite{Curtin:2021alk,Batell:2022tif} for a recent review.

The twin sector therefore features a smaller Fermi constant than the visible sector, $G_{F,B} = (v_A^2/v_B^2) G_{F,A}\simeq (v^2/f^2)G_{F,A}$.
Furthermore, the modified particle spectrum of the twin sector changes the running of the strong coupling, resulting in a confinement scale $\LQCDB$ that generally differs from its SM counterpart. This, in turn, alters the masses and interactions of the twin hadrons.
In our setup, the IR spectrum is further modified by allowing the non-top Yukawa couplings to depart from their SM values,
\begin{equation}
    \mathcal{L}_{X}\supset
    -y^u_{X}\bar{Q}_{X}\widetilde{H}_{X}u_{X}
    -y^d_{X}\bar{Q}_{X}H_{X}d_{X}
    + \mathrm{h.c.},
\end{equation}
where $Q$ denotes the left-handed quark doublet, $u$ and $d$ the right-handed up- and down-type quarks, and $X\in\{A,B\}$ labels the SM and twin sectors.
The departure of the twin Yukawa couplings from their SM values,
\begin{equation}
    y_A^f \neq y_B^f,
\end{equation}
is the primary source of $\mathbb{Z}_2$ breaking relevant for the phenomenology studied in this work. Generation indices are suppressed throughout for clarity; in practice, however, most of the phenomenology is controlled by the masses of the first-generation quarks.

The Lagrangian in Eq.~\eqref{eq: twin Lagrangian} generally predicts additional light degrees of freedom, most notably the twin photon and twin neutrinos. These contribute to the radiation density of the universe and therefore to the effective number of relativistic species, $N_{\rm eff}$, during BBN and at the time of the CMB. 
If the visible and twin sectors remained at the same temperature after decoupling, these additional relativistic species would contribute approximately $\Delta N_{\rm eff}\simeq 5.7$ \cite{Chacko:2016hvu,Craig:2016lyx}, in strong tension with the current bound $\Delta N_{\rm eff}\lesssim 0.17$ (95\% CL) \cite{AtacamaCosmologyTelescope:2025nti}.\footnote{See Refs.~\cite{Yeh:2026pil,Goldstein:2026iuu} for slightly stronger bounds by combining results from different surveys.} 
Throughout this work, we therefore assume that the twin sector is colder than the SM sector, suppressing the contribution of its light degrees of freedom to the late-time radiation density.\footnote{An alternative resolution is provided by the Fraternal Twin Higgs \cite{Craig:2015pha}, in which the light twin degrees of freedom are removed from the spectrum altogether through an explicit breaking of the $\mathbb{Z}_2$ symmetry. This would fundamentally alter the phenomenology which we study here and we do not explore this scenario further.} 
Such a temperature asymmetry can arise, for example, from late-time asymmetric reheating \cite{Chacko:2016hvu, Craig:2016lyx, Adshead:2016xxj, Alonso-Alvarez:2023bat}.

The cosmological evolution of the twin sector is controlled by only a small subset of the parameters appearing in the Lagrangian, primarily those governing the masses and interactions of the first-generation fermions.
The six quark-level Lagrangian parameters with the largest impact are:
\begin{itemize}
    \item $m_{u,B}$ and $m_{d,B}$: the masses of the twin up and down quarks,
    \item $\LQCDB$: the dark confinement scale,
    \item $m_{e,B}$: the twin electron mass,
    \item $v_B$: the vev of the twin Higgs, and
    \item $T_B/T_A$: the ratio of twin sector temperature to SM's.
\end{itemize}

In the IR, the fundamental parameters of the QCD Lagrangian are not directly accessible and are instead encoded in the properties of the low-energy hadronic spectrum. For example, the quark masses and confinement scale $\LQCD$ determine the average twin nucleon mass $m_{N}$, the twin neutron--proton mass splitting, and the twin pion mass. The precise mapping between the UV and IR parameters will be discussed in the following sections.

In our study of low-energy naDM phenomenology, we therefore parametrize the theory in terms of the IR quantities listed in \cref{tab: IR params}.\footnote{To clarify the notation, throughout this study, $m_N$ is reserved for the average of the proton and neutron mass, which are denoted by $m_p$ and $m_n$, respectively.} The table also summarizes the parameter ranges adopted in our numerical scans. These ranges are motivated by the requirements of addressing the coincidence and hierarchy problems, existing experimental and cosmological constraints, and theoretical correlations among the twin QCD observables. In particular, we focus on baryon masses close to their SM values, motivated by the observed similarity between the dark matter and baryonic energy densities.
The chosen parameter ranges encompass the exact $\mathbb{Z}_2$-symmetric limit, as well as the representative point $m_{N,B}\simeq 5\,m_{N,A}$, which realizes the observed DM-to-baryon energy density ratio with approximately $\mathbb{Z}_2$-symmetric baryon number densities.

\begin{table}
    \centering
    \renewcommand{\arraystretch}{1.4} 
    \begin{tabular}{|c|c|c|c|}
    \hline
   Parameter & Description & Minimum & Maximum \\
    \hline
       $m_{N,B}$  & Average twin nucleon mass & 0.5 GeV & 10 GeV 
       \\
    \hline
       $m_{\pi,B}$  & Twin pion mass & $\frac{m_{\pi,A}}{m_{N,A}}\,m_{N,B}$ & $0.56\,m_{N,B}$
       \\
    \hline
       $|\DeltMnp|$ & Twin neutron--proton mass splitting  &  $\frac{\Delta m_{np,A}}{2 m_{N,A}} \, m_{N,B}$ & $\sim 0.3\, m_{N,B} $ 
       \\
    \hline
       $m_{e,B}$ & Twin electron mass &  $\frac{m_{e,A}}{2 m_{N,A}}\, m_{N,B}$ & $\sim 0.3\, m_{N,B} $ \\
    \hline
       $G_{F,B}$ & Twin Fermi constant &  $10^{-2}\, G_{F,A}$& $G_{F,A}$
       \\
    \hline
       $\Delta N_{\rm eff}$ & Effective number of light new DOF 
       & 0.01& 0.15 \\
    \hline
    \hline
    $g_{a,B}$ & Effective nucleon axial coupling &  1.0  & 1.7\\
    \hline
    $\Delta m_{du,B}$ & Quark splitting & Eq. \eqref{eq:deltamnp_heavy} & Eq. \eqref{eq:deltamnp_light}\\
    \hline
    $B_D(m_{N,B}, m_{\pi,B})$ & Deuteron binding energy & Eq. \eqref{eq: binding energy deuteron} & Eq. \eqref{eq: binding energy deuteron} \\
    \hline
    $B_{He}(m_{N,B}, m_{\pi,B})$ & Helium binding energy & Eq. \eqref{eq: binding energy helium} & Eq. \eqref{eq: binding energy helium} \\
    \hline
    \end{tabular}
    \caption{The six parameters of the naDM IR framework  varied in this study  (\textbf{top}), and their explored ranges. Note that $m_{N,B} = 5~\mathrm{GeV}$ is our main benchmark solution to the matter coincidence problem.  We also consider four IR nuisance parameters of the model (\textbf{bottom}), which can in principle be determined from other microscopic parameters of the model, but theoretical uncertainties in their dependence motivate defining minimum and maximum values/expressions to bracket their possible range,  see the third and fourth columns. The upper end of our parameter space in $m_{\pi,B}$ is motivated specifically by the Froggatt--Nielsen Twin Higgs UV completion which we study in~\cref{sec: uv realization}.}
    \label{tab: IR params}
\end{table}

Additionally, we include in the table a number of nuisance parameters whose values are fixed to two extreme benchmark choices in order to bracket the theoretical uncertainties associated with determining them.
As in the SM, the cosmology of the twin sector is highly sensitive to several microscopic parameters, whose impact will be discussed in detail below. In particular, uncertainties in the determination of non-perturbative quantities can lead to qualitatively different cosmological histories and phenomenology. 
To account for these uncertainties conservatively, we consider benchmark extreme values for the relevant non-perturbative parameters in the twin sector, see the bottom half of \cref{tab: IR params}.

Let us now motivate the parameter ranges adopted in \cref{tab: IR params}.
The range of pion masses is primarily dictated by the availability of current lattice results for all quantities of interest (e.g. nuclear binding energies), although the upper end of the range is particularly motivated by the Froggatt--Nielsen UV completion of our naDM MTH setup sketched out in \cref{sec: uv realization}.
From Ref.~\cite{DeGrand:2019vbx}, the upper bound corresponds to a quark mass in QCD on the order of 200 MeV, well below the heavy quark regime typically associated with charm or bottom quarks.  In addition to constraints from lattice inputs, we prefer this lighter twin-quark regime in the context of the dark matter coincidence problem.  In the SM, the nucleon mass is predominantly set by the SM QCD confinement scale---so a twin nucleon mass similarly set by the twin QCD scale is the most natural solution, and further motivates us to avoid the heavy twin-quark regime.

For the twin neutron--proton mass splitting, the precise upper limit is ultimately determined by the range of viable underlying quark masses.  In the heavy-up-quark limit, letting $m_{d,B} \rightarrow 0$ at fixed $\DeltMnp$ gives the result $|\DeltMnp| / m_{N,B} \sim 2/3$. As discussed above, we do not approach this limit and so our maximal considered value $|\DeltMnp| / m_{N,B}\sim 0.3$ is somewhat lower.

For the twin electron mass and the twin neutron--proton mass splitting, we scan from values of $m_{e,B}/m_{N,B}$ and $|\DeltMnp| /m_{N,B}$ slightly below the SM ratios up to approximately $0.3$. 
This upper limit on the twin electron mass is motivated by the phenomenology of freeze-out. Once $m_{e,B}\gg |\DeltMnp|$, efficient annihilation and depletion of charged states, such as twin $\Delta^-$, becomes increasingly difficult. 
Under our assumptions, $m_{d,B}$ is the lightest twin quark, making the $\Delta^-$ the lightest of the twin $\Delta$ baryons. 
For sufficiently heavy pions, the decay
\begin{equation}
    \Delta^-_{B}\rightarrow \pi^-_{B}+ n_{B}
   \label{eq:Delta decay}
\end{equation}
can become kinematically forbidden, leaving the $\Delta_B^-$ to decay only through an off-shell pion to a twin electron. 
If the twin electron is also sufficiently heavy, the $\Delta_B^-$ becomes absolutely stable. 
In the upcoming calculations and solving Boltzmann equations, we assume $\Delta_B^-$ is unstable and decays promptly and neglect its effect on the governing Boltzmann equations.
Thus, our upcoming calculations are not applicable to the regime where $\Delta_B^-$ is stable or where the mass of the $\Delta_B^-$ is less than the mass of the twin neutron or proton. 
We therefore exclude regions of parameter space in which a stable $\Delta_B$ baryon is an unavoidable part of the late-time spectrum. The existence of such a stable charged relic could give rise to interesting phenomenology in this regime~\cite{Barbieri:2017opf} and is left for future work.

The range adopted for $G_{F,B}$ is motivated by the underlying Twin Higgs construction. With $f/v \lesssim 10$, as motivated by tuning arguments, this translates to
\[
G_{F,A} \gtrsim G_{F,B} \gtrsim \frac{G_{F,A}}{100},
\]
which ranges from the $\mathbb{Z}_2$-symmetric limit (excluded by Higgs coupling measurements in the MTH framework) to the boundary of the natural Twin Higgs regime.

For $\Delta N_{\rm eff}$, we consider two representative benchmark values, one of $0.15$ near current bounds \cite{AtacamaCosmologyTelescope:2025nti} and one of $0.01$ meant to represent something just below the cosmic-variance limited sensitivity of future measurements \cite{CMB-S4:2026mge}. 
Assuming all three twin neutrinos remain relativistic and the twin photon is massless, these benchmarks correspond to present-day temperature ratios $T_B/T_A$ of approximately 0.4 and 0.2.

Finally, we vary the axial coupling of the twin nucleon weak current $g_a$ around its SM value, which is measured to be equal to $1.2753 \pm 0.0013$ \cite{ParticleDataGroup:2024cfk} under the assumption that the vector current is conserved. 
This parameter controls the rate of weak interactions between twin neutrons and protons. Lattice results for the axial coupling were historically systematically low, due to a combination of systematic effects and the suppression of $g_a$ at heavy pion masses \cite{Chang:2018uxx}. 
Conversely, a non-relativistic large-$N$ quark model yields $g_a = 5/3 = 1.67$ \cite{Karl:1984cz}. This result is in apparent contradiction with the lattice results, which give lower values than the physical point for heavy pions. 
We find that our results remain stable as we vary $g_a$ within the range from \cref{tab: IR params}.
The remaining three nuisance parameters listed in \cref{tab: IR params} will be discussed in more detail in the following subsection.

Throughout this work, when we single out specific benchmark points in the IR parameter space, we 
use a combination of lattice results and heavy quark modelling to determine if 
this point can arise from a consistent set of UV parameters of the TH model.
The mapping between the IR and UV parameters is detailed in the upcoming sections.


\subsection{Dark Hadron Masses}\label{subsec: dark hadron masses}

In this section we establish the connection between the microscopic parameters of the twin sector and its low energy hadronic spectrum. While the late-time phenomenology is naturally described in terms of hadronic quantities, such as the twin nucleon and pion masses, several aspects of the early-universe evolution depend directly on the underlying quark masses and the confinement scale. 
For example, the twin quark masses are relevant for the cosmological history (discussed in \cref{sec: cosmo}), whereas direct detection observables (\cref{sec: terrestrial observables}) depend primarily on twin hadronic properties. 
Therefore, relating these UV and IR descriptions allows us to parametrize the model entirely in terms of IR quantities from \cref{tab: IR params}, while retaining the information needed to compute the cosmological evolution. 
Establishing this UV--IR mapping also allows us to identify regions of the nominal IR parameter space that are physically inaccessible, in the sense that no consistent choice of the underlying UV parameters can reproduce them.

Specifically, here we derive relations between the twin up- and down-quark masses and confinement scale, $(m_{u,B},m_{d,B},\Lambda_{\rm QCD,B})$, and the low-energy parameters $(m_{N,B},m_{\pi,B},\DeltMnp)$ that define our phenomenological parameter space. 
We combine lattice QCD results with expectations from the heavy quark limit to determine these relations and use these two as boundary cases to bracket the uncertainty in our estimates.

We begin by considering the limit of degenerate twin up and down quark masses, for which the twin proton and neutron are likewise degenerate up to twin electromagnetic corrections.  
We make extensive use of the analysis in Ref.~\cite{DeGrand:2019vbx}, which systematically reinterprets lattice QCD calculations at unphysical quark masses as predictions for QCD-like dark sectors. Following their approach, we perform numerical fits to lattice data for the average nucleon mass, $m_{N,B}$, and the $\Delta$ baryon mass, rescaling all dimensionful quantities by the ratio of the confinement scales in the two sectors, $\Lambda_B/\Lambda_A$, where, for brevity, we henceforth drop the ``QCD’’ label.\footnote{Note that the scale $\LQCD$ itself is not necessarily well-defined, as discussed in Ref.~\cite{DeGrand:2019vbx}; in practice we never make use of it explicitly, instead using the average nucleon mass $m_{N,B}$ and the ratio $m_{\pi,B}/m_{N,B}$ to set the scale of the twin QCD sector. However, $\Lambda_B / \Lambda_A$ is a useful shorthand for distinguishing the overall scale of the dark QCD sector relative to ordinary QCD.}
Further details on the numerical fits can be found in Appendix \ref{app: lattice fits}.

We estimate the twin pion mass using the Gell-Mann--Oakes--Renner (GMOR) relation~\cite{Gell-Mann:1968hlm,Gasser:1983yg},
\begin{equation}
\label{eq: pion mass GOR relation}
    m_{\pi,B}^2
    =
    \frac{2\langle \bar q q\rangle_B\, m_{q,B}}{f_{\pi,B}^2}
    =
    \left(\frac{\Lambda_B}{\Lambda_A}\right)
    \left(\frac{m_{q,B}}{m_{q,A}}\right)m_{\pi,A}^2 \, ,
\end{equation}
where $\langle \bar q q\rangle$ denotes the quark condensate and $f_\pi$ the pion decay constant. The second equality follows from the argument that $\langle \bar q q\rangle/f_\pi^2$ scales linearly with the confinement scale on dimensional grounds. In the presence of non-degenerate quark masses, the relation remains valid taking $m_{q,B}$ to be the average light quark mass. The GMOR relation breaks down for sufficiently heavy quark masses, but in the parameter space we study we do not approach this heavy-quark limit and we use this relation everywhere. In conjunction with the lattice results for the average nucleon mass, Eq.~\eqref{eq: pion mass GOR relation} can be inverted to determine the underlying UV parameters, $\Lambda_B/\Lambda_A$ and $m_{q,B}$, corresponding to a given pair of IR parameters $(m_{N,B},m_{\pi,B})$.

While the exact solutions we use for $\Lambda_B/\Lambda_A$ and $m_{q,B}$ are based on the extended fit to lattice data, a simple heuristic can be found by using the rough approximation $m_{N,B} \sim 1 \ \textrm{GeV} \left(\frac{\Lambda_B}{\Lambda_A}\right) + 3 m_{q,B}$, dividing the twin nucleon mass into the confinement and quark-mass contributions.  Combining this relation with Eq.~\eqref{eq: pion mass GOR relation} and expanding for small $m_{\pi,B}/m_{N,B}$:

\begin{equation}\label{eq: lambdaQCD dark}
     \left(\frac{\Lambda_B}{\Lambda_A}\right) \sim \frac{m_{N,B}}{\mathrm{GeV}}\left(1-0.5 \left(\frac{m_{\pi,B}}{m_{N,B}}\right)^2\right) \, .
\end{equation}
One can see that as the twin pion mass gets heavier with fixed twin neutron mass, the effective scale of QCD must drop. This relation is a useful shorthand when interpreting nuclear properties parameterized in terms of $\Lambda_B/\Lambda_A$, but we emphasize that it will not give numerically accurate results and should only be used as a way to qualitatively understand the scaling of other results.\footnote{We note that in this expression, heavier twin light-quark masses will increase $m_{\pi,B}/m_{N,B}$ and thus \emph{decrease} $\Lambda_B/\Lambda_A$.  This is not meant to indicate a renormalization-group relation between them; our expression is agnostic as to how $\Lambda_B$ arises and simply describes how our parameters are related to one another.}

We next consider adding the twin up-down quark mass splitting $\Delta m_{du,B}$, which has negligible impact on the estimated $m_{N,B}$ and $m_{\pi,B}$ but directly determines the twin neutron--proton mass splitting $\Delta m_{np,B}= m_{n,B}- m_{p,B}$. 
The mass splitting $\DeltMnp$ arises from two sources of isospin symmetry breaking: the light quark masses and their EM charges. The total mass splitting can be approximated as
\begin{equation}\label{eq:quark_splitting_general}
    \DeltMnp \simeq \Delta m_{m_q} +\Delta m_{EM} \, .
\end{equation}
We estimate the two contributions using two complementary approaches and use the resulting predictions to bracket the theoretical uncertainty associated with extrapolating away from the SM.

The first approach is based on the constituent quark model and is expected to be valid in the heavy-quark limit, where baryons can be treated as weakly bound states of non-relativistic constituent quarks. We follow the framework of Refs.~\cite{DeRujula:1975qlm,Manohar:1983md,Mitridate:2017oky,Asadi:2024bbq,Asadi:2024tpu}, adopting one of the variational ansatzes of Ref.~\cite{Asadi:2024tpu},
\begin{equation}
    \psi
    =
    \exp\!\left(
        -k\sum_{i<j}|\vec r_i-\vec r_j|
    \right),
\end{equation}
where $k$ is a variational parameter. This wavefunction is used to evaluate the expectation value of the constituent quark Hamiltonian \cite{DeRujula:1975qlm,Asadi:2024tpu}, with $k$ determined by minimizing the resulting baryon mass. 
In this model, the baryon mass is obtained by summing the constituent quark masses, the expectation value of the kinetic energy, and the pairwise strong and electroweak interactions. Since the strong interactions are flavour-blind, the leading contributions to the neutron-proton mass splitting arise only from (i) the constituent quark mass difference, and (ii) the electromagnetic part of the electroweak interactions, which depends on the quark charges and the spin-flavour wave functions of the proton and neutron. These two contributions directly map onto the two contributions in~\eqref{eq:quark_splitting_general}.

We computed the baryon masses for quark masses in the range
$10~\mathrm{GeV}<m_q<1000~\mathrm{GeV}$ for $\Lambda_B/\Lambda_A = (0.5,1,10)$. 
Following the variational method above in the quark model, we find that the leading contribution from the quark mass contribution is simply $\Delta m_{m_q}\simeq \Delta m_{du}$. The resulting EM contribution can be well approximated by:\footnote{The precise dependence has a weak scaling with $\Lambda_B/\Lambda_A$, which we neglect as it is sub-percent level.}
\begin{equation}
    \Delta m_{EM}^{heavy} \simeq -0.014 \,\alpha_{EM,B}\,m_{q,avg,B},
\end{equation}
which is subleading to the isospin contribution so long as $\Delta m_{du}/m_{q,avg}\gtrsim 10^{-4}$; thus,
\begin{equation}
    \DeltMnp \simeq \Delta m_{du,B}.
\end{equation}
This is in marked contrast to the SM limit, where $\Delta m_q/m_{q,avg}\sim 1$, and the two contributions are approximately equally weighted. This implies that as quarks get much heavier, eventually the EM correction becomes suppressed compared to the isospin breaking correction. Our parameter space never reaches these super-heavy limits, so this will always serve as an \textit{underestimate} of the EM correction. 

When $m_q\gg \Lambda_B$, we expect the dynamical scale that sets the electromagnetic contribution to be the quark masses. For $m_q \ll \Lambda_B$, we expect the relevant scale to be the confinement scale.\footnote{ Ref.~\cite{Gasser:1982ap} observes that the EM contribution to the mass splitting is roughly $\sim \alpha_{EM} \LQCD$. An estimate based on the electromagnetic form factor of a hadron in the static limit yields $\Delta m_{EM}= \alpha_{EM}5m/32$, which agrees with the observed mass splitting if $m\sim 800$ MeV, suggestive of the physical picture of vector mesons running in the hadron self-energy diagram loop.} 
For the light quark regime, we use lattice QCD to estimate the mass splittings. Ref.~\cite{BMW:2014pzb} computed the neutron--proton mass splitting on the lattice for pion masses between $200\ \textrm{MeV}\lesssim m_{\pi}\lesssim 400\ \textrm{MeV}$ to determine the relative contribution of the strong and electromagnetic mass corrections. 
Within lattice uncertainties, the strong isospin contribution $\Delta m_{m_q}$ was found to be equal to $\Delta m_{du,A}$ \cite{BMW:2014pzb}.
Under our assumed $\LQCD$ scaling, the electromagnetic contribution scales as 
\begin{equation}
    \Delta m_{EM}^{light}\simeq -137\ \textrm{MeV}\,\alpha_{EM,B} \left(\frac{\Lambda_B}{\Lambda_A}\right).
\end{equation}\label{eq:DeltaM_lattice}

\begin{figure}
    \centering
    \includegraphics[width=0.6\linewidth]{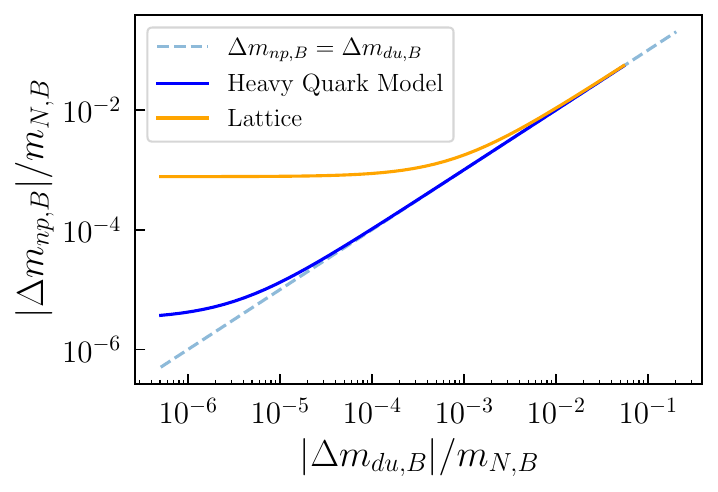}
    \caption{
    The baryon mass splitting as a function of the quark mass splitting for the two different regimes detailed in the text for the benchmark point $m_{N,B}= 5$ GeV and $m_{\pi,B}=1.7$ GeV. This implies $\Lambda_B/\Lambda_A\simeq 3.9$ and $m_{q,avg,B}\simeq150$ MeV. At large mass splittings, both estimates asymptote to the line $\DeltMnp = \Delta m_{du,B}$, while at small quark mass splittings they level off at the respective values of $\DeltMnp = \Delta m_{EM}$. 
    }
    \label{fig: DeltaMnp}
\end{figure}
This relation is expected to hold in the light-quark regime, where the nucleon masses are dominated by confinement dynamics. In this limit, the neutron--proton mass splitting receives comparable contributions from the strong isospin-breaking effects, and electromagnetic interactions. Our parameter space is much more similar to this regime than the above super-heavy limit, which we reach nowhere in our analysis. However, there is clear uncertainty in exactly which mass scale will rescale the EM contribution, and the turnover between the two regimes is not precisely known. Our final expressions are therefore:
\begin{align}
\label{eq:deltamnp_light}
    \DeltMnp^{light} &= \Delta m_{du,B} -\left( 137\ {\rm MeV}\right)\alpha_{EM,B}\left(\frac{\Lambda_B}
    {\Lambda_A}\right)
    ,
    \\
    \label{eq:deltamnp_heavy}
    \DeltMnp^{heavy} &= \Delta m_{du,B} -0.014\, \alpha_{EM,B}\,m_{q,avg}.
\end{align}

We use Eqs.~\eqref{eq:deltamnp_light} and \eqref{eq:deltamnp_heavy} as complementary estimates of the twin neutron--proton mass splitting as a function of the underlying quark mass splitting. 
Their predictions are compared in \cref{fig: DeltaMnp}. 
Note that since $\Delta m_{du,B}<0$ in our scenario, the two contributions have the same sign. For a fixed quark mass splitting, the lattice-based (light quark) estimate consistently provides a larger value of the twin neutron--proton mass splitting than the constituent quark model, indicating that the two methods approximately bracket the physically expected result.
In the following, we therefore use these two calculations to estimate the theoretical uncertainty in translating UV quark mass splittings into the corresponding IR hadron mass splittings.

Eqs.~\eqref{eq:deltamnp_light} and \eqref{eq:deltamnp_heavy}  can be trivially inverted to express the quark mass splitting in terms of the twin neutron--proton mass splitting. 
Combined with Eq. \eqref{eq: pion mass GOR relation} and the relations in \cref{tab: lattice fits}, this completes the mapping between the UV and IR descriptions, allowing us to determine the underlying twin quark masses and confinement scale, $\Lambda_{B}$, from the low-energy observables $(m_{N,B},m_{\pi,B},\Delta m_{np,B})$. Fortunately, because the exact twin quark masses have very small impacts on the cosmology, and the two estimates are in relatively close agreement at the large mass splittings of interest in later sections, we find the uncertainty in $\Delta m_{du,B}$ has little substantive impact on our conclusions. 

Finally, when the underlying quark masses and $\Lambda_B$ are modified substantially, the $\Delta_B$ baryon may have a mass comparable to the twin neutron and proton. 
Specifically, in the regime where $\Lambda_B\sim \textrm{few}\times m_{q,B}$ (equivalently, $m_{\pi,B}/m_{N,B}$ large) and the mass splitting between the two light quarks is large, $\Delta_B^-=ddd$ may have a mass similar to the heavy twin proton. 
To get an estimate of the $\Delta_B^{-}$ mass, we perform fits to the data in Ref.~\cite{DeGrand:2019vbx} but use the twin down quark mass in place of the average quark mass. 
While the $\Delta_B$ baryon being one of the low-lying baryonic states may be phenomenologically interesting, it is beyond the scope of this work and we simply exclude regions where $m_{\Delta^-,B}< m_{p,B}$ from our analysis.


\subsection{Dark Nuclear Binding Energies}\label{subsec: dark binding energies}

One of the primary goals of this work is to study the phenomenology of twin nuclear dark matter as a realization of aDM. To do so, we first develop a framework for calculating the properties of twin nuclei, with particular emphasis on their binding energies, which play a central role in determining the cosmological relic abundances.
In this section, we construct this framework by extrapolating existing lattice calculations of SM nuclear binding energies to the twin sector. 
For simplicity, here and throughout the remainder of this work, we give quantitative results only for twin deuterons and helium nuclei. 
We supplement this light-nucleus treatment in \cref{subsec: dark bbn} with an estimate of the largest atomic number $\Znum_{\max}$ that can be synthesized after the deuterium bottleneck, and only briefly comment on the possible role of heavier twin nuclei.

The SM deuteron is well-known to be very weakly bound, with a binding energy of approximately $2$ MeV. Since this is a fairly tuned system, predicting what will happen to the binding of a twin deuteron at larger than physical pion masses is highly challenging.

Lattice QCD calculations of di-nucleon binding energy exist, particularly results obtained at heavier-than-physical quark masses.  
Several early results indicated the presence of a bound di-nucleon state at such heavy masses \cite{NPLQCD:2012mex,Yamazaki:2012hi,NPLQCD:2013bqy,Yamazaki:2015asa,Berkowitz:2015eaa,NPLQCD:2020lxg}.  However, these are highly challenging calculations with subtle systematic effects.  More recent calculations in Refs.~\cite{BaryonScattering:2025ziz,Detmold:2026vjx,Detmold:2026shh,Wang:2026kab,Chakraborty:2026zeq,Zhang:2026cco,BaryonScattering:2026voj} have cast doubt on the earlier results, finding strong dependence on choice of operator basis and indicating that two-nucleon bound states may not actually exist at heavy quark masses. 
This would imply that as the twin pion-to-neutron mass ratio increases, the twin deuteron becomes unbound; in this case, twin BBN is cut off from proceeding.

In light of the unsettled status of lattice calculations for di-nucleon binding, we will proceed by studying both possibilities: that the dark deuteron is unbound, in which case dark BBN does not proceed, and that the dark deuteron is bound, in which case we will attempt to model its binding energy as described below.  Further study of the two-nucleon scattering problem in lattice QCD at unphysically heavy pion masses would be useful in shedding light on this question and making the naDM parameter space more robust.

Assuming the nuclei are bound, we need to estimate the nuclear binding energies in order to obtain the abundances of the lightest nuclei. The data that do exist have a wide scatter. Rather than try to parameterize the uncertainty, we bracket the available parameter space for the binding energies as a function of the pion mass. Given the large uncertainties already present in the binding energies of the deuteron, we do not model binding energies above helium.

For the deuteron, we use the expressions from Ref.~\cite{Chacko:2018vss}  based on data from Ref.~\cite{Savage:2015eya}:

\begin{equation}\label{eq: binding energy deuteron}
    \begin{aligned}
        B^{\rm min}_{D,B}  & = -0.66 \ \mathrm{MeV} \left( \frac{\Lambda_B}{\Lambda_A}\right)+ 0.021 \left(\frac{m_{\pi,B}}{\ \mathrm{MeV}}\right), \\
        B^{\rm max}_{D,B} & = - 9.2\ \mathrm{MeV}\left( \frac{\Lambda_B}{\Lambda_A}\right)+ 0.084 \left(\frac{m_{\pi,B}}{\ \mathrm{MeV}}\right). \\
    \end{aligned}
\end{equation}
Both of these expressions reduce to the physical value of $B_{D,A}= 2.2$ MeV at the SM point. Note that once $m_{\pi,B}$ and $m_{N,B}$ are selected, $\Lambda_B/\Lambda_A$ can be determined from the results of \cref{subsec: dark hadron masses}; thus this expression is implicitly a function of the pion and nucleon mass.  As noted above, in the present context of uncertainty from lattice results, the use of these fits (which we adopt as our baseline) applies only in the case where the dark deuteron is bound; we also discuss the alternative scenario of the unbound deuteron in the following studies.

We perform a similar fit to the data for the helium binding energy, drawn from Ref.~\cite{Kirscher:2015yda} for a combination of pionless EFT calculations and lattice calculations: 
\begin{equation}\label{eq: binding energy helium}
    \begin{aligned}
        B^{\rm min}_{He,B} & = 25.8\ \mathrm{MeV}\left( \frac{\Lambda_B}{\Lambda_A}\right)+0.018\left(\frac{m_{\pi,B}}{\ \mathrm{MeV}}\right), \\
        B^{\rm max}_{He,B} & =  11.7\ \mathrm{MeV}\left( \frac{\Lambda_B}{\Lambda_A}\right)+0.12 \left(\frac{m_{\pi,B}}{\ \mathrm{MeV}}\right). \\
    \end{aligned}
\end{equation}
As discussed above, we carry out our analysis using $B_{X,B}^{\rm min}$ and $B_{X,B}^{\rm max}$ as two benchmark choices that bracket the uncertainty associated with the nuclear binding energies.

Since twin neutrons are the primary DM candidate in our parameter space, we should also consider the possibility of multi-neutron bound states forming. These do not exist in the SM because the di-neutron state is marginally unbound, preventing the formation of heavier multi-neutron composites. If the binding energy of the deuteron increases with increasing pion masses, as assumed by the estimates above, then it is plausible that the di-neutron state may be bound in the twin sector. 
If the twin proton population is depleted, the relic twin neutron population could fuse into di-neutron and larger multi-neutron states.
Heavy neutron-rich twin isotopes are likely to be more stable than their SM cousins since the twin neutrons are not subject to beta decay, allowing potentially many neutrons to attach to nuclei with relatively small proton number $\Znum$. The limit to the number of twin neutrons is set by the balance of Pauli exclusion and the strong nuclear force. 
While we do not focus on these scenarios, the possibility has consequences for our phenomenology, which we comment on when relevant. 

As implied by this discussion and discussed further in the next sections, the uncertainty in these binding energies propagates into the predicted dark matter composition. In particular, the deuteron binding energy has the largest impact on the final abundance of naDM by determining the efficiency of deuteron formation. 
By contrast, the helium binding energy primarily controls how the nuclear component is partitioned between deuterons and helium (and potentially heavier twin elements). While this has a smaller effect on the overall DM composition (TNDM vs naDM), it can still lead to quantitatively different phenomenology discussed below.


\subsection{Nuclear and Atomic Stability}\label{subsec:stability}

The naDM composition depends on two distinct stability conditions. In the early universe, the most relevant criterion is the \textit{nuclear} stability; this determines if nucleosynthesis can form the nuclear state. 
However, if twin recombination is efficient, the late time  abundance will also depend on the \textit{atomic} stability.
The capture of a bound twin electron by a twin proton in the nucleus can cause the atom to decay; thus, twin atoms have additional stability conditions above those of their nuclei.
We discuss these important qualities in that order and explore their implications for DM phenomenology. Along the way, we will show that twin hydrogen atoms are never stable. 

We write the mass of a twin nucleus X with $\Znum$ protons and $\Anum-\Znum$ neutrons as $m_{X,B}= \Znum  m_{p,B}+ (\Anum-\Znum)m_{n,B} -B_{X,B}$. Note that throughout $\DeltMnp <0$, i.e. twin neutrons lighter than twin protons. 
A nucleus is stable if all kinematically accessible decays are forbidden. Two classes of decay are relevant: strong decays into light nuclei with the same total charge, and $\beta^+$ decay. The latter is forbidden if
\begin{equation}
    B_{X,B} - \max \left\{\sum_Y B_{Y,B}\right\} > -\DeltMnp - m_{e,B} \, ,
    \label{eq:stability_cond_beta}
\end{equation}
where $Y$ runs over all daughter nuclei in a particular decay channel. This condition has to be checked for all possible decay routes for the $X$ nucleus. 

For the twin proton, ($B=0$, $Y=n_B$), this reduces to 
\begin{equation}
    m_{e,B}> -\DeltMnp \, ,
    \label{eq:stability_cond_p}
\end{equation}
i.e. the proton is stable when the twin electron is heavier than the mass splitting between the twin protons and neutrons. 
Deuterons have an additional energy barrier from the nuclear binding energy, thus they are stable to beta decay if
\begin{equation}\label{eq:stability_cond_D}
    B_{D,B}> -\DeltMnp - m_{e,B} \, .
\end{equation}

Twin alpha particles (He nucleus) could, in principle, undergo strong decay into triton nuclei (nnp) or helium-3 nuclei (ppn) and the appropriate baryon. Helium-3 is expected to be subdominant since helium-3 is more weakly bound due to the additional Coulomb repulsion compared to triton. Strong stability of alpha particles then requires both 
\begin{equation}
\begin{aligned}
        B_{He,B} &> B_{T,B},
        \\
        B_{He,B} &> 2 B_{D,B} \, .
    \label{eq:stability_cond_He}
\end{aligned}
\end{equation}
with $B_{T,B}$ the nuclear binding energy of the triton nuclei. The results of Ref.~\cite{Kirscher:2015yda} imply that $B_{He,B}/B_{T,B}\sim 2$, and the decay to triton does not occur. Thus the only decay pathway we expect to be relevant is decay to two deuterons.

Alpha particles may also decay weakly via $\beta^+$ decay. The stability condition requires 
\begin{equation}\label{eq:stability_cond_He_beta}
    B_{He,B} -\max(B_{D,B},B_{T,B})> -\DeltMnp - m_{e,B}
\end{equation}
This is implied by deuteron stability, Eq.~\eqref{eq:stability_cond_D}, provided that $B_{He,B}-\max(B_{D,B},B_{T,B})>B_{D,B}$. If $B_{T,B}<B_{D,B}$, this follows directly from stability against strong decays, Eq~\eqref{eq:stability_cond_He}. 
In the limit that $B_{He,B}\simeq 2 B_{T,B}$, the same logic applies. If the proportionality constant is $<2$ as implied by a few results in Ref.~\cite{Kirscher:2015yda}, this condition may mildly differ from the deuteron case. In practice, we take the $\beta^+$ stability of the alpha particle to follow from that of the deuteron, and do not attempt a detailed modelling of the triton binding energy.

After the twin recombination epoch, these nuclei can capture twin electrons to form twin neutral atoms, which have their own stability conditions. 
Even if nucleosynthesis proceeds and twin nuclei form, these nuclei may still not survive until today, as they may decay via the process of \textit{electron capture}. 
Simple estimates show that if this decay channel is open, the decay rate can be fast compared to the Hubble time at the time of recombination, see below. 

After the formation of the twin atom with atomic number $\Znum$, one of the twin protons in the nucleus can capture 
an orbital electron and decay to another twin atom with atomic number $\Znum - 1$ through the process ${}^{A}_{Z}\mathrm{X}_B+e_B^-\to{}^{A}_{Z-1}\mathrm{Y}_B+\nu_B$. 
As the twin neutron is lighter than the twin proton, the process 
\begin{equation}
    p_B^+ + e_B^- \rightarrow n_B+\nu_B 
\end{equation}
is always exothermic, implying that a bound state of a twin electron and proton is always unstable in the parts of parameter space we study. 
Specifically, denoting the electronic binding energy of the twin hydrogen atom by $B^e_{H,B}$, the criteria for twin hydrogen atoms to be stable is 
\begin{equation}
    B_{H,B}^e > - \DeltMnp + m_{e,B}.
\end{equation}
In the SM, the quantity $\Delta m_{np,A}=1.29$ MeV is positive and greater than the electron mass, so the right hand side is negative and hydrogen is stable. In our scenario, the mass splitting $\DeltMnp<0$, thus the right hand side is always greater than zero. 
As the electronic binding energy can never exceed the electron mass, see Eq. \eqref{eq: hydrogen binding} below, the \textit{twin hydrogen atom is always unstable to decay.} 

The same is not necessarily true of heavier twin elements, as their nuclear binding energies counterbalance the process. Treating the electronic binding energy as negligible, twin deuterium atoms are stable if
\begin{equation}\label{eq:stability_cond_D_EC}
    B_{D,B}> -\DeltMnp + m_{e,B}.
\end{equation}

The case for the helium atomic stability is slightly more involved.
As in the nuclear case, helium has multiple decay pathways: ${}_2^4\textrm{He}_B+e_B^-\to {}_1^2\textrm{D}_B+2n_B+\nu_B$ and $^4_2\textrm{He}_B+e_B^-\to {}_1^3\textrm{T}_B+n_B+\nu_B$. For the electron capture decays to be closed, we must have 
\begin{equation}\label{eq:stability_cond_He_EC}
        B_{He,B}-\max (B_{D,B},B_{T,B})  > -\DeltMnp + m_{e,B}. 
\end{equation}
As in the case of twin helium beta decay of Eq.~\eqref{eq:stability_cond_He_beta}, we find that if the twin alpha particle is stable to decay to deuterons, and twin deuterium atoms are stable, then twin helium atoms are also stable via this decay path. As in the beta decay case, we use the deuterium criteria in our analysis, as the triton criterion is unlikely to substantively differ. We do not consider double electron capture in this work.

The twin helium atomic stability discussion above illuminates the general stability condition for heavier twin atoms denoted by X:
\begin{eqnarray}
    B_{X,B}-\max \left\{\sum_Y B_Y\right\} >  -\DeltMnp + m_{e,B}
 \end{eqnarray}
where $Y$ runs over all possible combinations of daughter nuclei and $\max$ is taken over all possible decay routes. This includes the exotic possibility of twin multineutron bound states, discussed in~\cref{subsec: dark binding energies}.

The many possible reaction pathways and unknown binding energies preclude a quantitative treatment beyond helium. Nevertheless, some qualitative conclusions can be drawn. 
Generically, nucleosynthesis produces a hierarchical composition dominated by a single nuclear species. The post-recombination naDM composition is then governed primarily by the fate of this dominant species: whether \textit{it recombines efficiently} and whether \textit{the resulting twin atoms are stable} against electron capture.

First suppose that recombination is efficient. For a neutral parent that remains a single nucleus after capture, the daughter is also neutral: capturing one of the \(\Znum\) electrons reduces both the nuclear charge and the number of electrons to \(\Znum-1\). An unstable daughter may undergo further captures, producing a cascade that terminates at the first atomically stable state or, if none exists, in free twin neutrons (or stable multineutron bound states, if present). Parametrically, the capture rate is
\begin{equation}
    \Gamma_{\rm EC}\sim G_{F,B}^{\,2}Q^2|\psi_e(0)|^2
    |\mathcal{M}_{\rm nuc}|^2
    \label{eq:elec_capture_rate}
\end{equation}
where \(Q\) is the energy release, $\psi_e(0)$ the electron wavefunction at the origin (nucleus), and \(\mathcal{M}_{\rm nuc}\) is the nuclear transition matrix element. For hydrogen, the electron capture rate is approximately
\begin{equation}
     \Gamma_{\mathrm{ EC},H} \sim \frac{1}{\pi^2} G_{F,B}^2 m_{e,B}^3 \alpha_B^3 (-\Delta m_{np,B} + m_{e,B})^2.
\end{equation} 
Although the rate for nuclei can be suppressed near threshold or by nuclear selection rules, this indicates that open channels are typically much faster than cosmological timescales in our parameter space and therefore do not leave cosmologically stable relics. Three outcomes are then possible:
\begin{itemize}
    \item If the twin atoms of the dominant species produced by nucleosynthesis are stable against twin electron capture, they remain the dominant late-time naDM component.

    \item If the dominant twin atoms are unstable but their capture cascade reaches a stable lower-\(\Znum\) atom, that endpoint dominates the late-time naDM abundance.

    \item If the cascade contains no stable atomic endpoint, the naDM component ultimately converts into free twin neutrons or, if stable multineutron bound states exist, into those states; the late-time atomic naDM fraction then vanishes.
\end{itemize}
Importantly, stable twin atoms heavier than deuterium or helium can survive even where atomic deuterium and helium are unstable. Subdominant primordial species undergo analogous cascades and may leave residual naDM populations if they reach stable atomic endpoints.

Unless otherwise stated, our baseline assumption is that recombination is efficient: nuclei formed during twin BBN rapidly capture the available twin electrons and initially form neutral twin atoms. 
Electron capture may subsequently destabilize some of these twin atoms and alter the late-time composition. 

If recombination is inefficient, which is the case for very heavy twin electron masses, the dominant twin nuclear species may persist as ions even when the corresponding twin atoms would be unstable.\footnote{Ref.~\cite{Kaplan:2009de} found that if atomic DM makes up 100\% of DM, for SM-like $\alpha_{EM,B}$ and $m_{p,B}= 10$ GeV, greater than 10\% of electrons will survive recombination as free ions for $m_{e,B}\gtrsim 1$ GeV, however, recombination may become inefficient for lighter twin electrons as well if the initial twin proton fraction is small. The rates may also be somewhat modified for heavier nuclei.}
Ions may also arise after electron capture.
The sudden change in nuclear charge and the recoil of the daughter atom (or atoms, in the case of fragmentation) can eject bound electrons. Fragmented daughters may also redistribute bound electrons unevenly, resulting in ionized states.
At the low post-recombination densities, the resulting ions may fail to recombine, and survive as charged relics. 
A detailed treatment of the ionization history through recombination, electron-capture cascades, and structure formation is beyond the scope of this work. 

Given these uncertainties, particularly in the binding energies of nuclei heavier than helium, we restrict our quantitative atomic-stability analysis to deuterium and helium. Where both are unstable to electron capture, we remain agnostic about the surviving non-neutron 
heavy twin element ``X'' component, which may contain ions and stable heavy twin atoms. Determining this late-time composition and its phenomenology warrants a dedicated future study. Within this restricted scope, we can turn to the mapping of the possible states in our parameter space.


\subsection{Phase Diagram of naDM}\label{subsec: phase diagram}

In this section, we put the results of the previous subsections together to determine which regions of the IR parameter space could admit a non-negligible naDM component today. 
We track twin neutrons (TNDM), as well as twin atoms D, He and heavier elements X (X=Li or heavier), through twin nucleosynthesis and recombination. 
Following the discussion from the previous section, depending on the stability of various twin nuclei (post-BBN) and twin atoms (post-recombination), we can end up with different scenarios of DM candidates today.
The most relevant IR parameters controlling this phase structure are the twin neutron--proton mass splitting, $\DeltMnp$, the twin nuclear binding energies, and the twin electron mass. 
Together, these quantities determine the stability of twin nuclei and atoms, allowing us to identify seven distinct phases in the theory (summarized in~\cref{tab: phase diagram regions}).
As we discuss in the following sections, cosmological and astrophysical constraints likely require the twin neutron to constitute the dominant dark matter component in all viable regions of parameter space. Nevertheless, depending on the stability of twin atoms, a subdominant naDM component may also be present, leading to qualitatively different phenomenology. 
As we mentioned above, our default assumption throughout this work is that recombination is efficient, though we discuss some effects of residual ions (which can also arise as a result of electron capture cascades) as well.

Depending on the stability of intermediate nuclei, heavy twin nuclei may also be synthesized in the primordial universe. Their relative abundances depend sensitively on their binding energies and on the relevant nuclear reaction rates. To streamline our analysis, we explicitly treat D and He, and later estimate in~\cref{subsec: dark bbn} whether nucleosynthesis can continue to heavier nuclei. The complete reaction network and isotope abundances are beyond the present scope.

The different possible phases of the theory, depending on whether various twin nuclei and atoms are stable or not, are collected in \cref{tab: phase diagram regions}. 
Since twin hydrogen atoms are always unstable to electron capture, we never have a H-naDM component.
Late-time ionic relics are possible in most regions of parameter space, as discussed in~\cref{subsec:stability}---whether due to atomic instability to electron capture or inefficient recombination. 
We will highlight them only in scenarios where no atomic candidates are possible, and leave their possible presence in other regions implicit. Charged relics from inefficient recombination are denoted \Ions, while charged relics from partial fragmentation of heavy twin atoms (due to the electron capture process) are denoted \XIons. 
We will proceed case-by-case through the scenarios in~\cref{tab: phase diagram regions}. 

\begin{table}[t]
\centering
\resizebox{\columnwidth}{!}{
\begin{tabular}{|c|c|c|c|l|}
\hline
\textbf{Deuteron} & \textbf{Alpha particle} & \textbf{Deuterium} & \textbf{Helium} & \textbf{DM candidates} \\
\hline
Unstable & \multicolumn{3}{c|}{--} & \TNDM\ (+ $p_B^+$) \\ \hline
\multirow{6}{*}{Stable}
  & \multirow{2}{*}{Unstable}
    & Unstable & -- & \TNDM\ (+ \Ions) \\ \cline{3-5}
  & & Stable   & -- & \TNDM + \DnaDM\ (+ \Ions)\\ \cline{2-5}
  & \multirow{4}{*}{Stable}
    & \multirow{2}{*}{Unstable} & Unstable & \TNDM + \XnaDM \ (+ \Ions) + \XIons \\ \cline{4-5}
  & &                           & Stable   & \TNDM + \HenaDM\ (+ \Ions) + \XIons \\ \cline{3-5}
  & & \multirow{2}{*}{Stable}   & Unstable & \TNDM + \DXnaDM\ (+ \Ions) + \XIons \\ \cline{4-5}
  & &                           & Stable   & \TNDM + \DHenaDM (+ \Ions) + \XIons \\ \hline
\end{tabular}
}
\caption{ 
The seven distinct regions of our parameter space. When a twin nucleus is unstable, it impedes the production chain of heavier twin elements.
When the corresponding atom is unstable, that state cannot appear as naDM, even if it forms in the early universe. 
If the twin proton is stable, it is possible for it to remain in ion form to late times, but hydrogen will always decay. 
The two different ion contributions are from the two processes discussed in the main text which may result in late time ion populations: in brackets, ion relics from inefficient recombination (possible at least in principle in all scenarios) and without brackets to account for the possibility that heavier twin atoms are unstable to electron capture and decay into lighter, stable ions. 
}
\label{tab: phase diagram regions}
\end{table}

If the twin deuteron is unstable, no heavier naDM states can form and the dark sector consists only of \TNDM. 
Even if heavier twin nuclei are stable in this region, in the absence of an intermediate deuteron state in the spectrum, they are not expected to form with an appreciable abundance. 
(Alternative production channels would require several twin nucleons to come together simultaneously and are therefore extremely suppressed.) 
There is only one possible charged relic, a stable twin proton. 
Since deuteron decay is proton decay with an additional energy barrier, a stable twin proton relic requires the twin deuteron is not just unstable but \textit{unbound} and that recombination be inefficient.
If the twin proton is unstable no ionized relics are possible.

If the twin deuteron is stable, but the twin alpha particle is unstable, then twin nucleosynthesis terminates at deuterons. The twin proton lifetime can strongly affect the final deuteron abundance. 
If the lifetime is too short, the twin protons decay before an appreciable amount of deuterons can form during twin BBN. By contrast, if the twin protons are sufficiently long-lived, a larger fraction can be incorporated into twin deuterons, where nuclear binding stabilizes them in analogy with the stabilization of neutrons inside SM nuclei.
If the twin deuterium atom is unstable, then this scenario can yield only \Ions\, whereas if twin deuterium is stable, then TNDM may be accompanied by a deuterium naDM component, denoted by \DnaDM. 

If both the deuterons and alpha particles are stable, we will have four additional different scenarios (see the bottom four rows in \cref{tab: phase diagram regions}).
The situation grows even richer with the possibility that the fusion chain proceeds above alpha particles. 
In the SM, heavier nuclei can be formed by incrementally adding free protons to a nucleus because there are more free protons than there are neutrons. In the twin sector, the inverse is true, and after the initial twin deuteron formation, the population of free twin protons will be greatly suppressed as the majority are incorporated into nuclei. 
Thus, direct fusion of two nuclei of the same species will dominate the production mechanism. 
If a twin element in the required formation chain is unbound, the entire sequence of reactions leading to heavier nuclei may be interrupted.
The nuclear stability thus determines the spectrum of twin nuclei that can be formed during twin BBN.

In the scenarios with stable deuterons and alphas, the composition of the surviving twin atoms after recombination depends on the atomic stability. 
If both twin deuterium and helium are unstable, then the only possible atomic components are heavier twin atoms (Li or heavier), denoted by \XnaDM. As a reminder, we remain agnostic about whether such stable heavier twin elements exist and postpone their detailed study to future works---here we merely point out the existence of the possibility.
If twin helium is stable but twin deuterium is not, then D-naDM cannot form, so the scenario is denoted \HenaDM, i.e., twin helium or heavier naDM are possible. 
If twin deuterium is stable but twin helium is not, then we may have deuterium or heavier naDM, but not twin helium. We denote this scenario \DXnaDM, but note that it cannot appear in our phase diagrams, since we approximate the twin helium stability using only the deuteron decay chain. If electron capture decays to triton were more favored, it is possible that the twin helium atom is unstable, but a relic twin deuterium population remains; see \cref{subsec:stability}. 
Finally, if both twin helium and deuterium are stable, then the full spectrum of atomic states is possible, \DHenaDM, including the two light elements and potentially heavier stable twin atoms.

An intriguing consequence of any naDM phase that contains heavy twin elements X is the possible formation of \emph{primordial twin dust} via crystallization after twin-recombination. 
Twin dust may evade late-time self-interaction constraints, as its  geometric cross section per unit mass decreases with grain size; see \cref{sec: bullet cluster}. However, DAO constraints will still apply. This possibility requires further dedicated study.

Throughout most of the parameter space, the binding energies are sufficiently well-separated that a single nuclear species is likely to dominate the dark matter abundance. 
Near phase boundaries, however, where nuclear stability conditions are only marginally satisfied, multiple species may coexist with comparable abundances. A quantitative calculation of these mixed-abundance regions requires solving the full nuclear reaction network and is beyond the scope of this work.

\begin{figure}
    \centering
    \includegraphics[width=0.7\linewidth]{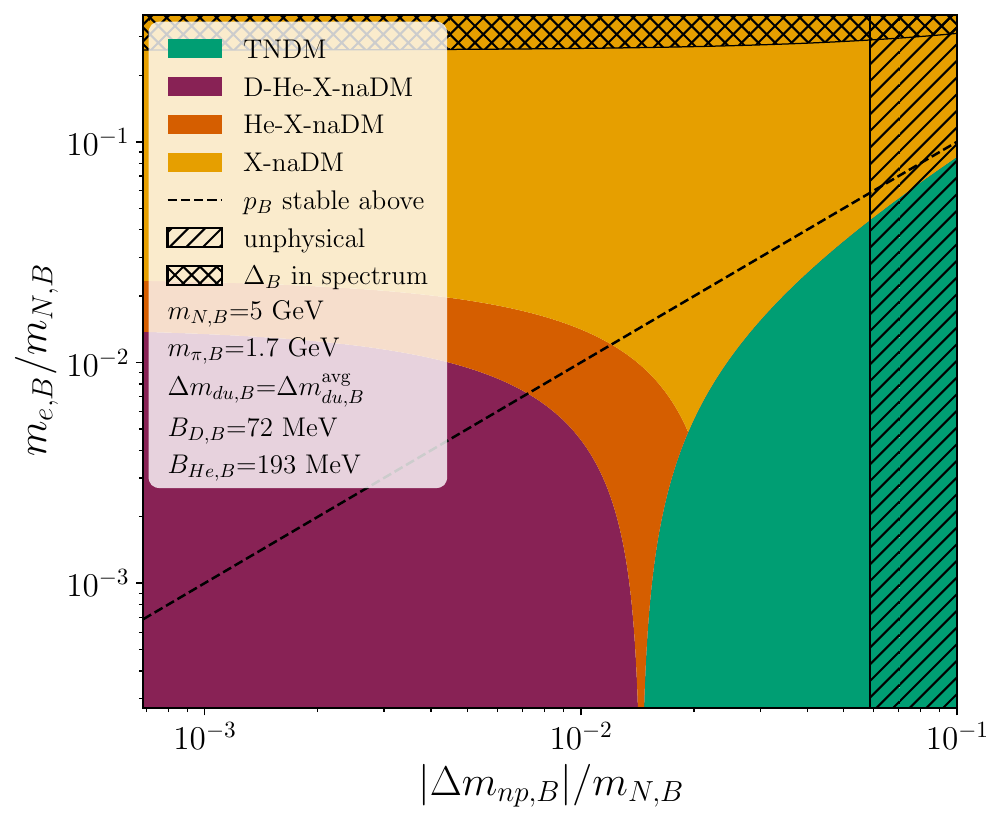}
    \caption{Different phases of the naDM model (coloured regions) as a function of the IR parameters defined in~\cref{tab: IR params} in the plane of twin neutron--proton mass splitting and twin electron mass, for our \emph{canonical naDM benchmark point} with $m_{N,B} = 5~\mathrm{GeV}$ and other parameters chosen as the average of the minimum and maximum of their respective considered ranges. 
    For small $|\DeltMnp|+m_{e,B}$ (the \textcolor{D_He_naDM_new}{burgundy} region), twin deuterium and helium atoms are stable post-recombination and can be viable naDM candidates. In the \textcolor{He_naDM_new}{orange} region, only twin helium or heavier atoms are viable naDM candidates. For larger $|\DeltMnp|+m_{e,B}$ in the \textcolor{X_naDM_new}{yellow} region, only potential heavy twin elements X could form stable twin atoms. If the twin deuteron is unstable, no nuclei form and the scenario is pure TNDM (the \textcolor{N_SIDM}{green} region). The hatched regions are either unphysical or outside the scope of our approximate BBN treatment. Above the dashed black line the twin proton is stable.
    }
    \label{fig: phase diagram benchmark}
\end{figure}

Using the stability conditions of Eqs.~\eqref{eq:stability_cond_p}--\eqref{eq:stability_cond_He_EC}, together with the binding-energy fits of Eqs.~\eqref{eq: binding energy deuteron}--\eqref{eq: binding energy helium}, we can determine the different TNDM--naDM phases listed in \cref{tab: phase diagram regions} for different choices of the naDM IR parameters in \cref{tab: IR params}.
We demonstrate this in~\cref{fig: phase diagram benchmark}, where we show the phase structure of naDM in the plane of $|\DeltMnp|/m_{N,B}$ vs. $m_{e,B}/m_{N,B}$ for our \textit{canonical naDM benchmark}, with $m_{N,B} = 5~\mathrm{GeV}$ and other parameters defined by the average of the minimum and maximum values (or expressions) for each parameter in \cref{tab: IR params}.

As discussed before, for large enough $m_{e,B}/m_{N,B}$ values the $\Delta_B^-$ baryon becomes stable, and our analysis which neglects their abundance breaks down. We simply label this region in figures as ``$\Delta_B$ in spectrum" and leave its analysis for a future study.

Achieving large values of $|\DeltMnp |$ while keeping $m_{\pi,B}$ and $m_{N,B}$ fixed requires increasingly large up--down quark mass splittings. Beyond a certain point, this cannot be realized with physical quark masses ($m_{q,B} > 0$), and we indicate these unphysical regions with hatched regions labelled ``unphysical'', as discussed in \cref{subsec: dark hadron masses}.

We  can see that increasing the twin electron mass eventually stabilizes the twin proton; free twin protons are a viable ionic candidate above the proton stability line. 
For fixed $m_{e,B}/m_{N,B}$, increasing $|\DeltMnp|/m_{N,B}$ makes it progressively more favourable for twin protons to decay into twin neutrons rather than remain bound inside a nucleus, as dictated by Eqs.~\eqref{eq:stability_cond_D},~\eqref{eq:stability_cond_He_beta}. 
Beyond a critical value of $|\DeltMnp|-m_{e,B}$, the deuteron becomes unstable, impeding the formation of heavier nuclei and leaving a purely TNDM scenario. 
Over the parameter space considered here, the transition occurs at $|\DeltMnp |/m_{N,B} \sim 10^{-2}$ for $m_{e,B}/m_{N,B} \lesssim 10^{-3}$, and can be pushed to $|\DeltMnp |/m_{N,B} \sim 10\%$ for $m_{e,B}/m_{N,B} \sim 10\%$.

Increasing $|\DeltMnp|+ m_{e,B}$ makes electron capture in twin atoms increasingly favourable. At our benchmark, deuterium becomes unstable for $m_{e,B}/m_{N,B}\gtrsim 10^{-2}$, while helium is stable for electron masses slightly larger. Above this value of the electron mass, the only possible naDM candidates are hypothetical heavy stable atoms (denoted by X).

\begin{figure}
    \centering
    \begin{tabular}{c}
     \includegraphics[width=\linewidth]{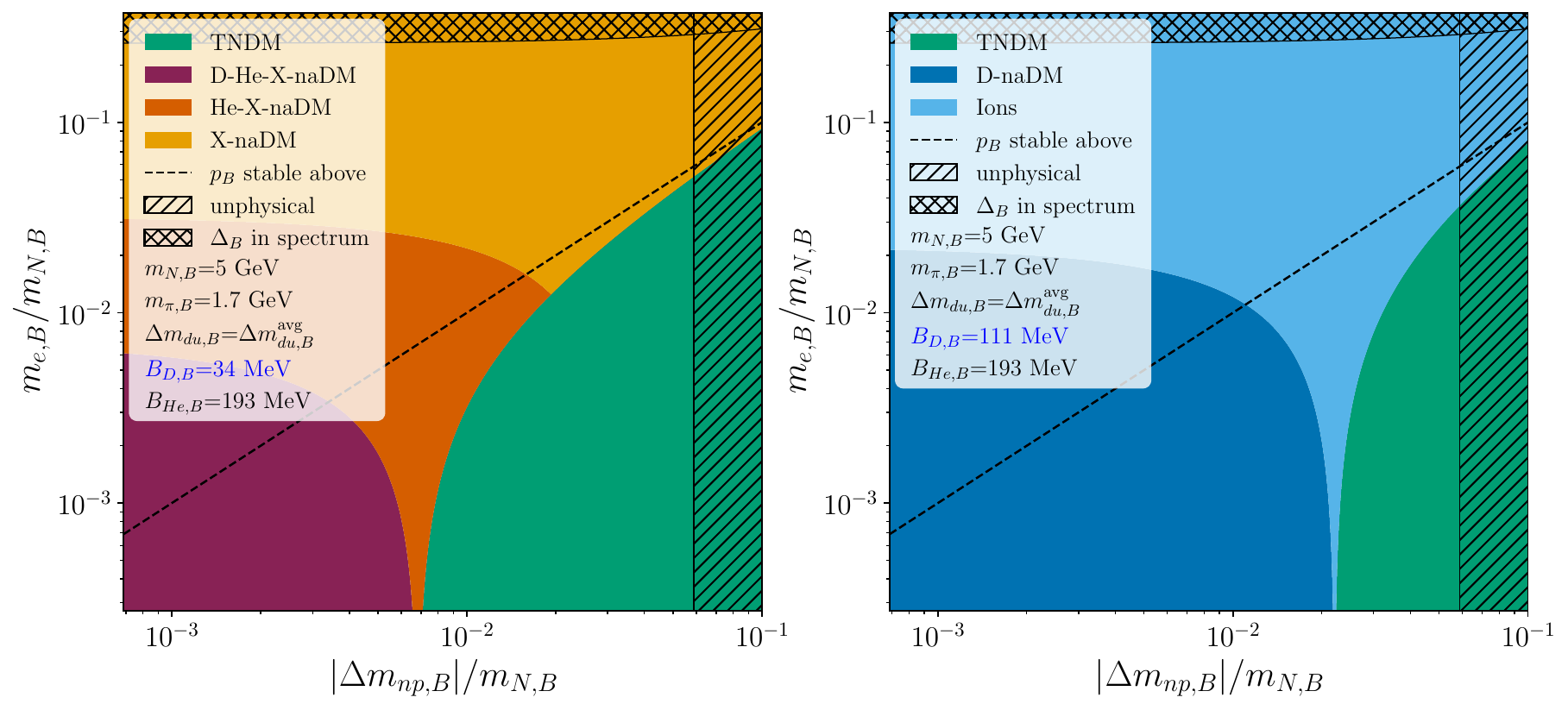}  
    \end{tabular}
    \caption{
    Different phases of naDM as defined by the IR parameters of \cref{tab: IR params} as we vary $B_{D,B}$ compared to the canonical benchmark of \cref{fig: phase diagram benchmark}: $B_{D,B}$ given by $B_{D,B}^\mathrm{min}$ (\textbf{left}) or by $B_{D,B}^\mathrm{max}$ (\textbf{right}) of Eq.~\eqref{eq: binding energy deuteron}. 
    \textbf{Left:} Lowering $B_{D,B}$, i.e. shallower twin deuteron bound states, keeps twin alpha particles stable against breakup into deuterons, Eq.~\eqref{eq:stability_cond_He}, and increases $B_{He,B}-B_{D,B}$, so that twin helium atoms remain stable up to larger $|\DeltMnp|+m_{e,B}$, see Eq.~\eqref{eq:stability_cond_He_EC}. 
    \textbf{Right:} Increasing $B_{D,B}$ stabilizes twin deuterons for larger $|\DeltMnp|$, Eq.~\eqref{eq:stability_cond_D}, and destabilizes twin alpha particles and prevents BBN from proceeding. Twin deuterium atoms remain stable up to larger $|\DeltMnp|+m_{e,B}$, see Eq.~\eqref{eq:stability_cond_D_EC}.
    }
    \label{fig: phase diagram BD}
\end{figure}

\begin{figure}
    \centering
    \begin{tabular}{c}
    \includegraphics[width=\linewidth]{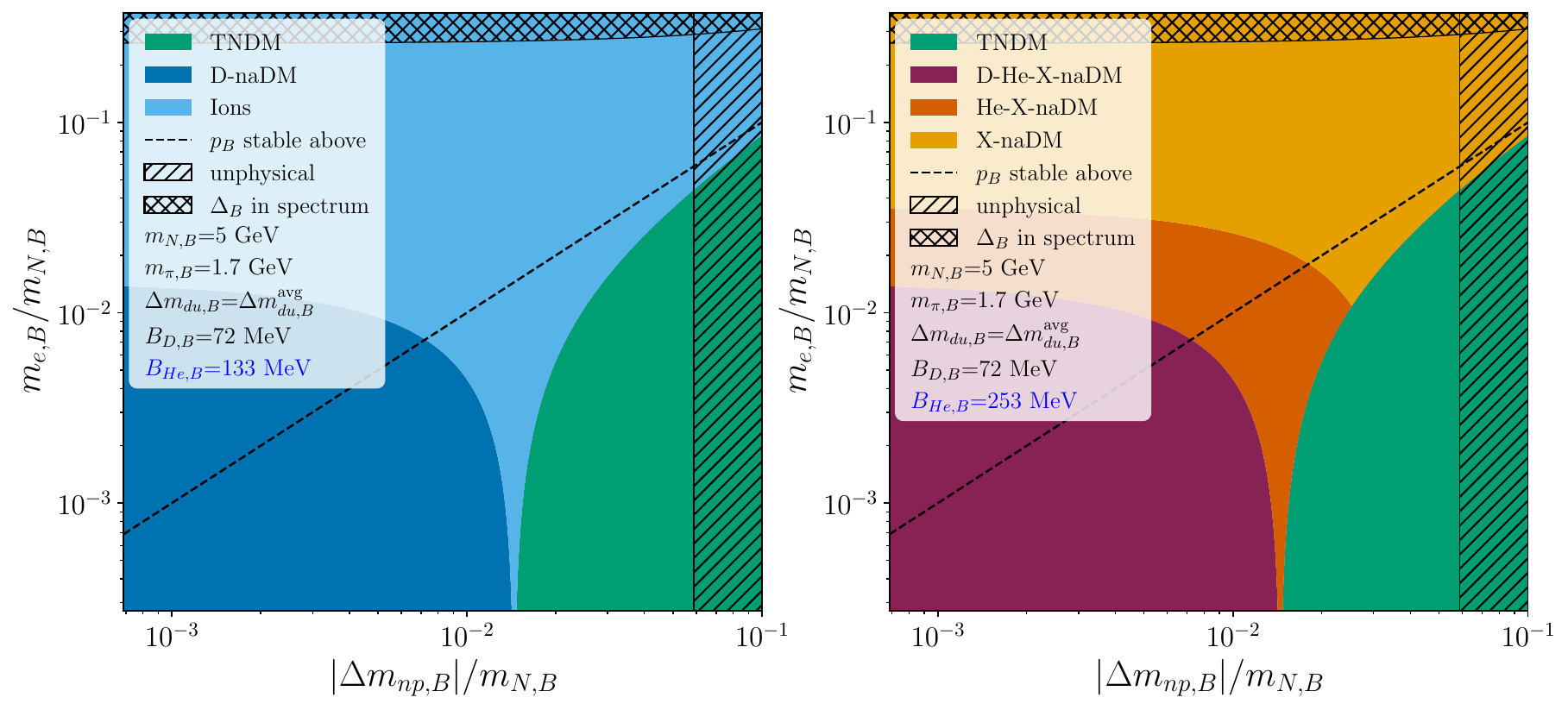}
    \end{tabular}
    
    \caption{Different phases of naDM as defined by the IR parameters of \cref{tab: IR params} as we vary $B_{He,B}$ compared to the canonical benchmark of \cref{fig: phase diagram benchmark}: $B_{He,B}$ given by $B_{He,B}^\mathrm{min}$ (\textbf{left}) or by $B_{He,B}^\mathrm{max}$ (\textbf{right}) of Eq.~\eqref{eq: binding energy helium}. 
    \textbf{Left:} Decreasing $B_{He,B}$ destabilizes twin alpha particles against break up into deuterons, Eq.~\eqref{eq:stability_cond_He}, preventing BBN from proceeding. 
    \textbf{Right:} Increasing $B_{He,B}$ stabilizes twin alpha particles, Eq.~\eqref{eq:stability_cond_He}, and twin helium in a large region, Eq.~\eqref{eq:stability_cond_He_EC}. 
    Note that the region in which twin deuterium is stable remains the same in both panels. 
    }
    \label{fig: phase diagram BHe}
\end{figure}

\begin{figure}
    \centering
    \begin{tabular}{c}
    \includegraphics[width=\linewidth]{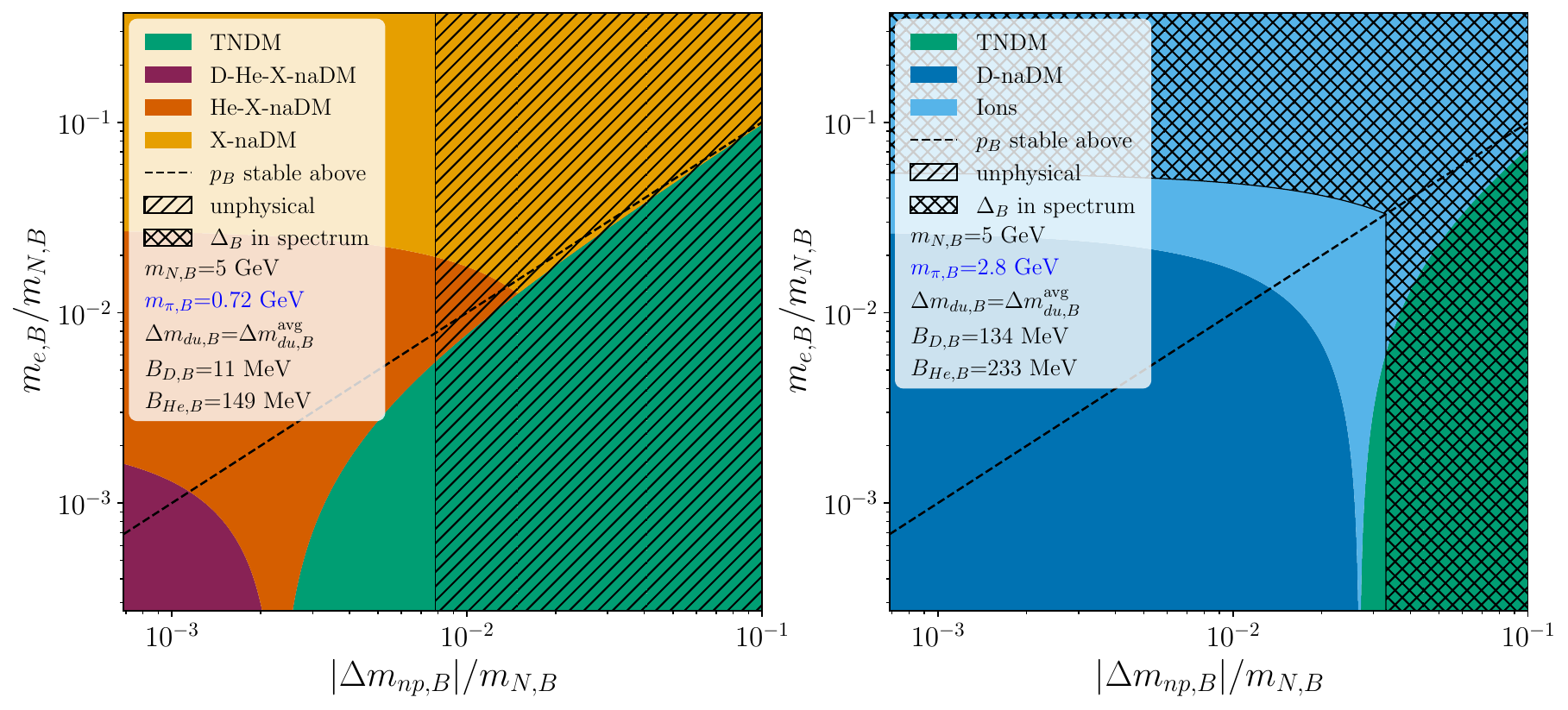}
    \end{tabular}
    
    \caption{
    Different phases of naDM as defined by the IR parameters of \cref{tab: IR params} as we vary $m_{\pi,B}$ compared to the canonical benchmark of \cref{fig: phase diagram benchmark}: $m_{\pi,B}/m_{N,B} = 0.14$ (\textbf{left}) or by $m_{\pi,B}/m_{N,B} = 0.56$ (\textbf{right}). Note that because the binding energies are functions of the twin pion and neutron masses, varying $m_{\pi,B}$ has the effect of changing the binding energies, but they remain fixed to the average of the two expressions in Eqs.~\eqref{eq: binding energy deuteron},~\eqref{eq: binding energy helium}.
    \textbf{Left:} 
    For small $m_{\pi,B}$, large $|\DeltMnp|$ implies unphysical twin quark masses (hatched region), while the relatively small $B_{D,B}$ allows stable twin helium in a large region. 
    \textbf{Right:} 
    As $m_{\pi,B}$ increases, $B_{D,B}$ grows more rapidly than $B_{He,B}$, rendering twin alpha particles unstable to decay to deuterons and shifting the naDM--TNDM boundary to larger $|\DeltMnp|$. 
    For sufficiently large $m_{\pi,B}$, the $\Delta_B^-$ decay channel of Eq.~\eqref{eq:Delta decay} becomes less energetically favourable, whereas for sufficiently large $|\DeltMnp|$ the twin down quark is so light that the $\Delta_B^-$ has a mass comparable to $p_B$, so a complete treatment in either case requires extending the Boltzmann equations to include the $\Delta_B^-$ abundance. This panel with its large value of $m_{\pi,B}$ corresponds most closely to the parameters favored by the UV completion in \cref{sec: uv realization}.
    }
    \label{fig: phase diagram mpi}
\end{figure}

To develop intuition on the effect of different parameters of the model, we present the resulting phase diagrams while varying several key IR parameters, one at a time, compared to the canonical benchmark of \cref{fig: phase diagram benchmark}: $B_{D,B}$ (\cref{fig: phase diagram BD}), $B_{He,B}$ (\cref{fig: phase diagram BHe}), and $m_{\pi,B}$ (\cref{fig: phase diagram mpi}).
The impact of each parameter on the naDM phase in the plane of $|\DeltMnp|/m_{N,B}$--$m_{e,B}/m_{N,B}$ is discussed in the corresponding caption.
Parameters in \cref{tab: IR params} that are not varied have imperceptible impact on the phase diagram.
We note that the choice of model for the mapping between $\Delta m_{du,B}$ and $\DeltMnp$ sets the cutoff at which the apparent underlying down quark mass becomes negative (and thus unphysical), thus in principle modifying the allowed parameter space. However, this cutoff occurs at large mass splittings where the two estimates agree within less than a percent, and so in practice the impact is minimal.

\begin{figure}
    \centering
    \includegraphics[width=0.7\linewidth]{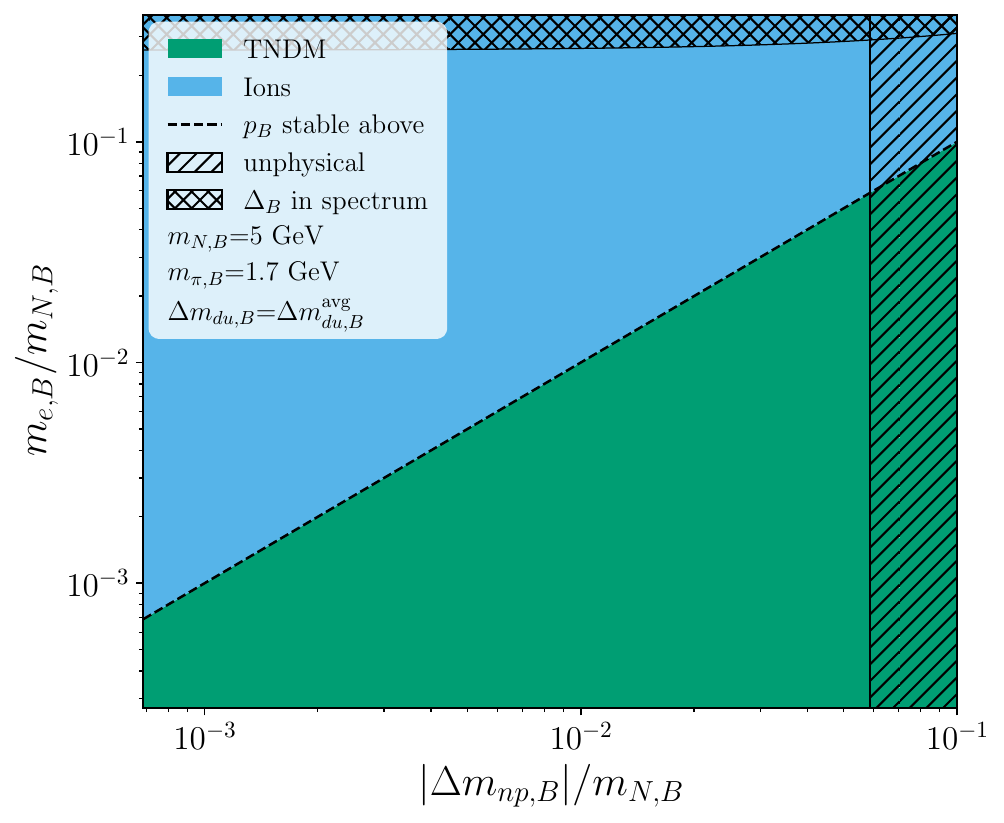}
    \caption{Different phases of our model if we instead use the lattice results of Refs.~\cite{BaryonScattering:2025ziz,Detmold:2026vjx,Detmold:2026shh,Wang:2026kab,Chakraborty:2026zeq,Zhang:2026cco} (as opposed to results used in deriving Eq.~\eqref{eq: binding energy deuteron}) which conclude the twin deuteron is unbound in the entirety of our parameter space. 
    At sufficiently large $m_{e,B}$, the twin proton is stable and could in principle survive in ionic form if recombination is inefficient. 
    The two phases only depend on the twin proton stability.
    }
    \label{fig: unbound D phase diagram}
\end{figure}

We also remind the reader that all phase diagrams presented so far have been obtained using the deuteron binding-energy fit of Eq.~\eqref{eq: binding energy deuteron}. As discussed in \cref{subsec: dark binding energies}, some recent lattice calculations indicate that the deuteron may instead remain unbound throughout our parameter space.  In this alternative scenario, only two phases remain viable: pure TNDM when the twin proton is unstable, and TNDM + twin proton ions when the twin proton is stable (see \cref{tab: phase diagram regions}). The corresponding phase diagram for the canonical benchmark is shown in \cref{fig: unbound D phase diagram}. 
The resulting phenomenology is qualitatively different from that obtained using Eq.~\eqref{eq: binding energy deuteron}. This sensitivity to the deuteron binding energy underscores the importance of resolving the current disagreement between lattice calculations, providing strong motivation for further studies of two-nucleon systems.

All in all, using the deuteron binding-energy fits of Eq.~\eqref{eq: binding energy deuteron}, we find that light twin electrons and small twin neutron--proton mass splittings allow a rich spectrum of naDM sub-components. As $|\DeltMnp|/m_{N,B}$ increases to $\mathcal{O}(1\%-10\%)$, the deuteron becomes unstable and the model transitions to a pure twin neutron DM scenario. Larger fractional mass splittings are unphysical for the range of $m_{\pi,B}$ and $m_{N,B}$ considered in this work.


\section{Cosmological History and naDM Fraction}\label{sec: cosmo}

In this section, we calculate the relic abundance of twin nuclear species across our parameter space, in the scenario where 100\% of DM is made up of twin baryons. 
We denote $\fnaDM$ as the fraction of the total DM energy density that is made up of twin \textit{atoms}. 
As discussed in~\cref{subsec:stability} and~\cref{subsec: phase diagram}, thanks to incomplete recombination or instability of twin atoms to electron capture, there is also the possibility that an ionized constituent remains. We will take as our default scenario the atomic case (efficient recombination and atomic stability post-recombination), but comment on the exotic possibilities if these assumptions are removed.

A precise determination of the relative abundances of the different twin elements would require detailed knowledge of their binding energies and nuclear reaction rates. 
It also requires a proper understanding of the efficacy of the recombination.
Because the final composition is highly sensitive to these poorly known quantities, we restrict our analysis to the total naDM abundance rather than attempting to predict the abundance of each individual twin element. 
Our treatment is merely the initial step in studying the rich physics of a twin BBN process to identify the parameter space with non-negligible naDM fraction, while many further detailed studies are left for future works.

A complete calculation would require solving the coupled Boltzmann equations for twin protons, neutrons, and nuclei of different mass numbers, as well as for the $\Delta$ baryon wherever it is stable. 
Since the equilibrium abundances depend exponentially on both the binding energies and the temperature, uncertainties in the reaction rates translate into large uncertainties in the final nuclear mass fractions
\begin{equation}
X_{\Anum,B} \equiv \Anum\,\frac{n_{\Anum,B}}{n_{b,B}}\, ,
\label{eq:def_X_A}
\end{equation}
where $\Anum$ is the atomic mass number of the nucleus, $n_{\Anum,B}$ is its number density, and $n_{b,B}$ is the total twin baryon number density which is determined by requiring that twin baryons comprise all of DM:\footnote{Carefully, the average mass here should be the average baryon mass weighted by relative abundance, however, in the range of mass splittings we consider we verified numerically that the corrections to the resulting abundances are sub percent level, and therefore negligible at our level of precision. }
\begin{equation}\label{eq: dark baryon number density}
n_{b,B}(T) = \eta \, n_{\gamma,A}(T)\,\frac{\Omega_{\rm DM}}{\Omega_b}\,\frac{m_{N,A}}{m_{N,B}}\, .
\end{equation}
Here $\Omega_{\rm DM}$ and $\Omega_b$ denote the present-day DM and SM baryon density fractions, respectively, with
\begin{equation}
\eta \equiv \frac{n_{b,A}}{n_{\gamma,A}} \simeq 6 \times 10^{-10},
\end{equation}
denoting the measured SM baryon-to-photon ratio \cite{ParticleDataGroup:2024cfk}.

We adopt a simpler strategy aimed at determining only the total naDM fraction. 
We begin by solving the Boltzmann equations for the twin proton and neutron abundances. At early times, the high entropy per baryon favours unbound nucleons, and the nuclear abundances are negligible, justifying this simplified treatment. 
We neglect the $\Delta_B$ baryon in this calculation; accordingly,  as in the previous section, we identify the regions in which it becomes stable or light and regard our analysis as inapplicable there.

If the twin deuteron is unbound, no further nuclear calculation is required. 
Depending on whether the twin proton is unstable or stable, the final state is then either pure TNDM or a mixture of TNDM and twin proton ions, with twin hydrogen inherently unstable, as discussed above. 
If the twin deuteron is bound, we supplement the nucleon Boltzmann equations with the nuclear equilibrium relation for deuterons and determine the temperature at which deuteron photodissociation becomes inefficient. 
Once this bottleneck is overcome, binding into twin deuterons is energetically favoured, and we approximate all surviving twin protons as being rapidly incorporated into deuterons and hence into the naDM component. 
When the free twin proton is unstable, this binding also prevents its subsequent decay into a twin neutron, thereby preserving the corresponding naDM abundance.

We do not attempt to follow the subsequent redistribution of this abundance among heavier nuclei, since doing so would require precise knowledge of uncertain nuclear reaction rates. 
In the absence of cosmic expansion, nuclear reactions would drive the naDM toward the stable nucleus with the largest binding energy per nucleon, provided a chain of fusion reactions connects it to the initial constituents. In an expanding universe, however, each fusion step remains efficient only while its rate exceeds the Hubble rate; once it falls below, that step freezes out and the chain toward more tightly bound nuclei is cut off. We therefore assume that the naDM in each region is ultimately dominated by the most tightly bound stable nucleus that can be reached before freeze-out, i.e., the heaviest stable nucleus up to a maximum set by the fusion rates (see~\cref{eq: heavy nuclei fusion rate} and below for more detailed discussion). Thus, in the D--naDM region, most of the naDM is
expected to reside in deuterium, while in the D--He--X naDM region it is expected to reside predominantly in X, with only a subleading fraction of other twin elements. The different possible outcomes are summarized in~\cref{tab: phase diagram regions}.

We find that, over a non-negligible portion of the parameter space, naDM can constitute an $\mathcal{O}(10\%)$ fraction of the present-day DM abundance. 
As discussed in the following section, even such a subdominant component can lead to rich and potentially observable phenomenology.

In \cref{subsec: dark bbn}, we describe the calculation of the relic twin proton population and the resulting estimate for the deuterium and helium naDM abundances. We also show an estimate for the maximum attainable heavier-nucleus state from twin nucleosynthesis, denoted by its atomic number $\Znum_{\max}$.
In \cref{subsec: naDM abundances}, we present the final naDM fraction over the same $|\DeltMnp|/m_{N,B}$ versus $m_{e,B}/m_{N,B}$ planes considered in the previous section and discuss its dependence on the model parameters. 
Throughout, the calculation is performed against the background evolution of the expanding and cooling SM and twin sectors, reviewed in Appendix~\ref{app: cosmo BG}.

\subsection{Big Bang Nucleosynthesis: From Twin Baryons to Twin Nuclei}\label{subsec: dark bbn}

Let us start by writing down the Boltzmann equation governing the twin proton abundance. 
Assuming the twin proton and neutron to be the only relevant baryons (i.e. neglecting the twin $\Delta$ baryon), we parametrize their abundances with their relative number density
\begin{equation}\label{eq: p/n relative number density}
    X_{i,B} = \frac{n_{i,B}}{n_{b,B}}, ~~~ i=p,n \, .
\end{equation}
Relevant interactions appearing in their Boltzmann equations include weak scatterings 

\begin{equation}\label{eq: freeze-out processes}
    p_B + \bar{\nu}_B \leftrightarrow n_B + e_B^+, \qquad p_B + e_B^- \leftrightarrow n_B + \nu_B.
\end{equation}
and potential twin proton decay
\begin{equation}
    p_B \rightarrow n_B + e_B^+ + \bar{\nu}_B \, .
\end{equation}
Assuming that the entire net twin baryon number resides in twin protons and neutrons, we have $X_{n,B}=1-X_{p,B}$. 
The Boltzmann equation governing the twin proton fraction is then
\begin{equation}\label{eq: np boltzmann evolution}
    \frac{dX_{p,B}}{dT_A}
    =
    -\left(\frac{dT_A}{dt}\right)^{-1}
    \left[
        \left(
            X_{p,B}
            -
            \left(1-X_{p,B}\right)e^{-|\DeltMnp|/T_B}
        \right)
        \Gamma_{p\rightarrow n,B}(T_B)
        +
        X_{p,B}\Gamma_{p,B}
    \right]\, ,
\end{equation}
where $\Gamma_{p \rightarrow n,B}$ ($\Gamma_{p,B}$) is the scattering rate for the 2-to-2 processes (twin proton decay rate) and are calculated in Appendix \ref{app: np freeze-out} following the treatment of Ref.~\cite{Mukhanov:2003xs}.
This neglects the twin-$\Delta_B$ baryon as discussed above.
These rates depend on $T_B$, whereas the evolution variable in the equation above is the SM temperature $T_A$. 
We must therefore determine $T_B$ as a function of $T_A$, as detailed in Appendix \ref{app: cosmo BG}.

At sufficiently high temperatures, essentially all twin protons and neutrons remain free, since the abundant high-energy photons efficiently photodissociate any nuclei that form. 
As the universe expands and cools, the number density of photons energetic enough to dissociate deuterons becomes exponentially suppressed, allowing the deuteron abundance to build up. 
We estimate the temperature at which this bottleneck is overcome using the condition
\begin{equation}
    n_{D,B}(T_D)
    \sim
    n_{\gamma,B}(E>B_{D,B};T_D),
    \label{eq:bottle_neck_T}
\end{equation}
where $n_{\gamma,B}(E>B_{D,B}; T)$ denotes the number density of twin photons with energies above the deuteron binding energy at temperature $T$. 
This equation defines $T_D$, the twin-sector temperature at which photodissociation becomes inefficient and deuterons begin to accumulate.

We determine $T_D$ by using the equilibrium deuteron abundance and solving Eq.~\eqref{eq:bottle_neck_T} numerically. 
The equilibrium mass fraction of a nucleus with mass number $\Anum$ and atomic number $\Znum$ is given in Ref.~\cite{Kolb:1990vq}
\begin{equation}
\label{eq: nuclei eq abudance}
    X_{\Anum,B}
    =
    g_\Anum
    \Anum^{3/2+1}
    2^{-\Anum}
    \left(
        \frac{2\pi}{m_{N,B} T_B}
    \right)^{\frac{3(\Anum-1)}{2}}
    n_{b,B}^{\Anum-1}
    X_{p,B}^{\Znum}
    X_{n,B}^{\Anum-\Znum}
    e^{B_{\Anum,B}/T_B}\, .
\end{equation}
Here, $X_{\Anum,B}$ is the mass fraction defined in Eq.~\eqref{eq:def_X_A}, $m_{N,B}$ is the average twin nucleon mass,
$g_\Anum$ is the number of internal degrees of freedom of the nucleus, $n_{b,B}$ is the net asymmetric twin baryon number density, and $B_{\Anum,B}$ is its nuclear binding energy.
Solving Eq.~\eqref{eq:bottle_neck_T} numerically using Eq.~\eqref{eq: nuclei eq abudance}, we find that $T_D$ is well approximated by rescaling the SM value with the dark sector deuteron binding energy,
\begin{equation}
    \frac{T_{D,B}}{B_{D,B}}
    \simeq
    \frac{T_{D,A}}{B_{D,A}}\, .
    \label{eq:T_bottle_from_SM}
\end{equation}
Nevertheless, because the resulting abundances are exponentially sensitive to $T_D$, we use the full solution to Eq. \eqref{eq:bottle_neck_T} in the calculations below.

We then assume that, once the deuterium bottleneck\footnote{In light of the nuclear vs. atomic stability discussion in~\cref{subsec:stability}, we might more pedantically refer to this as the \textit{deuteron} bottleneck, since it is the formation of the nuclear state and not the atomic state that causes the bottleneck. We will use the conventional \textit{deuterium bottleneck} throughout for simplicity. } is overcome at $T_B=T_D$, all available twin protons are efficiently incorporated into deuterons and subsequently processed into heavier nuclei whenever such nuclei are stable. 
This determines the total naDM abundance. 
By far the most consequential inputs to this equation are the nuclear binding energies ($B_{\Anum,B}$). As discussed in \cref{subsec: dark binding energies}, we consider nuclear binding energies at the upper and lower range of what is allowed by lattice data, and we allow for both bound and unbound twin deuterons. 
If the twin deuteron is unbound at a given point in our parameter space, then  we assume twin BBN does not proceed.

This uncertainty in the binding energy of light twin nuclei, reflected in the uncertainty of whether or not deuterons are bound at all, makes quantitatively proceeding with BBN calculations beyond $D_B$ untenable. 
However, this uncertainty extends in the other direction as well: it is possible that the twin nuclei in parts of our parameter space are deeply bound, removing the bottlenecks present in dark BBN and allowing the formation of large twin nuclei and/or multi-twin-neutron bound states \cite{Krnjaic:2014xza, Hardy:2014mqa, Hardy:2015boa, Gresham:2017cvl, Gresham:2017zqi, Gresham:2018anj}.

Under the assumption that the deuteron binding energy follows \cref{eq: binding energy deuteron}, we note that this implies the twin deuteron binding energy can be orders of magnitude higher than the SM. This allows the twin deuteron to form at a much higher temperature, when nuclear fusion reactions may be efficient. As has previously been observed in, e.g., Refs.~\cite{Krnjaic:2014xza, Hardy:2014mqa}, when the Coulomb barrier is reduced, BBN naturally proceeds to saturating the heaviest stable state. While both of these works assumed dark electromagnetism was either weakly coupled or not present, in our case with $\alpha_B \approx \alpha_A$, raising the temperature of twin nucleosynthesis can still partly overcome the Coulomb barrier and lead to the formation of heavier twin nuclei during BBN. 

If the formation of bound states is Coulomb limited, we can estimate the formation rates for heavy nuclei. In our scenario $X_{p,B}<X_{n,B}$, thus once the deuterium bottleneck resolves we expect that the majority of twin protons are processed into nuclei, leaving a negligible free twin proton fraction. This means that the capture of free twin protons will not be a substantial formation pathway for twin nuclei. 
The fusion rate for a particle with atomic mass $\Anum_1$ to find and combine with another particle of mass number $\Anum_2$ into a new nucleus of mass number $\Anum_f = \Anum_1 + \Anum_2$, is given by \cite{Krnjaic:2014xza, Hardy:2014mqa}
\begin{equation}\label{eq: heavy nuclei fusion rate}
    \Gamma_{\Anum_1 + \Anum_2 \to \Anum_f} =  n_{\Anum_i }\sigma_{\Anum_i } v_{\Anum_i }e^{-\sqrt{E_G/E}}
    \simeq
    \left(2X_{p,B}n_{b,B}\sigma_1v_1\right){\Anum}_i^{-5/6}e^{-\sqrt{E_G/E}} \, ,
\end{equation}
where $E$ is the typical energy of the incoming particles.
Here $n_{\Anum_i }$ is the number density of initial state nuclei with atomic mass number $\Anum_i$, $\sigma_{\Anum_i}$ is the interaction cross section for two initial state nuclei, and $v_{\Anum_i}$ is the average velocity of the initial state nuclei. In writing the second equality, we used 
\begin{equation}
    n_{\Anum_i}\simeq  \frac{X_{p,B} n_{b,B}}{\Znum_i}\simeq \frac{2 X_{p,B}n_{b,B}}{\Anum_i}, 
    \quad\qquad 
    \sigma_{\Anum_i}\simeq \Anum_i^{2/3} \sigma_1, 
    \quad\qquad 
    v_{\Anum_i}\simeq \frac{v_1}{\sqrt{\Anum_i}}.
\end{equation}
Here $\Znum_i$ is the proton number of the nucleus and is approximated as $\Anum_i/2$. 
We have assumed that the cross section scales geometrically with nuclear radius and define
\begin{equation}
    \sigma_1\simeq \pi L_B^2 \, 
    \label{eq:sig1_LB}
\end{equation}
as the cross section for free nucleons, with $L_B$ representing the characteristic scattering radius for free nucleons (for more discussion see~\cref{sec: bullet cluster}). 
Rather than performing the full thermal average of $\langle \sigma v \rangle$, we obtain an estimate by evaluating the rate at the mean relative speed of a Maxwell--Boltzmann distribution, $v_1\simeq \sqrt{8 T/ (\pi \mu_1) }\simeq \sqrt{16 T/ (\pi m_{N,B})}$, where $\mu_1 =m_{N,B}/2$ is the reduced mass of two free nucleons.
The relative velocity for heavier nuclei is assumed to scale as $v_{\Anum_i}^2\sim T/ \mu_{\Anum_i}$, with $\mu_{\Anum_i}=\Anum_i m_{N,B}/2$ the reduced mass of two identical nuclei.\footnote{In the limit where Coulomb repulsion is large, the rates are dominated by the high velocity tail of the MB distribution, in the so-called Gamow peak (see e.g. \cite{Wagoner:1966pv}). Using the mean velocity thus yields an underestimate of the rate and thus an \textit{overestimate} of the freeze-out temperature, however, since we are interested in the case where sufficient reactions occur that a majority of the reactants are processed, we take this as a more conservative and applicable estimate.}
We have additionally assumed that a single reactant species $\Anum_1=\Anum_2=\Anum_i$ ($\Znum_1=\Znum_2=\Znum_i$) comprises the majority of the naDM fraction. The Gamow energy \cite{Maoz:2016textbook}
\begin{equation}
     E_G = 2\mu_{\Anum_i}(\pi \alpha_B \Znum_1\Znum_2)^2 \sim  m_{N,B}\alpha_B^2 \Znum_i^5, 
\end{equation}
accounts for the probability of tunneling through the Coulomb barrier, with the reduced mass $\mu_{\Anum_i}\sim \Znum_i m_{N,B}$.

At this point, we note that both the scaling arguments and Gamow factor imply that the fusion rates decrease with increasing $\Anum_i$ and $\Znum_i$. Therefore at a fixed temperature, fusion rates for heavier nuclei are slower than for lighter nuclei and an upper limit must exist on the heaviest nuclei that can form at temperature $T$. 

The fusion process freezes out when $\Gamma_{\Anum_i} \sim H$. 
Assuming that the SM dominates the Hubble evolution after asymmetric reheating, i.e. $H\sim T_A^2/\Mpl$, and using the previous equations, this freeze-out takes place when the twin photon bath reaches the temperature
\begin{equation}\label{eq: fusion freezeout temp approx}
        T_{\rm FO}^{\Anum_f} \sim    3 \times 10^{-8}\ \textrm{GeV}
    \Anum_f^5
    \left(\frac{\alpha_B}{\alpha_A}\right)^2
    \left(\frac{m_{N,B}}{5\ \textrm{GeV}}\right)
    \left[1 - 0.07 \ln \Anum_f - \mathfrak{L} \right]^{-2}.
\end{equation}
To derive this expression we use $n_{b,B}$ from Eq. \eqref{eq: dark baryon number density} in Eq.~\eqref{eq: heavy nuclei fusion rate}.
This expression utilizes the simplification $\Anum_f= 2\Anum_i$ and $\Znum= \Anum/2$. It is thus inaccurate for $\Anum\lesssim3$, but gives reasonable agreement with a numerical solution for $\Anum\in (4,12)$. 
The logarithmic corrections $\mathfrak{L}$ are 
\begin{equation}
     \mathfrak{L} \simeq 0.09
    \ln\left(
    \left(\frac{g_{\star,S}}{10.75}\right)^{1/2}
    \left(\frac{m_{N,B}}{5\ \textrm{GeV}}\right)
    \left(\frac{\xi}{0.2}\right)
    \left(\frac{0.1}{X_p}\right)
    \left(\frac{1 \ \textrm{GeV}^{-1}}{L_B}\right)^{2}
    \left(\frac{ \textrm{MeV}}{T_{\rm FO}^{\Anum_f}}\right)^{3/2}
    \right),
\end{equation}
where increasing $\mathfrak{L}$ increases the freeze-out temperature, disfavouring heavy nuclei. 
As expected, increasing the twin proton fraction or the interaction radius favours heavy nuclei because the density of the incident nuclei is higher and they interact more frequently. 
Increasing the twin baryon mass disfavours heavy nuclei, since it must correspond with a decrease in twin number density.  
Note that, interestingly, increasing the temperature of the dark sector via $\xi = T_B/T_A$ actually \textit{disfavours} heavy nuclei. 
The twin baryon density is fixed by the asymmetry and therefore tracks the SM temperature $n_{b,B}\propto T_A^3$, as does the Hubble rate. 
At fixed twin sector temperature $T_B$, larger $\xi$ corresponds to lower SM temperature $T_A$, and hence to a more dilute twin baryon population. 

Twin nucleosynthesis begins once twin deuteron photodissociation becomes inefficient at the twin-photon temperature $T_D$. 
A fusion step can then proceed only if it is still active, i.e., if its rate exceeds the Hubble rate at $T_D$:
\[
\Gamma_{\Anum_i\rightarrow\Anum_{f}}(T_D)\gtrsim H(T_D).
\]
Since each fusion reaction is active above its freeze-out temperature, this requires $T_{\rm FO}^{\Anum_f}\lesssim T_D$. 
We define $\Anum_{\max},\,\Znum_{\max}$ as the mass and atomic numbers of the heaviest nucleus whose production reaction can still operate efficiently at \(T_D\). 
The reaction producing this nucleus satisfies
$T_{\rm FO}^{\Anum_{\max}} \lesssim T_D$, while the next step in the chain with $\Anum_f > \Anum_{\max}$ has
$T_{\rm FO}^{\Anum_f} \gtrsim T_D$.
This criterion provides an optimistic estimate of the heaviest nucleus that can form, since it assumes that intermediate nuclei are bound and sufficiently abundant.
Using the estimate from Eq. \eqref{eq:T_bottle_from_SM} for $T_D$ and Eq.~\eqref{eq: fusion freezeout temp approx} for $T_{\rm FO}$, we find an approximation for the heaviest twin element that can be created in this fusion chain\footnote{We verified that for the SM parameters, this estimate predicts $\Znum_{\max} \sim 2$, in agreement with BBN measurements.} 
\begin{equation}\label{eq: heavy nuclei Zmax}
    \Znum_{\max }\sim 5 \,
    \left(\frac{B_{D,B}}{100 \ \textrm{MeV}}\right)^{1/5}
    \left(\frac{\alpha_B}{\alpha_A}\right)^{-2/5}
    \left(\frac{m_{N,B}}{5\ \textrm{GeV}}\right)^{-1/5} \left[1 - 0.07 \ln \Anum_{\max} - \mathfrak{L} \right]^{2/5}\, .
\end{equation} 
This implies primordial formation of $\Znum_{\max}\sim5$, i.e. twin boron. 
This criterion can be straightforwardly understood from the Gamow factor, using $E\propto T$ and $T_D\propto B_D$. Increasing the binding energy triggers fusion to start earlier, when fusion reactions of heavier nuclei are still favoured. 
Similarly, decreasing the Coulomb barrier via decreasing the twin fine structure constant favours formation of heavier twin elements. 
Increasing the twin nucleon mass disfavours heavy nuclei: at fixed $T$ the incident states move more slowly and are less likely to tunnel through the Coulomb barrier.
Note that we expect that neutron-rich isotopes will dominate, due to the larger twin neutron population.

We emphasize that this represents an \textit{upper bound} on the possible nuclei that form, and the heaviest nuclei that actually form may be somewhat lighter. The full nuclear reaction network depends on many parameters whose values for a rescaled twin sector are poorly known at present. 
For instance, the $\Anum=5$ states in the SM ($^5$He and $^5$Li) are unbound, thus the formation of $\Anum=10$ would have to proceed via combinations of other nuclei, which generically will not exist in comparable numbers.

We therefore find that across a wide range of our parameter space, our models support the possibility of heavier nuclei than helium forming in significant abundance.  
While we spend most of our analysis on the more concrete realizations of naDM with the light nuclei (D and He) where quantitative results exist, we comment where relevant on the different phenomena that may be possible if the dark sector is instead dominated by heavy nuclei, such as primordial twin dust, heavy nuclei in direct detection, and metallic mirror stars.

A final comment is in order about the possibility of twin multi-neutron bound states forming. Per the discussion in \cref{subsec: dark binding energies}, it is possible that twin multi-neutron bound states are stable and twin BBN would populate these states even if all of the twin protons were depleted. The fusion rates in this case would not be subject to the Coulomb barrier, and their formation would  proceed to the heaviest stable bound state almost unimpeded, as found in Refs.~\cite{Krnjaic:2014xza, Hardy:2014mqa}. The proliferation of these aggregates could qualitatively change some of the TNDM behaviours. Heavy neutron-rich isotopes could continue to grow even after the fusion processes that determine $\Znum_{f}^{\max}$ freeze-out.


\subsection{naDM Abundance and Composition}\label{subsec: naDM abundances}

In this section, we show the result of our calculation for $\fnaDM$, the fraction of DM that is made of twin nuclei, over the IR parameter space of the naDM scenario demarcated by the ranges in \cref{tab: IR params}. 
We will highlight the parameters with the most impact on this fraction ($\DeltMnp$, $m_{e,B}$, $m_{\pi,B}$, $m_{N,B}$, $G_{F,B}$, and $B_{D,B}$), separating them from those with 
marginal to negligible impact on the naDM sub-component fraction ($\Delta N_\mathrm{eff}$, $g_{a,B}$, $\Delta m_{du,B}$, and $B_{He,B}$). 
The range we scan for each parameter was reviewed in the previous section, see \cref{tab: IR params}; here we discuss how and why varying them affects the fractional relic abundance of naDM. 

We use the approximation outlined in the previous section to calculate the relic abundance: we first solve the twin proton Boltzmann equation, then determine the twin deuterium bottleneck temperature $T_{D,B}$, and finally assume an instantaneous conversion of all surviving protons into naDM, assuming that the heaviest stable nucleus becomes the dominant naDM component, as it is the relic with the largest binding energy. 
As in our investigation of the naDM phase space in Figs.~\ref{fig: phase diagram benchmark}--\ref{fig:  phase diagram mpi}, we present our results in the $|\DeltMnp|/m_{N,B}$--$m_{e,B}/m_{N,B}$ plane, since these parameters have the largest impact on the naDM abundance. 
They are also directly related to our envisioned UV realization of \Ztwo\ breaking, in which the SM and twin fermions couple differently to their respective Higgs fields.

We additionally present the value of $\Znum_{\max}$ implied by the freeze-out of the heavy nuclei fusion reactions Eq. \eqref{eq: heavy nuclei fusion rate} by solving the resulting expression for the freeze-out temperature numerically. Rather than the naive identification $\Znum= \Anum/2$ used in \eqref{eq: heavy nuclei Zmax}, we use $\Znum= \lfloor\Anum/2\rfloor$ for any twin element and $\Anum_1 = \lfloor \Anum_f/2 \rfloor$, $\Anum_2 = \Anum_f -  \Anum_1$, where the choice of floor identifies the neutron-rich isotopes most plausible in a neutron dominated dark sector. We set the length scale $L_B$ to the nuclear radius $1.2$ fm and rescale with $\Lambda_A/\Lambda_B$ on dimensional grounds. We find this refined criterion agrees reasonably well with the approximation Eq. \eqref{eq: heavy nuclei Zmax} at $X_{p,B}=0.1$ for nuclei above helium where the approximation is expected to be better, and correctly identifies processes which are not Coulomb suppressed.

\begin{figure}
    \centering
    \includegraphics[width=0.8\linewidth]{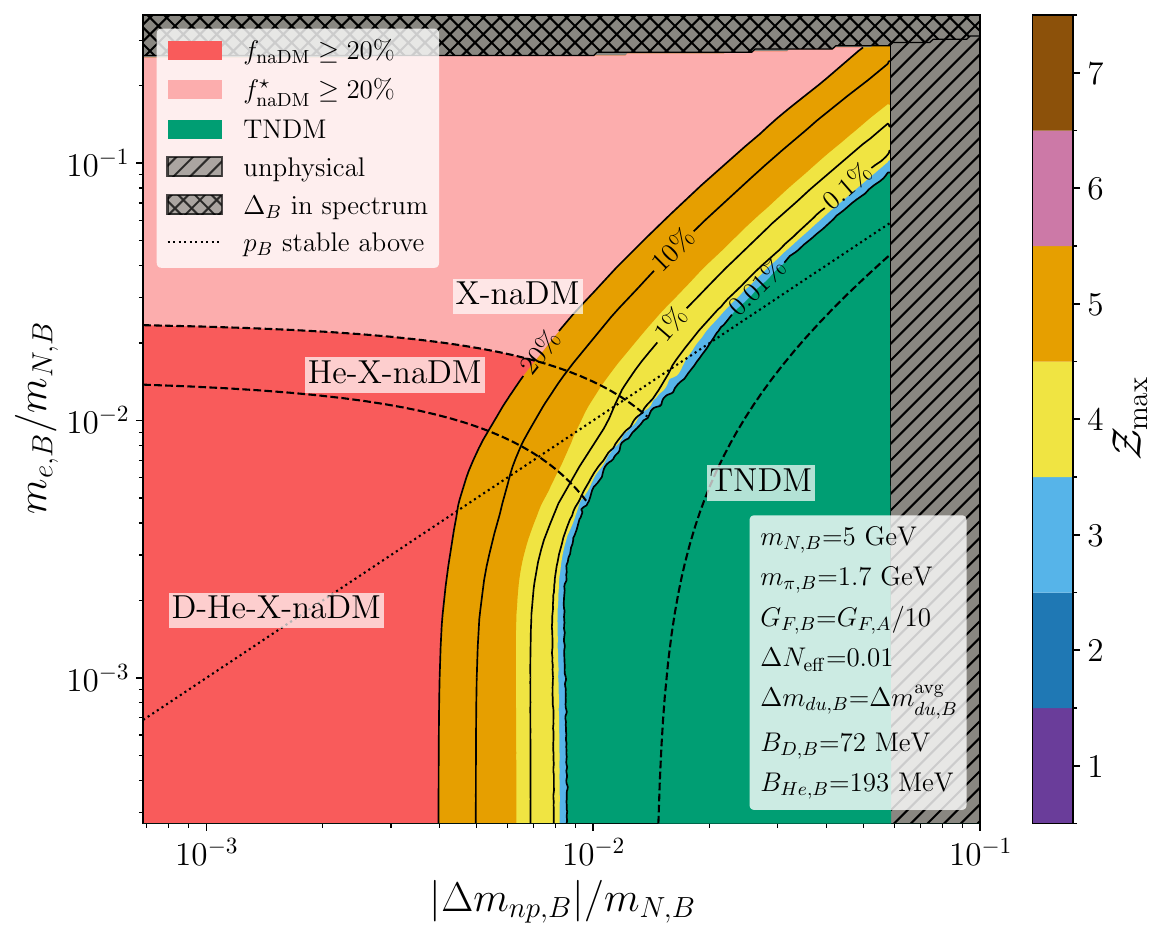}
    \caption{  
    Contours show the twin nucleon DM fraction $\fnaDM$ for the naDM canonical benchmark with $m_{N,B} = 5~\mathrm{GeV}$. Overlaid with dashed lines are the phase space boundaries of
    \cref{fig: phase diagram benchmark}. \textcolor{red}{Red} regions have $\fnaDM > 20\%$, in likely conflict with various self-interaction bounds, while \textcolor{N_SIDM}{green} regions have $\fnaDM < 0.01\%$, likely rendering the naDM sub-component astrophysically and cosmologically inert. The \textcolor{red!50}{pink} regions are labelled $\fnaDM^\star > 20\%$ to indicate that the fraction of \textit{atomic} DM is uncertain.
    In the intermediate region, we indicate with different colours the value of $\Znum_{\max}$ (highest possible twin atomic number) implied by our analysis. 
    $B_{He,B}$ is set to the average of the two estimates in Eq.~\eqref{eq: binding energy helium}.
    }
    \label{fig: nadm abundance benchmark}
\end{figure}

\begin{figure}
    \centering
    \includegraphics[width=\linewidth]{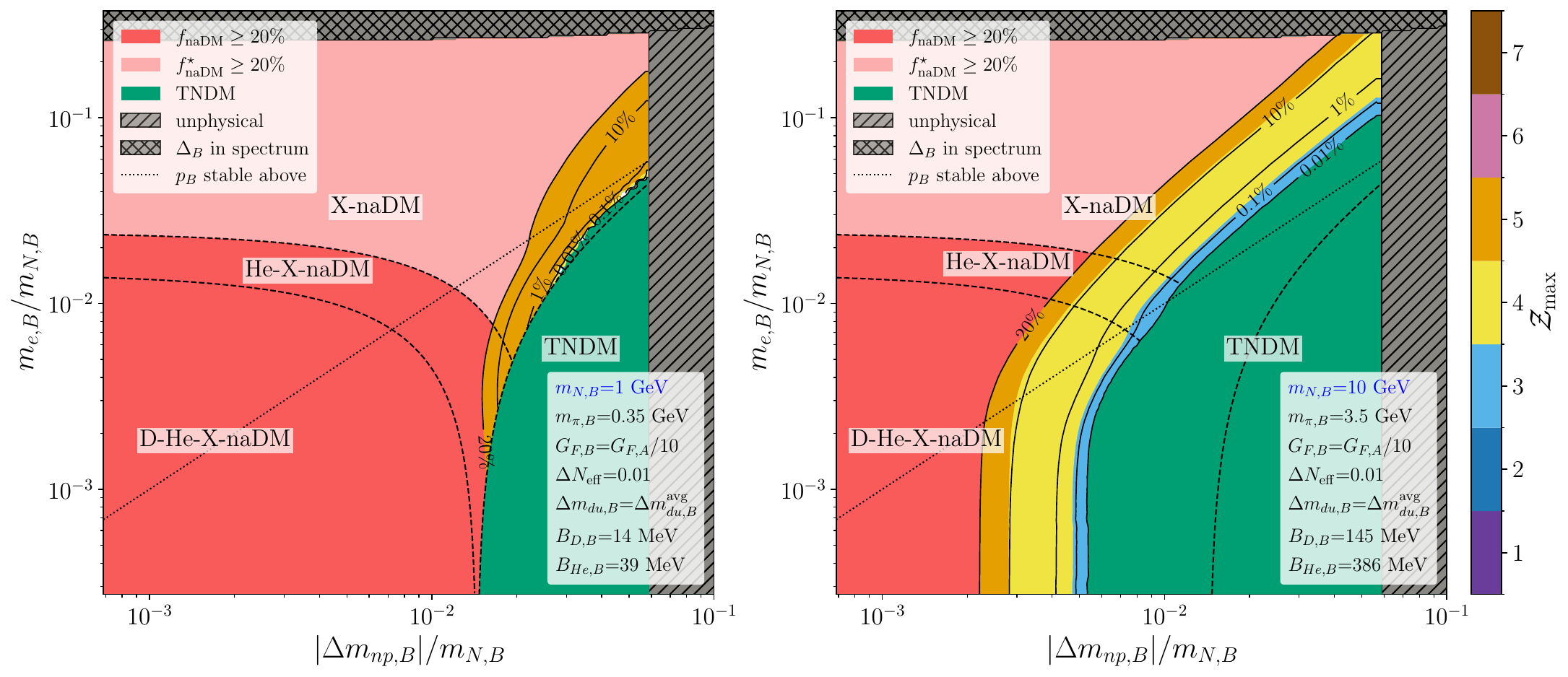}
    \caption{
    Total naDM fraction as a function of $|\DeltMnp|/m_{N,B}$ and $m_{e,B}$, shown for 
    $m_{N,B} = 1~\mathrm{GeV}$ (\textbf{left}) and 10 GeV (\textbf{right}) versions of the naDM canonical benchmark, compare to \cref{fig: nadm abundance benchmark}.
    The naDM phase boundaries identified in \cref{fig: phase diagram benchmark} are overlaid as dashed lines.
    Increasing $m_{N,B}$ at fixed $m_{\pi,B}/m_{N,B}$ implies $\Lambda_B\propto m_{N,B}$ and $m_{\pi,B}\propto m_{N,B}$, so $B_{D,B}$ grows linearly with twin nucleon mass, see Eq.~\eqref{eq: binding energy deuteron}. 
    Twin deuterons can therefore form at a higher temperature, when there is a larger relic twin proton population, generating larger naDM abundances.
    The trend is clearest at fixed $\DeltMnp$. Note that comparing equal $|\DeltMnp|/m_{N,B}$ across panels can be misleading, since the normalization ($m_{N,B}$) differs. 
    $\Znum_{\max}$ tends to be higher for $m_{N,B}=1$ GeV,
    because the interaction radius $L_B$ (from Eq.~\eqref{eq:sig1_LB}) is a factor of 10 larger than $m_{N,B}=10$ GeV, increasing the nuclear fusion rate.
    }
    \label{fig: nadm abundance mn}
\end{figure}

\begin{figure}
    \centering
    \includegraphics[width=\linewidth]{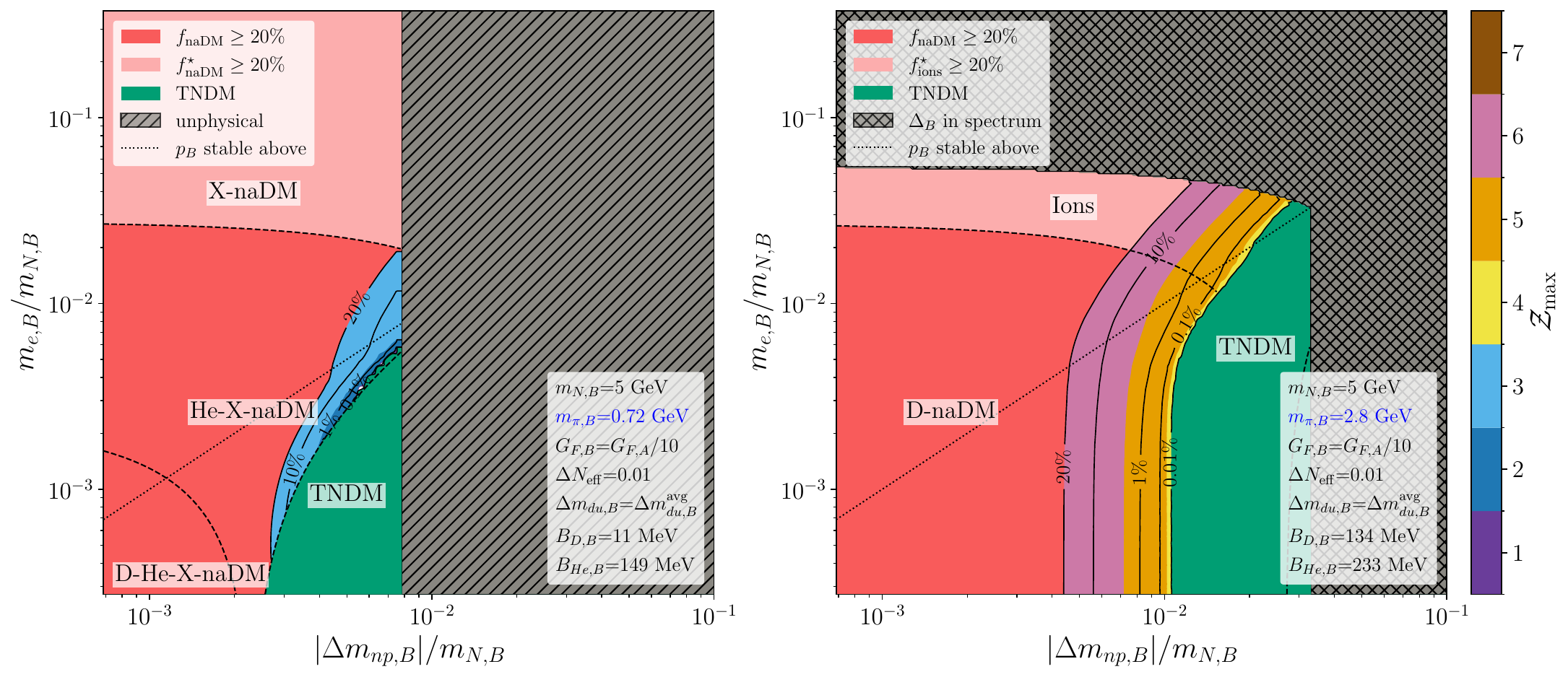}
    \caption{
    Total naDM fraction as a function of $|\DeltMnp|/m_{N,B}$ and $m_{e,B}$ for the canonical naDM benchmark with the minimum (\textbf{left}) and the maximum (\textbf{right}) value of $m_{\pi,B}/m_{N,B}$ from \cref{tab: IR params}. 
    The naDM phase boundaries identified in \cref{fig: phase diagram mpi} are overlaid as dashed lines.
    All other parameters are fixed to the canonical benchmark values specified in the text.
    Increasing $m_{\pi,B}/m_{N,B}$ increases $B_{D,B}$, thus increasing the temperature at which twin deuterons begin to accumulate, $T_D$.
    Consequently, the final naDM fraction increases with $m_{\pi,B}/m_{N,B}$. 
    The same increased $T_{D}$ means that nuclear fusion reactions are more efficient, leading to higher $\Znum_{\max}$ at large twin pion mass.
    We note that a large value of $m_{\pi,B}$ is preferred by the UV completion in \cref{sec: uv realization}, implying that such models could generate twin carbon.}
    \label{fig: nadm abundance mpi}
\end{figure}

\begin{figure}
    \centering
    \includegraphics[width=\linewidth]{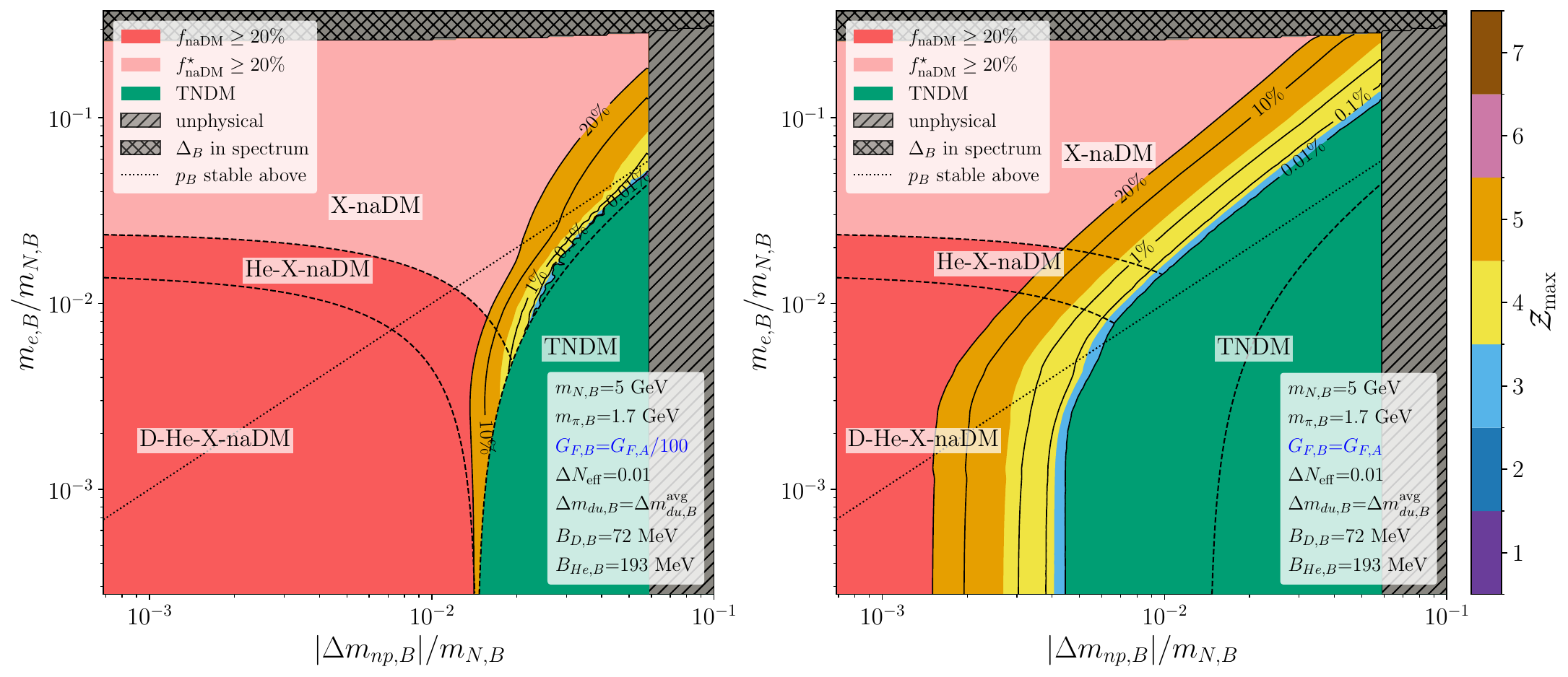}
    \caption{
    Total naDM fraction as a function of $|\DeltMnp|/m_{N,B}$ and $m_{e,B}$ for the canonical naDM benchmark with the minimum (\textbf{left}) and the maximum (\textbf{right}) value of $G_{F,B}$ from \cref{tab: IR params}. 
    The naDM phase boundaries identified in \cref{fig: phase diagram benchmark} are overlaid as dashed lines.
    Increasing $G_{F,B}$ strengthens the twin weak interactions and delays their decoupling. 
    This allows twin proton-to-neutron conversion to remain efficient for longer, further suppressing the relic proton abundance and thereby reducing the final naDM fraction. 
    For the same $\fnaDM$, the same $\Znum_{\max}$ is found, since this formation is independent of $G_{F,B}$.}
    \label{fig: nadm abundance GF}
\end{figure}

\begin{figure}
    \centering
    \includegraphics[width=\linewidth]{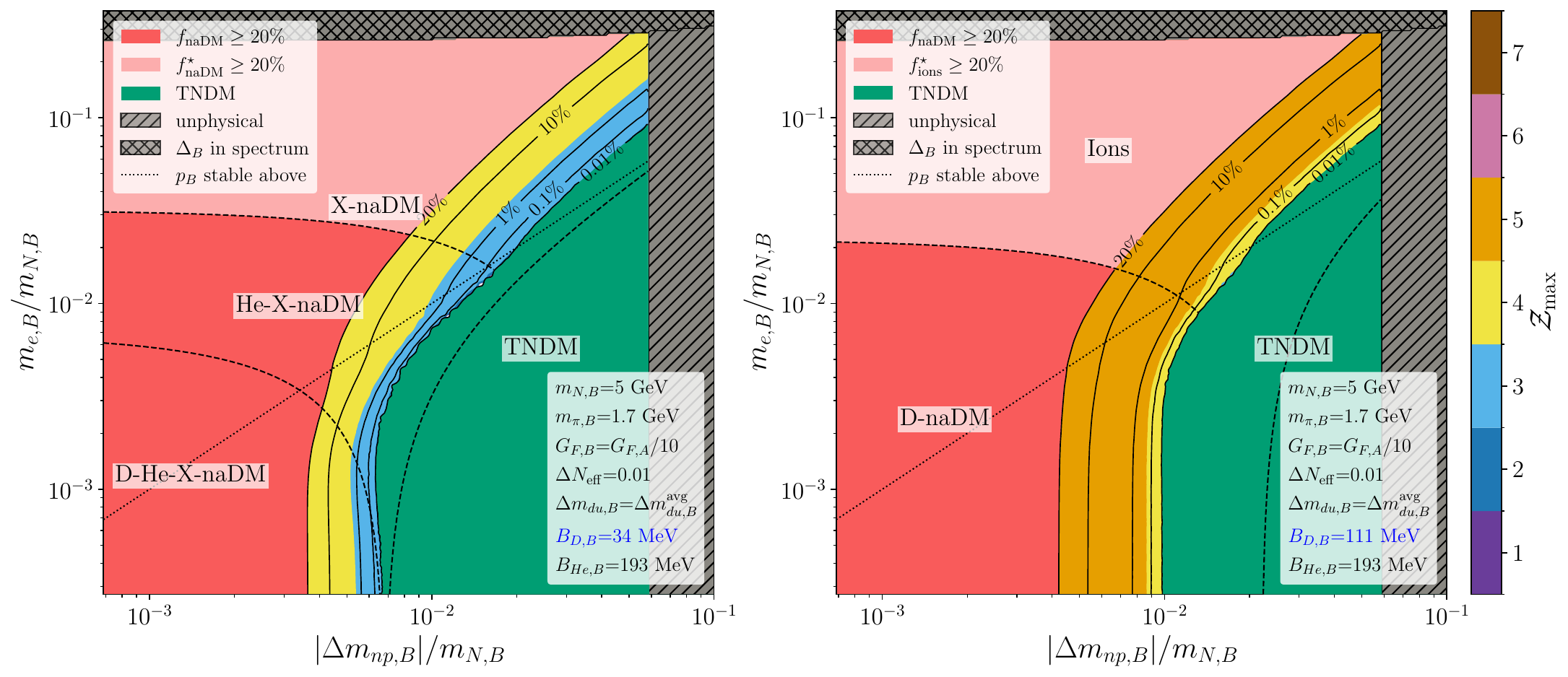}
    \caption{
    Total naDM fraction as a function of $|\DeltMnp|/m_{N,B}$ and $m_{e,B}$ for the naDM canonical benchmark with  the minimum (\textbf{left}) and the maximum (\textbf{right}) value of $B_{D,B}$ from \cref{tab: IR params}. 
    The naDM phase boundaries identified in \cref{fig: phase diagram BD} are overlaid as dashed lines.
    Increasing the twin deuteron binding energy raises the temperature at which deuterons begin to accumulate, $T_D$. 
    Consequently, more twin protons remain available when the formation of twin deuterons begins, leading to a larger final naDM fraction. 
    Over the range of twin nuclear binding energies considered here, the overall effect on $\fnaDM$ is relatively small.
    However, the increased $T_D$ means nuclear fusion reactions are more efficient, leading to higher $\Znum_{\max}$.}
    \label{fig: nadm abundance BD}
\end{figure}

\begin{figure}
    \centering
    \includegraphics[width=\linewidth]{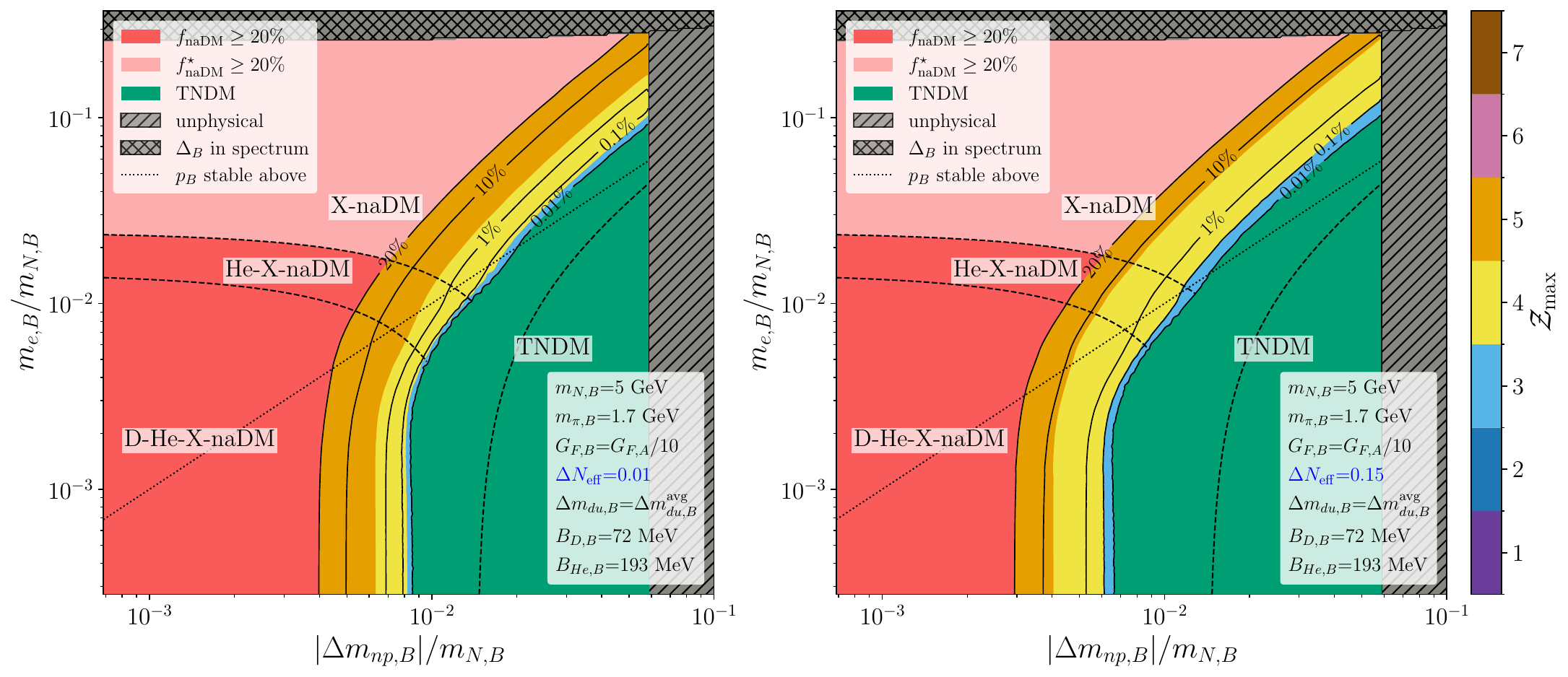}
    \caption{
    Total naDM fraction as a function of $|\DeltMnp|/m_{N,B}$ and $m_{e,B}$ for the naDM canonical benchmark with  the minimum (\textbf{left}) and the maximum (\textbf{right}) value of $\DNeff$ from \cref{tab: IR params}. 
    The naDM phase boundaries identified in \cref{fig: phase diagram benchmark} are overlaid as dashed lines.
    As anticipated above, $\DNeff$ has only a minor effect on the final abundance. 
    For the larger value, $\DNeff=0.15$, the twin sector is hotter and the twin neutron--proton conversion processes remain in equilibrium for longer, resulting in a smaller relic twin proton fraction. 
    As with variations in $B_{D,B}$, however, this produces only a modest change in the total naDM abundance.
    However, smaller $\DNeff$ (larger temperature hierarchy) allows for the formation of heavier nuclei, as discussed below Eq. \eqref{eq: fusion freezeout temp approx}.}
    \label{fig: nadm abundance Dneff}
\end{figure}

The twin nucleon abundance for the naDM canonical benchmark with $m_{N,B} = 5~\mathrm{GeV}$ is shown in \cref{fig: nadm abundance benchmark}.
We again vary one parameter at a time and present the resulting naDM fraction for varying $m_{N,B}$ in \cref{fig: nadm abundance mn}, $m_{\pi,B}/m_{N,B}$ in \cref{fig: nadm abundance mpi}, $G_F$ in \cref{fig: nadm abundance GF}, $B_{D,B}$ in \cref{fig: nadm abundance BD}, and $\DNeff$ in \cref{fig: nadm abundance Dneff}. 
In each figure, we also overlay the different phases identified in \cref{subsec: phase diagram}. 
In all the plots, we fix the alpha binding energy to the average of the estimates in Eq.~\eqref{eq: binding energy helium}, since this quantity does not enter the twin proton relic calculation and only modifies the phase space boundaries.

Several common features are immediately apparent and clarify the dependence on $m_{e,B}$ and $\DeltMnp$. 
First, we distinguish between the case where twin deuterium or helium atoms are stable ($\fnaDM$) and the X-naDM scenario ($\fnaDM^\star$), where the star indicates the early-time abundance, which may be larger than the late-time $\fnaDM$ in this scenario. In the latter case, the late-time composition of the naDM is highly uncertain, and may be comprised of heavy twin atoms, ions, or pure TNDM.

Constraints on dissipative and self-interacting DM require that $\fnaDM\lesssim 10\%-20\%$. 
In the starred regions the present-day bounds apply only to whatever fraction survives to late times, and may therefore be weakened. Bounds from DAOs (see~\cref{subsec:dao}), on the other hand, are imprinted before electron capture can reduce the nuclear population, and therefore constrain the early-time fraction $\fnaDM^\star$ directly. We therefore expect a viable scenario to require $\fnaDM^\star\lesssim 10\%-20\%$ interacting fraction in these regions as well, up to the dependence of the DAO constraints on the twin recombination history.
In regions where the only possible candidates are relic ions, we indicate them with $f^\star_{\mathrm{ions}}$, with the star again used to denote the uncertainty in the late-time composition. We will focus our attention primarily on the deuterium and helium regions, where we can be more concrete about the possible outcomes.

Requiring the naDM fraction to remain below approximately $10\%$--$20\%$,
while simultaneously restricting to physical twin-quark masses, selects values of $|\DeltMnp |/m_{N,B}$ near the percent level. 
As this ratio increases, the twin proton depletion reactions in \cref{eq: freeze-out processes} and proton decays become increasingly favoured. The twin proton abundance is therefore depleted sufficiently early that no appreciable naDM abundance survives, leaving a pure TNDM scenario.
The twin electron may be arbitrarily lighter than the nucleons, but values $m_{e,B}\gtrsim |\DeltMnp|$ stabilize the twin proton sufficiently to produce an excessively large naDM abundance except where the mass splitting is large. 
Although varying the remaining parameters shifts the precise viable ranges of $m_{e,B}/m_{N,B}$ and $|\DeltMnp|/m_{N,B}$, the qualitative structure remains unchanged: 
for small $|\DeltMnp|/m_{N,B}$, including SM-like values of this ratio, naDM constitutes a large fraction of DM, which is excluded. For percent-level twin neutron--proton mass splittings, TNDM dominates while naDM constitutes a small but astrophysically and cosmologically interesting sub-component. 
For even larger splittings, twin protons (and twin deuterium) are unstable and dark matter is entirely made up of twin neutrons.

For the benchmark in \cref{fig: nadm abundance benchmark}, $\fnaDM\sim 10\%$ implies the formation of twin boron may be possible, as anticipated in Eq. \eqref{eq: heavy nuclei Zmax}. 
The value of $\Znum_{\max}$ decreases with decreasing $\fnaDM$, for the simple reason that the number density of the nuclei is lower, so the total fusion rate is lower. 
The dependence on each additional parameter is discussed in the caption of the corresponding figure.

Let us now examine the shape of the contours of constant naDM fraction in more detail. 
The figures indicate that, for sufficiently light twin electrons, the freeze-out abundance is approximately independent of $m_{e,B}$, whereas for heavier twin electrons it is approximately constant along contours of fixed $m_{e,B}/|\DeltMnp|$. 
This behavior reflects two distinct mechanisms by which the twin weak interactions decouple.

When the twin electron is light, decoupling proceeds analogously to the SM case. 
Using
\begin{equation}
    H
    \sim
    \frac{T_A^2}{\Mpl}
    =
    \frac{T_B^2}{\xi^2\Mpl}\, ,
    \qquad
    \Gamma
    \sim
    G_{F,B}^2T_B^5\, ,
\end{equation}
we obtain
\begin{equation}
    T_{B,\mathrm{dec},1}
    \sim
    \left(
        \frac{1}{\Mpl G_{F,B}^2\xi^2}
    \right)^{1/3}\, .
\end{equation}
Here, $\xi\equiv T_B/T_A$ is related to $\DNeff$ at CMB times through Eq.~\eqref{eq: Dneff in terms of temp ratio}. 
The resulting twin proton freeze-out abundance scales approximately as
\begin{equation}
    X_{p,B}
    \sim
    \exp\left(
        -\frac{|\DeltMnp|}{T_{B,\mathrm{dec},1}}
    \right).
\end{equation} 
and the phenomenologically interesting regime where $\fnaDM\sim 10\%$ is approximately 
\begin{equation}
    |\Delta m_{np,B}|\sim 50\ \textrm{MeV} \left(\frac{G_{F,B}}{G_{F,A}/10}\right)^{-2/3} \left(\frac{\xi}{0.2}\right)^{-2/3}\,.
\end{equation}

If instead the twin electron is heavy, the nucleon-conversion reactions become inefficient once the electrons and positrons become nonrelativistic and their equilibrium number densities are Boltzmann suppressed. 
We parametrize the corresponding decoupling temperature as
\begin{equation}
    T_{B,\mathrm{dec},2}
    \sim
    \frac{m_{e,B}}{x_e}\, ,
    \qquad
    x_e\sim \mathcal{O}(5\text{--}10)\, ,
\end{equation}
and use $x_e\simeq 6$ as a representative estimate.\footnote{
This decoupling occurs before the process $e^+e^-\leftrightarrow\gamma\gamma$ fully freezes out. 
The electron--positron bath remains close to chemical equilibrium while its equilibrium pair density becomes Boltzmann suppressed. 
Consequently, the familiar estimate $T\sim m/20$ for the freeze-out of a massive relic does not apply here.}
The resulting twin proton abundance scales parametrically as
\begin{equation}
    X_{p,B}
    \sim
    \exp\left(
        -\frac{|\DeltMnp|}{T_{B,\mathrm{dec},2}}
    \right)
    \sim
    \exp\left(
        -x_e\frac{|\DeltMnp|}{m_{e,B}}
    \right),
\end{equation}
It is therefore approximately constant along contours of fixed
$m_{e,B}/|\DeltMnp|$ and is comparatively insensitive to $G_{F,B}$ and
$\DNeff$ in this regime. 

In order for the reactions to remain in equilibrium the weak interaction must be faster than Hubble and there must be sufficient electrons to initiate the equilibrium reactions in \eqref{eq: freeze-out processes}.\footnote{While the weak scattering process with a neutrino in the initial state is not explicitly shut off by electron depletion, in practice this interaction is subleading to the process with an electron in the initial state, especially for larger electron masses, where the initial state neutrino needs to carry sufficient energy to create the final state electron.}
Therefore the physical decoupling temperature is approximately the larger of the two estimates:
\begin{equation}
    T_{B,\mathrm{dec}}
    \sim
    \max\left\{
        T_{B,\mathrm{dec},1},
        T_{B,\mathrm{dec},2}
    \right\},
\end{equation}
which determines the location of the turnover between the two regimes and thus the knee in the constant-abundance contours. The turnover occurs when the two temperatures become comparable. Their ratio is
\begin{equation}
    \frac{T_{B,\mathrm{dec},1}}{T_{B,\mathrm{dec},2}}
    \sim \left(\frac{x_e}{6}\right)
    \left(\frac{G_{F,A}/10}{G_{F,B}}\right)^{2/3}
    \left(\frac{100\ \textrm{MeV}}{m_{e,B}}\right)
    \left(\frac{0.2}{\xi}\right)^{2/3}.
\end{equation}
When this ratio is much larger than unity, decoupling is SM-like, whereas if it is much smaller than unity it is controlled by the Boltzmann suppression of the twin-electron abundance.

This estimate is only parametric. As the temperature decreases, the scaling of the relevant weak rates can change from $G_{F,B}^2T_B^5$ to
$G_{F,B}^2T_B^3|\DeltMnp|^2$ \cite{Mukhanov:2003xs}, while threshold and Boltzmann-suppression effects also become important. 
The precise location of the turnover and the $\fnaDM=10\%$ curve can therefore depend on $m_{e,B}$, $\DeltMnp$, $G_{F,B}$, and $\xi$ (or equivalently $\DNeff$).
Since $\xi$ varies only modestly over the range considered here, its effect is the least pronounced. This effect can be seen in \cref{fig: nadm abundance Dneff}.

The remaining model parameters ($ g_{a,B},\ \Delta m_{du,B},\ B_{He,B}$) have minimal effect on the fractional naDM abundance. 
The axial coupling $g_{a,B}$ modifies the weak interaction rates by an $\OfOrder{1}$ factor and $\Delta m_{du,B}$ modifies the exact quark masses and thus the underlying cosmology, but neither effect is significant within the allowed ranges. 
$B_{He,B}$ determines whether twin helium (and therefore possible heavier nuclei) is an accessible stable state, but does not affect the calculation of the total naDM abundance.

\begin{figure}
    \centering
    \includegraphics[width=0.7\linewidth]{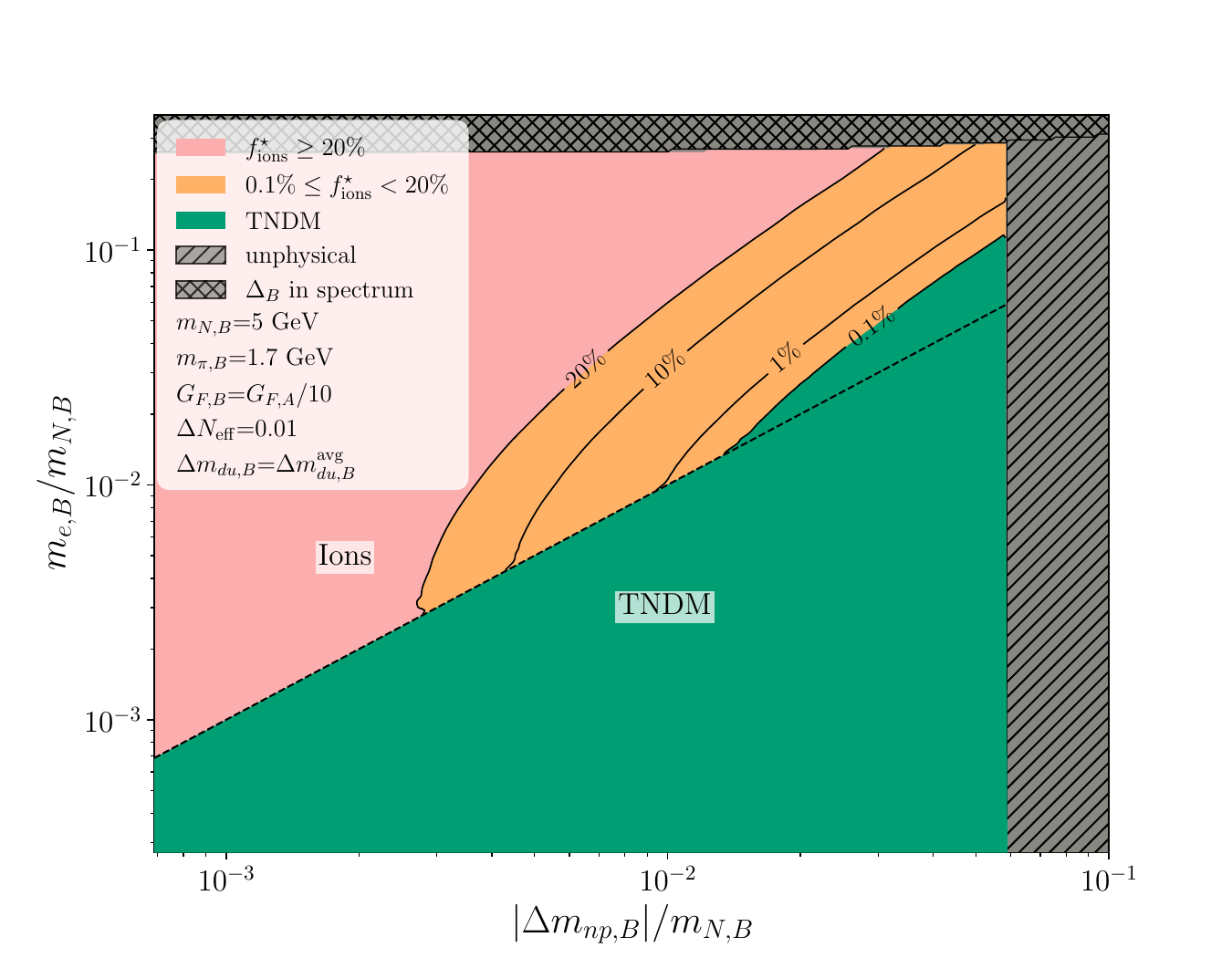}
    \caption{
    Possible relic twin proton ion fraction as a function of $|\DeltMnp|/m_{N,B}$ and $m_{e,B}$ for the canonical benchmark if the twin deuteron is unbound. 
    Here the \textcolor{orange}{orange} region is used to indicate naively viable fractions of relic ions.
    The naDM phase boundaries identified in \cref{fig: unbound D phase diagram} are overlaid as dashed lines. If the twin proton is stable, it is possible, though highly uncertain, that a relic ion population could survive until late times.
    }
    \label{fig: nadm abundances unbound deuteron benchmark}
\end{figure}

In \cref{fig: nadm abundances unbound deuteron benchmark} we show the twin proton abundance for the scenario where the twin deuteron is unbound (see \cref{fig: unbound D phase diagram}). In the regions of parameter space where the twin proton is stable, atomic states cannot form, since twin hydrogen is unstable. We therefore identify the only other DM candidate population as $f^\star_{\rm ions}$, where once again the star indicates that if recombination is efficient, the population of ions can be eliminated or severely reduced via electron capture at recombination. 
If the twin proton is unstable, no relic population can persist until late times. 

The largest uncertainty in our results remains the binding energy of nuclei, which sets the final abundance. However, for a particular set of twin nuclear assumptions, our results highlight the values of underlying parameters that could give rise to a phenomenologically viable cosmology. 
Depending on the details of the twin sector, our setup can yield either pure neutron DM or a mixed scenario containing a subdominant naDM component. 
The naDM component arises within a relatively narrow region of parameter space, typically around $|\DeltMnp|/m_{N,B}\sim 1\%$. 
Importantly, twin DM with a mass of order $5~\mathrm{GeV}$, as motivated by the matter coincidence problem in an IR \Ztwo-breaking framework, can still contain a substantial naDM fraction in parts of the parameter space. 
The presence of such a sub-component leads to a rich phenomenology to which we now turn.


\section{Cosmological and Astrophysical Observables}\label{sec: observables}

We now consider the cosmological and astrophysical observables of the naDM scenario. 
The most distinctive discovery opportunity is provided by a \textit{two-stage DAO} (\cref{subsec:dao}): in the first stage the entire DM population oscillates, followed by an epoch where only the naDM fraction oscillates.
Although only a subdominant fraction of the DM is charged under long-range twin electromagnetism, strong scattering between twin nuclei and the dominant TNDM at early times can acoustically drag the entire dark sector. 
The resulting suppression and oscillatory structure in the primordial matter power spectrum
could therefore remain observable at near-galactic scales in future surveys,
even for percent-level naDM abundances.
Once the twin strong nuclear interactions freeze out, only the charged component oscillates, allowing the smaller-amplitude second-stage DAOs to persist at larger scales, which can be  constrained or discovered in cosmological observations. 

We argue in~\cref{sec: bullet cluster} that the self-interaction of twin neutrons (and twin atoms when $\fnaDM \lesssim 0.1$) is likely too weak to show up in Bullet Cluster observations, but it may be discoverable in future higher-precision probes. 
We then briefly discuss the dynamics of naDM in the galaxy in~\cref{subsec: cosmo ionization}, which sets the stage for our discussions of direct detection in~\cref{sec: terrestrial observables}, as well as the possible formation of mirror stars and other compact objects, see~\cref{sec: mirror stars}. These can be highly exotic given the rich twin nuclear physics of the naDM dark sector and the effect of nuclear and atomic stability conditions, such as the absence of twin hydrogen atoms. 
We close in~\cref{subsec: phase transition} by briefly explaining how gravitational waves from cosmological phase transitions may arise in the naDM scenario. 
We focus in this section on the effects of light twin atoms (D-He) but comment, where relevant, on the possible effects of heavier states where appropriate.  

In discussing the cosmological and astrophysical signatures, we again take efficient recombination as our baseline. 
When relevant, we also comment on how inefficient recombination (and post-recombination instability of twin atoms) can affect the potential signatures.


\subsection{Dark Acoustic Oscillations}\label{subsec:dao}

Because the naDM is in thermal contact with the dark photons, in the tight coupling limit it will undergo DAOs with the dark photons in the same way that SM baryons oscillate with the SM photon. This leads to a suppression of structure on scales that enter the horizon prior to dark recombination and gives rise to oscillatory features in the matter power spectrum. 
Novel to the naDM framework are the twin neutron-proton interactions, which drag the TNDM alongside the naDM component at early times, resulting in the entire DM population initially participating in DAOs. This manifests as maximal oscillatory suppression of the matter power spectrum, but only at relatively small galactic scales. 
After the twin strong interactions decouple, only the naDM undergoes oscillations.

Provided that dark recombination occurs at low enough temperatures, these effects can be detected in CMB measurements~\cite{Bansal:2022qbi, Barron:2026nks}, the UV luminosity function (UVLF) of galaxies \cite{Barron:2025dys}, Lyman Alpha signatures  \cite{Bose:2018juc, Yuan:2026zlr}, and 21-cm signatures~\cite{Munoz:2019hjh}.
Existing studies find that the strongest current constraints on DAOs at these scales arise from measurements of the high-redshift UVLF~\cite{Barron:2025dys}. 
When all of the dark matter participates in the acoustic oscillations, these observations require the onset of DAO suppression to occur at wavenumbers \(k_{\rm DAO}\gtrsim 50\,h/{\rm Mpc}\)~\cite{Yuan:2026zlr}. 
When instead only a subdominant naDM fraction participates, the resulting suppression is weaker and may begin on larger spatial scales. 
In particular, for \(f_{\rm naDM}\lesssim7\%\), DAOs on any scale are allowed~\cite{Barron:2025dys, Yuan:2026zlr}.

It should be noted that the DAO signal arises from the population of naDM in the early universe (pre-recombination), whereas the competing self-interaction bounds, e.g. at the Bullet Cluster (to be discussed in \cref{sec: bullet cluster}), are probing the late-universe (post-recombination) abundance of naDM. 
The early and late time naDM populations can be substantially different, due to the different criteria for atomic and nuclear stability, see \cref{subsec:stability} for details.

To determine how cosmological measurements are sensitive to DAO signals in the naDM scenario, it is necessary to estimate the temperature at which dark recombination occurs. 
(Unless otherwise stated, we adopt efficient recombination as our baseline and return below to the modified DAO history that can arise from residual ions when recombination is incomplete.)
The bound-electron-capture channel that destabilizes some twin atoms is absent before recombination and is therefore neglected in estimating the recombination temperature \(T_{\mathrm{rec}}\). As discussed in \cref{subsec:stability}, its rate, estimated in \cref{eq:elec_capture_rate}, can be rapid on cosmological timescales while remaining slow compared to the atomic formation process; for example, the twin-hydrogen lifetime is typically \(\mathcal{O}(10^2)\) years in our parameter space. Thus, for determining the onset of recombination and the associated DAO scale, we may treat nuclei that are stable at that epoch as forming metastable atoms, including twin hydrogen. Electron capture can subsequently alter the post-recombination composition without appreciably shifting \(T_{\mathrm{rec}}\), giving rise to the range of late-time phases summarized in \cref{tab: phase diagram regions}.

Beginning with the case of twin hydrogen, the relative fraction of free twin electrons is
\begin{equation}
    X_{e,B}= \frac{n_{e,B}}{n_{p,B} + n_{H,B}},
\end{equation}
with $n_{p,B}$ ($n_{H,B}$) the number density of free twin protons (atomic hydrogen). Note that in SM calculations it is conventional to define $X_e=n_e/n_b$, but when the twin protons are a small sub-component, it is convenient to use the form above. In equilibrium we can find $X_{e,B}$ using the Saha equation, using $n_{p,B} + n_{H,B} = X_{p,B} n_{b,B}$ with $n_{b,B}$ from Eq. \eqref{eq: dark baryon number density}:
\begin{equation}
    X_{e,B} = \frac{2}{1+ 
    \sqrt{1+16\left(\frac{2}{\pi}\right)^{1/2}\zeta(3) \eta X_{p,B}  \left(\frac{T_A}{m_{e,B}\xi}\right)^{3/2}\left(\frac{m_{N,A}}{m_{N,B}}\right)\left(\frac{\Omega_{\rm DM}}{\Omega_b}\right)
    e^{B^e_{H,B}/T_A\xi}
   }}\, ,
\end{equation}
where $B^e_{H,B}$ is the electron binding energy of the twin hydrogen and we use the superscript $e$ to distinguish these electronic binding energies from nuclear binding energies. 
The electronic binding energy depends on the Bohr radius of a twin hydrogen atom, which is given by $a_{0,B} = \left( m_{e,B} \alpha_B \right)^{-1}$. This assumes that the central nucleus is much heavier than the electron and can be approximated as stationary, and the substitution of the reduced mass of the twin electron-proton system $m_{e,B}\rightarrow \mu_{red,B}$ should be made for heavy twin electrons. The electronic binding energy of twin hydrogen is 
\begin{equation}\label{eq: hydrogen binding}
    B^e_{H,B} = \frac{1}{2 a_{0,B}^2 \mu_{red,B}} = \frac{\mu_{red,B} \alpha_B^2}{2 } \, .
\end{equation}

We define the recombination temperature $T_{\rm rec}$ as the temperature (expressed in terms of the SM photon temperature $T_A$) at which only 10\% of dark electrons remain free. This can be approximated as the solution to the implicit equation
\begin{equation}
    \frac{B^e_{H,B}}{T_{\rm rec}\xi}\simeq 47 + \ln\left[ \left(\frac{m_{N,B}/m_{N,A}}{5}\right)\left(\frac{m_{e,B}\xi/T_{\rm rec}}{10^6} \right)^{3/2} \left(\frac{0.1}{X_p} \right)\right] \;.
    \label{eq:Trec_full}
\end{equation}
The logarithmic term is small and does not vary much over the parameter space of interest.
Using the hydrogen binding energy of \eqref{eq: hydrogen binding}, we can estimate the temperature of the SM when twin hydrogen recombines as
\begin{equation}
    \label{e.TrecB}
    T_{\rm rec} \sim \frac{B^e_{H,A}}{47}\left(  \frac{\alpha_B^2m_{e,B}}{\alpha_A^2 m_{e,A}}\right)\frac{1}{\xi}\simeq
  1.5 \, \mathrm{eV} \left(  \frac{\alpha_B^2m_{e,B}}{\alpha_A^2 m_{e,A}}\right)\left(\frac{0.2}{\xi}\right)\,.
\end{equation}
This assumes $\alpha_B$ is not so tiny that $e_B$ and $p_B$ decouple from photons even before they form twin atoms.
Here $B^e_{H,A}/47 \simeq 0.3$ eV corresponds to the recombination temperature of SM hydrogen. 
Note that the dependence on $\xi$ is inverted, which has the effect of making recombination occur earlier in the dark sector. However, since $\xi$ only varies by a factor of 2 over our parameter space the effect is small. 
The scale corresponding to this decoupling is approximately 
\begin{equation}\label{eq:DAO_recombination}
k_{\rm rec}\sim
  0.1\,h \,\mathrm{Mpc}^{-1}\left(  \frac{\alpha_B^2m_{e,B}}{\alpha_A^2 m_{e,A}}\right)\left(\frac{0.2}{\xi}\right)\,,
\end{equation}
which is well within the region constrained by DAO observations, and thus requires that the acoustic fraction at this time is $\fnaDM\lesssim 7\%$. 
For most of the parameter space of interest, the resulting oscillatory features in the matter power spectrum can then be probed by cosmological observations.

For heavy twin electrons ($m_{e,B}\gtrsim 200$ MeV), or very large dark fine structure constants ($\alpha_B \gg \alpha_A$), dark recombination can occur at $T_{\rm rec}\sim 0.6$ keV and $k_{\rm rec}\gtrsim 50 \,h \,\mathrm{Mpc}^{-1}$, naively loosening the limits on $\fnaDM$. However, in this regime recombination may not be efficient and a more detailed analysis would be required to set precise bounds. 
While Eq.~\eqref{eq:Trec_full} estimates the equilibrium recombination scale, it does not by itself demonstrate that the plasma tracks the Saha equilibrium. We use this estimate under our baseline assumption that the relevant recombination rates remain faster than the Hubble rate. If this assumption fails, a residual ionized fraction survives and a modified DAO history arises, discussed below.

If a heavier twin atom constitutes most of the naDM, the relative abundance of free twin electrons \textit{per proton} is given by $X_{e,B}=n_{e,B}/\Znum_B(n_{atom,B}+n_{ion,B})$, where $\Znum_B$ represents the atomic number of the species and $n_{atom,B}$ ($n_{ion,B}$) the number density of atomic (ionized) nuclei. In the limit that this single species constitutes all of naDM, 
\begin{equation}
    n_{atom,B}+ n_{ion,B} = \frac{\fnaDM}{\Anum_B}n_{b,B}\;.
\end{equation}
Changes in the prefactors of $\OfOrder{1}$ due to the inclusion of $\Anum$ and $\Znum$ will have little effect relative to the change in binding energy, and the recombination temperature can be well approximated by 
\begin{equation}
    \label{e.Trecatom}
    T_{\mathrm{rec},atom}\sim \frac{B^e_{\Anum,A}}{47}\left(  \frac{\alpha_B^2m_{e,B}}{\alpha_A^2 m_{e,A}}\right)\frac{1}{\xi}.
\end{equation}
The helium ionization energy is a factor of a few higher than the hydrogen, and the recombination temperature will be correspondingly higher. However, the resulting signals are still expected to be within reach of cosmological observations in most parts of the parameter space. If heavier nuclei form, they are likely to (at least partially) recombine at even higher energies.
A precise computation of the matter power spectrum and DAO signatures for general naDM scenarios is left for future work.

Thus far, we have only considered DAO signals arising from oscillations of the naDM sub-component of DM. However,
at early times, the dark neutrons are in equilibrium with the naDM through the weak and strong interactions. The dark neutrons experience drag from the naDM and undergo DAOs alongside the naDM. 
Before $n_B-p_B$ decoupling, the charged component can drag essentially \textit{all of DM} into acoustic oscillations. This makes the DAO feature potentially observable even when $\fnaDM \lesssim 10\%$.
To understand this effect, we now estimate the temperature at which the twin neutrons decouple from the twin protons.

It is easy to see that, just as in the SM, the weak interactions in the dark sector decouple at temperatures above an MeV for the range of $G_{F,B}$ we consider, and are thus safe from a DAO perspective.
The strong interactions in the dark sector are a different story. Below $T_B\sim $ GeV, the dark baryons will be non-relativistic. Based on the discussion in \cref{sec: bullet cluster} below, when the dark baryons are non-relativistic their self-interactions can be parameterized in terms of two scattering lengths, $a_{0,B}$ and $a_{1,B}$. In the limit that $a_{0,B}\simeq a_{1,B} \equiv a_B$ and $a_B\ll 1/(m_{N,B} v_0)$, the strong interaction cross section scales as $\sigma_s \simeq \pi a_B^2$. In the limit that $a_B\gg 1/(m_{N,B} v_0)$ we find $\sigma_s \simeq \pi/(m_{N,B} v_0)^2$. We use the late time twin nucleon number density, Eq. \eqref{eq: dark baryon number density}. Letting $L_B^2 = \min \left[a_B^2,(m_{N,B} \xi T_A)^{-1} \right]$ and using $v_0^2 \sim T_B/m_{N,B}$, this yields an approximate rate for twin neutrons to scatter off of twin protons:
\begin{equation}
    \Gamma_{s,B} = n_{p,B} \angl{\sigma v}
    \sim  X_{p,B} \eta \,T_A^3\,\frac{\Omega_{\rm DM}}{\Omega_b}\,\frac{m_{N,A}}{m_{N,B}}
    L_B^2\sqrt{\frac{\xi T_A}{m_{N,B}}}.
\end{equation}
Equating this to the Hubble rate, we can estimate the temperature of the SM photon bath at which twin neutrons and twin protons decouple as
\begin{equation}
    T_{n_Bp_B,dec} \sim \frac{m_{N,B}\,  \gstarS^{1/3} }{\left(\eta X_{p,B} L_B^2 m_{N,A} \Mpl (\Omega_{\rm DM}/\Omega_b) \xi^{1/2}\right)^{2/3}} .
\end{equation} 
Inserting the measured values of $\eta, \ m_{N,A},\ \Mpl,\ \Omega_{\rm DM}/\Omega_b$, late time $\gstarS=3.94$, and including the numerical prefactors neglected above, we find
\begin{equation}
    T_{n_Bp_B, dec}\simeq 
    0.6\ \mathrm{keV} 
    \left(\frac{m_{N,B}}{5\ \mathrm{GeV}}\right)
    \left(\frac{0.05}{X_{p,B}} \right)^{2/3}
    \left(\frac{10 \ \mathrm{GeV}^{-1}}{ L_B}\right)^{4/3}
    \left(\frac{0.2}{\xi}\right)^{1/3}\, .
\end{equation}
Here $X_{p,B}=0.05$ would correspond to $\fnaDM=0.1$ if nuclei heavier than hydrogen formed. $L_B \sim 10\ \mathrm{GeV}^{-1}$ is in the fm range, as would be expected for a strong nuclear force. Note the extremely weak dependence on $\DNeff \propto \xi^{4}$.
The scale today corresponding to this decoupling temperature is
\begin{equation}
    k_{n_Bp_B, dec} \sim 
     50 \,h \,\mathrm{Mpc}^{-1}
    \left(\frac{m_{N,B}}{5\ \mathrm{GeV}}\right)
    \left(\frac{0.05}{X_{p,B}} \right)^{2/3}
    \left(\frac{10 \ \mathrm{GeV}^{-1}}{ L_B}\right)^{4/3}
    \left(\frac{0.2}{\xi}\right)^{1/3}.
    \label{eq:k_npdec}
\end{equation}

The imprint left on the matter power spectrum strongly depends on the order of $T_{\rm rec,B}$ and $T_{n_Bp_B,\rm dec}$.
The DAO history is controlled by two successive links in the momentum-transfer chain: scattering between the dominant neutral TNDM and the charged nuclear component, and the coupling of that charged component to twin photons. Under our baseline assumption of efficient recombination, photon drag on naDM becomes ineffective at \(T_{\rm rec}\). 
If 
\begin{equation}
    T_{n_Bp_B,{\rm dec}}>T_{\rm rec} \, ,
    \label{eq:Tdecrec_baseline_order}
\end{equation}
the \(n_B-p_B\) interaction decouples first: essentially all DM undergoes acoustic oscillations until \(T_{n_Bp_B,{\rm dec}}\), after which only the naDM component continues to oscillate until recombination. This is the baseline \textit{two-stage} DAO history which holds in almost all of our parameter space.
The resulting matter transfer function, therefore, contains two characteristic scales: modes with \(k\gtrsim k_{n_Bp_B,\rm dec}\) experience an approximately all-DM acoustic suppression, whereas modes entering between \(p_B\!-\!n_B\) decoupling and dark recombination experience only the weaker, naDM-component suppression.

For the strong-DAO benchmark with \(100\%\) interacting dark matter, Ref.~\cite{Barron:2025dys} finds \(k_{\rm DAO}\simeq 50\, h\,{\rm Mpc}^{-1}\). (The Lyman-\(\alpha\) analysis of Ref.~\cite{Yuan:2026zlr} places comparable bounds on this scenario.)
If we approximately identify the scale generated by \(p_B\!-\!n_B\) decoupling with this physical DAO scale, then for \(L_B=10\,{\rm GeV}^{-1}\), Eq.~\eqref{eq:k_npdec} implies bounds of \(\fnaDM\lesssim{\cal O}(5\!-\!10\%)\).
A precise limit requires calculation of the two-stage matter transfer function.

Following \(p_B\! - \!n_B\) decoupling, the residual naDM-only oscillations are subject to the usual fractional-DAO limits. These limits depend jointly on the interacting fraction and the DAO scale: for example, Ref.~\cite{Barron:2025dys} obtains \(k_{\rm DAO}> 20\, h\,{\rm Mpc}^{-1}\) for an interacting fraction \(f_{\rm naDM}=0.1\), while no meaningful UVLF bound on \(k_{\rm DAO}\) remains for \(f_{\rm naDM}\lesssim0.07\). Thus, the second stage need not be observable when \(\fnaDM\) is sufficiently small, even though the earlier all-DM stage can leave a substantial small-scale imprint. These bounds are derived for a simpler oscillation history than described here, however, so long as the decoupling scales $T_{\rm rec}$ and $T_{n_Bp_B,\rm dec}$ are well separated, diffusion damping effects at the two stages should not interfere and the constraints of Ref.~\cite{Barron:2025dys} should still approximately apply.

In a small region of our parameter space, we could instead have
\begin{equation}
    T_{\rm rec} > T_{n_Bp_B,{\rm dec}} \, .
    \label{eq:Tdecrec_alternative_order}
\end{equation}
In this region, efficient recombination removes the charged scatterers while the neutral and charged nuclear components are still coupled, terminating the acoustic motion of essentially all DM in a single stage at $T_{n_Bp_B,\mathrm{dec}}$.

Both the $T_{\rm rec} < T_{n_Bp_B,{\rm dec}}$ and the $T_{\rm rec} > T_{n_Bp_B,{\rm dec}}$ scenarios are modified by inefficient recombination. In this case, oscillations of the electrically charged component are terminated when Compton-scattering of the twin electrons freezes out at $T_\mathrm{Com,dec}$. Qualitatively, this just replaces $T_\mathrm{rec} \to T_\mathrm{Com,dec}$ in both scenarios. Notably, this may lead to the second stage of DAOs persisting for much longer than they would if recombination was efficient. CMB constraints on DAOs on large scales may then constrain $f_\mathrm{naDM} \lesssim \mathrm{O}(1\%)$~\cite{Bansal:2022qbi}.\footnote{One could imagine a maximally complicated and scenario where recombination had $\mathcal{O}(1)$ efficiency, leading to a three-stage DAO signature with $T_{n_B p_B,\mathrm{dec}} > T_\mathrm{rec} > T_\mathrm{Com, dec}$.}

The possible scenario with twin multi-neutron bound states may also modify this picture. A pure multi-neutron state can still be dragged along with the twin protons (if they are not depleted to the point that DAOs do not occur), though the characteristic interaction length will be modified compared to the single nucleon expectation.  If neutron-rich isotopes form, the naive identification of $\fnaDM \sim 2X_{p,B}$ can be violated. In regimes where there are a few surplus neutrons in each nucleus, the fraction undergoing oscillations is correspondingly higher. In this case, two-stage DAOs can still occur, but the fraction of DM participating in each stage is modified. In the extreme limit, if every twin neutron is bound in a state which includes at least one twin proton, then the effective fraction of DM that undergoes oscillation is 100\% until recombination.

The coupling of the cold TNDM component to naDM is therefore an amplifier of the DAO signal: a small charged fraction can imprint acoustic structure on nearly the entire dark sector, giving rise to a cutoff and oscillatory features in small-scale structures.
Our limited understanding of the twin nuclear interaction length $L_B$ is a significant bottleneck in precise predictions, but for an approximate expected scale of $L_B \sim \mathcal{O}(1~\mathrm{fm})$, the predicted feature lies near the scales currently reached by UVLF and Lyman-$\alpha$ measurements. 
In the absence of these oscillatory features, the same measurements bound the interaction length and naDM fraction.


\subsection{Bullet Cluster Self-Interaction Constraints}\label{sec: bullet cluster}

An important probe of dark matter self-interactions arises from observations of the Bullet Cluster \cite{Markevitch:2003at} and other systems. The Bullet Cluster system is comprised of a subcluster which has collided with a larger cluster. The gas in the two clusters is highly collisional and is slowed, while lensing observations show that the bulk of the mass density (the stars and the DM) pass through essentially without collision. 
These observations preclude a fraction of DM which experiences long range interactions larger than $\sim 10\%$.
Nevertheless, given current uncertainties \cite{Fan:2013yva}, smaller fractions of naDM are still viable. 
We therefore focus on the possibility of discovering DM self-interaction of the dominant twin-neutron component.
Simulations of Bullet Cluster mergers compared to observations constrain the DM self-interaction cross-section to be $\sigma/m<1.25\ \rm{cm}^2/g$ \cite{Randall:2008ppe}. This is slightly smaller than SM neutron-neutron interactions. Thus, further observations and simulations of clusters~\cite{Harvey:2018uwf, Andrade:2020lqq, Adhikari:2022sbh,Silich:2023mvd,ODonnell:2025pkw} could provide a promising avenue for discovery of our setup if twin neutrons have self-interactions close to their SM counterparts. 

SM nucleon interactions are known to be enhanced at physical pion masses \cite{Chen:2010yt}, due to the deuteron binding energy being unusually small. Explicitly, at low energy, the nucleon-nucleon scattering amplitude can be expanded in terms of the effective range expansions \cite{Bethe:1949yr, Blatt:1949zz,Chu:2019awd}

\begin{equation}
    \mathcal{M}_{s,B} = \frac{4 \pi}{m_{N,B}(-i p -a_{s,B}^{-1} + \frac{1}{2} r^0_{s,B}p^2 + \OfOrder{p^4})}.
\end{equation}
Here $a_{s,B}$ are scattering lengths and $r^0_{s,B}$ the effective range parameters in the singlet $s=0$ and triplet $s=1$ total spin channels. 
The scattering length, in particular, parametrizes how close the system lies to a pole in the S-matrix (the bound state of the deuteron, in the SM case).  If the scattering length is large relative to $1/(m_{N,B} v)$, the cross-section will approach the hard sphere scattering limit $\sigma \sim 1/(m_{N,B} v)^2$. If the scattering length is small, the cross section tends to $\sigma \sim a_B^2$ and $a_B$ can be interpreted as a `geometric' radius. Based on the discussion in \cref{subsec: dark binding energies}, it is likely that in much of our parameter space the deuteron will be less tuned, and thus the scattering length will be smaller than the SM. 

Since the nucleons are very non-relativistic in the late universe, we expect the dependence on $r^0_{s,B}$ and the higher $p$ terms to be negligible and focus only on the scattering lengths. The cross section is
\begin{equation}
    \sigma = \sum_{s=0}^1(2s +1) \frac{m^2_N |\mathcal{M}_{s,B}|^2}{16 \pi} \, ,
\end{equation}
and depends on the relative velocity of the dark matter via the non-relativistic momentum relation $p = m_{N,B} v/2$ where the factor of two comes from the reduced mass. Following the treatment of Ref.~\cite{Cline:2022leq}, we integrate the cross section over a Maxwell Boltzmann distribution and find 

\begin{equation}
    \begin{aligned}
        \angl{\sigma} &= \sum_s (2s +1) \frac{a_{s,B}}{m_{N,B}}F(b_{s,B}),\\
        F(b_{s,B}) & = 2\pi b_{s,B}
        \left[1 
        - |b_{s,B}|
        \sqrt{\frac{\pi}{2}} 
        e^{b_{s,B}^2/2}
        \left(1- \erf|b_{s,B}|\right) \right], \\
        b_{s,B} &\equiv \frac{2}{m_{N,B} a_{s,B} v_0} \, .
    \end{aligned}
\end{equation}
Taking the approximation that $a_{0,B}\approx a_{1,B}$, we find that for most of our parameter space Bullet Cluster considerations require $|a_{i,B}|\lesssim 100$ fm. The SM value for neutron--proton scattering in the singlet channel, $a_{0,A} = -23.7148(43)$ fm \cite{Hackenburg:2006qd}, is already anomalously large compared to the naive expectation $a_0\sim \Lambda^{-1} \sim \mathrm{fm}$ due to the shallowly bound deuteron state.  All other scattering lengths in the SM (triplet neutron--proton and singlet and triplet neutron--neutron) are smaller than the singlet channel and closer to this expectation. Therefore, it is likely that for most of our parameter space Bullet Cluster observations have no constraining power for purely strong-force mediated nuclear interactions, unless the dark sector deuteron system is even more tuned than the SM version. 

The scattering length of heavy nucleons can be estimated by multiple methods. Ref.~\cite{Chen:2010yt} computed the pion mass dependence of the scattering length in a ``pionless'' nuclear EFT. While this expansion is not valid for the large values of $m_{\pi,B}/m_{N,B}$ considered in this work, it is notable that they found that as $m_{\pi,B}$ increases above the SM point, $1/a_{0,B}$ crosses zero, and so the scattering length appears to diverge and the cross section saturates the unitarity bound $\sigma \sim 1/k^2 \sim 1/m_{N,B}^2 v_0^2$. This occurs for a very narrow range of pion masses a few MeV larger than the SM value. Models existing near this point would in fact exhibit the tuning described above and likely be excluded by Bullet Cluster considerations. However, we note that the final numerical result of Ref.~\cite{Chen:2010yt} was found by fitting to values of the scattering length computed on the lattice for heavier than physical pion masses. Lattice calculations of two-nucleon observables are currently still subject to large debates (see the discussion in \cref{subsec: dark binding energies}), so the exact location (and possibly the existence) of the pole in $a_{0,B}$ is not concretely known.

For pion masses above the range of the EFT, we have only lattice calculations to draw from, subject to the same uncertainties.  Recent lattice calculations at $m_{\pi,B} \sim 700-800$ MeV, although still with significant systematic uncertainties, point to scattering lengths $a_Bm_{\pi,B} \sim 3-4$ \cite{BaryonScattering:2025ziz}, or in the case of Ref.~\cite{Detmold:2026vjx}, a bound $a_Bm_{\pi,B} \gtrsim 10$ (or in fm, $a_B \gtrsim 2.5$ fm using $m_{\pi,B} = 800$ MeV.) Based on these results, we conclude that the vast majority of our parameter space will likely remain safe from Bullet Cluster constraints from purely strong interaction, though there is a regime of near exact mirror models which may be excluded by more precise calculations of the scattering length. 

Finally, we comment on the effects of heavier states, focusing on neutral states which could in principle form a large component of DM. In the geometric limit, $\sigma/m$ decreases with mass for composite states. For twin multi-neutron states with $R \sim \Anum^{1/3}$ and $m\sim \Anum$, $\sigma/m \sim \Anum^{-1/3}$ and the bounds are slightly weaker for heavier twin elements. 
Twin \textit{atoms} will have a larger geometric radius than nuclei and are thus more strongly constrained, however, heavier twin atoms are heavily suppressed relative to twin deuterium; as in the SM, we expect that the twin atomic radii remain roughly constant as atomic mass number increases, so $\sigma/m\sim \Anum^{-1}$. By this argument, a twin metal atom will be more weakly constrained by the Bullet Cluster than twin deuterium.

A heavy twin atomic component crystallizing into twin dust would realize an even more extreme version of this effect.
For a dust grain with radius $r$ and density $\rho$, $\sigma/m \sim r^2/\rho r^3= 1/\rho r$. Thus at fixed density, larger grains are less constrained by Bullet Cluster interactions. Additionally, for most of our parameter space $m_{e,B}>m_{e,A}$, and thus the twin atomic radius is smaller than that of a SM atom of the same $\Anum$ and the twin crystals would be correspondingly denser, possibly by multiple orders of magnitude. These arguments suggest that naDM clumped into grains of dust could make up a larger fraction of DM than naive interpretations of the long range Bullet Cluster bounds ($\sim10\%$) imply; the DAO bounds of \cref{subsec:dao} would still constrain the nuclear fraction and could provide a pathway for discovering our naDM in this scenario. 
On the other hand, in the \Ions\ phase from \cref{tab: phase diagram regions}, the Bullet Cluster bounds on naDM directly apply to our scenario and the abundance of twin atoms ought to be limited to below $\sim 10\% - 20\%$.

\subsection{naDM Ionization and Distribution in the Milky Way}\label{subsec: cosmo ionization}

In \cref{subsec: naDM abundances}, we determined the fraction of DM that can reside in the naDM component across the parameter space. 
This calculation, however, does not establish whether these relics remain ionized or efficiently recombine with twin electrons to form neutral twin atoms. 
In this section, we outline an approximate criterion for identifying the regions of parameter space in which the present-day configuration of dark atoms in the Milky Way halo is likely to be ionized or atomic. 
We emphasize that the small $\fnaDM < 0.2$ fractions we consider would render even ionized naDM exempt from most self-interaction bounds.

The present day dark ionization fraction will depend on the history of the dark sector between the end of BBN and the present day. The atomic component of the dark sector could be subject to gravitational collapse and shock heating, comparable to SM baryons. The atomic component may remain in a halo if sufficiently dilute or may collapse into a disk. A detailed calculation of this history is beyond the scope of this work. In our phenomenology results, we will instead consider a few boundary cases, following the treatment of Ref.~\cite{Chacko:2021vin}, and here we briefly comment on the likely present day equilibrium state of the ionization.

A first order test is whether the virial temperature of the halo is sufficiently large that collisions have a large probability of dissociating a bound twin electron from the nucleus. 
To determine this, we should compare the kinetic energy 
\begin{equation}
    T_{vir} \approx \frac{1}{2}m_{N,B} v_{avg}^2 \, ,
\end{equation}
to the electron binding energy in the dominant naDM component. 
For the purpose of our rough estimates, we use the electron binding energy in hydrogen $B^e_{H,B}$, see \cref{eq: hydrogen binding}, as a proxy for the electron binding energy in twin atoms. 
For $T_{vir}\gg B^e_{H,B}$, we expect the gas to approach complete ionization, while for $T_{vir}\ll B^e_{H,B}$ we expect the bound state to dominate. 

The ratio of the binding energy to the virial temperature is 
\begin{equation}
    \frac{T_{vir}}{B^e_{H,B}} \approx \frac{m_{N,B} v_{avg}^2}{m_{e,B} \alpha_B^2} 
    \approx 10^{-2} 
    \left(\frac{v_{avg}}{220\ \mathrm{km/s}}\right)^2
    \left(\frac{\alpha_A}{\alpha_B}\right)^2
    \left(\frac{m_{N,B}}{m_{e,B}}\right) \, .
\end{equation}
Thus for heavy electrons $m_{e,B}\rightarrow m_{N,B}$, $T_{vir}\ll B^e_{H,B}$, and it is possible that the halo will be primarily atomic. For the lightest electrons, this ratio will exceed unity, and the ionization fraction could be relevant. The analysis for the deuterium and helium is similar up to $\OfOrder{1}$ prefactors in the masses and atomic binding energies. We find that, for the \DHenaDM\ nuclei bound to the lightest twin electrons in our analysis, (small $m_{e,B}/m_{N,B}$ in figures in previous sections), the thermal motion of the DM halo may be sufficient to ionize the halo. 

Recombination of twin atoms is inefficient at large electron masses $m_{e,B}\sim $ GeV \cite{Kaplan:2009de,Cyr-Racine:2012tfp}. However, during cosmological evolution, there are many processes that could cause the ionization fraction to change. Shock heating during gravitational collapse is liable to cause additional ionization, and if mirror stars form (see \cref{sec: mirror stars}), the resulting dark photon emission could reionize the atomic states and result in a primarily ionized population. Cooling during the formation of a dark disk could result in the formation of additional bound states, even if recombination is inefficient at early times. 
A detailed calculation of the complete ionization evolution and simulation of the galactic naDM distribution is beyond the scope of this work.

Heavier nuclei would complicate this picture further. We leave a detailed treatment to future investigations, but note a few qualitative features. Because $T_{vir}$ scales with halo mass, high- and low-mass galaxies may have different ionization states. 
These differences could be substantially amplified when heavy nuclei dominate $\fnaDM$. In hot, massive galaxies, the heavy naDM nuclei will remain largely ionized and behave as a standard plasma. In cooler, low mass galaxies, the naDM would instead form atoms, which may further crystallize into twin dust. Such galactic-scale crystallization will modify the naDM self-interactions far more than recombination into free twin atoms, as would be expected for, e.g., helium.

In general, the naDM scenario could have significant effects on galaxy formation and dynamics. In certain regions of parameter space, the dissipative nature of the naDM sub-component may allow it to cool rapidly, so that its current distribution in the Milky Way is in the form of a dark disk~\cite{Fan:2013yva,Ghalsasi:2017jna, Chang:2018bgx, Huo:2019yhk, Shen:2021frv, Shen:2022opd, Roy:2023zar, Roy:2024bcu}. In addition, the sizable self-interactions between the twin neutrons mean that the primary DM component exhibits the features of SIDM~\cite{Spergel:1999mh}, though the self-interaction rate is likely too low for detection in the foreseeable future~\cite{Tulin:2017ara}.

The most dramatic impact on galactic dynamics likely arises from the dissipative naDM component.
Cooling processes in the dark sector that contribute to modified structure formation scale as $\alpha_B^p/ m_{e,B}^s$ with $p, \, s>0$ \cite{Rosenberg:2017qia}, corresponding to faster cooling for larger values of $\alpha_B$ and smaller 
dark electron masses. When $\alpha_B$ and $m_{e,B}$ take similar values as in the SM, the naDM will collapse into a disk~\cite{Roy:2023zar, Gemmell:2023trd}, which can leave significant imprints on galactic structure and the distribution and properties of subhalos. 
Characterizing those signatures systematically, as a function of (n)aDM parameters is an outstanding problem due to the high computational cost of conducting the corresponding simulations. Such a study is highly motivated, since it would not only supply new constraints on (n)aDM, but also elucidate the detailed discovery signatures one could read off galactic dynamics observables to potentially discover such a dissipative DM sub-component.

It should be noted that in our parameter space, $\alpha_B$ is close in value to $\alpha_A$ while $m_{e,B}> m_{e, A}$, resulting in a reduced cooling rate compared to the SM. 
The naDM in our Milky Way is therefore likely to still be in an uncooled ionized halo distribution, which naively minimizes its impact on galactic dynamics. Fortunately, this is also the regime where direct detection (see~\cref{sec: terrestrial observables}) is most sensitive, demonstrating the complementarity of astrophysical and terrestrial probes. Furthermore, even in this slowly cooling regime, detailed galactic simulation studies are highly motivated, since inhomogeneities in the halo may lead to localized cooling, collapse and compact object formation despite the average ionized state today, which could imply discovery opportunities that are not currently understood.


\subsection{Mirror Stars}
\label{sec: mirror stars}

As a concrete realization of aDM with twin nuclear physics, naDM gives rise to the full gamut of mirror star~\cite{Blinnikov:1982eh, Khlopov:1989fj, Foot:2000vy, Curtin:2019ngc,Curtin:2019lhm} signatures, and represents an attractive target for dedicated studies of dark stellar physics and associated signatures, especially in the predictive $m_{N,B} \approx 5~\mathrm{GeV}$ benchmark case where naDM solves the DM coincidence problem.

Like any aDM scenario, naDM leads to the formation of mirror stars in large parts of parameter space where naDM in our galaxy cools enough for dark gas clouds to collapse to stellar-scale objects that shine in dark photons.
The existence of twin nuclei means that such mirror stars could maintain dark fusion processes in their cores to dramatically extend their main sequence lifetimes compared to the Kelvin-Helmholtz timescale. 
In the presence of a small kinetic mixing between the dark and SM photons~\cite{Holdom:1985ag, Gherghetta:2019coi}, these mirror stars would capture SM gas from the interstellar medium, resulting in a ``nugget'' of SM gas that is heated to $\sim 10^4$~K by the mirror star core, resulting in distinctive optical and X-ray signatures~\cite{Curtin:2019ngc,Curtin:2019lhm, Armstrong:2023cis,Cabral:2026tjb}. 
For $m_{N,B}\approx 5~\mathrm{GeV}$ and dark electrons which are significantly heavier than SM electrons, we would expect such mirror stars to have core densities that are a few orders of magnitude higher than SM stars \cite{Ryan:2022hku}. The nuggets captured by such stars would populate a region of the HR diagram just below the white dwarf band~\cite{Cabral:2026tjb}, though their local abundance within the $\sim 100-1000$~pc of our Sun where such mirror stars are observable is very difficult to predict and depends on the local ionization history, see \cref{subsec: cosmo ionization}.

Mirror star stellar physics is extremely rich and exotic in the naDM scenario.
For instance, if the naDM population is dominated by heavier nuclei than helium, the evolution of these \emph{metal-dominated}  mirror stars could differ substantially from the H- and He-dominated evolution of SM stars, but would still be broadly recognizable from our experience with SM stars.
On the other hand, if the twin nuclei are stable, and twin atoms are unstable, and the naDM fraction never recombines efficiently, the ionic naDM fraction could still efficiently cool and form mirror stars. Twin electron capture, subsequent decay of twin nuclei, and twin neutron stability would then significantly impact mirror star evolution. For example, twin electron capture could serve as an energy source similar to nuclear fusion, and the mirror star would become enriched in twin neutrons. Interestingly, the end product of this `fusion process' has lighter nuclear weight than the initial fuel, assuming the naDM ions are not solely $p_B^+$ and twin neutrons do not form large bound states. 
Their nuclear interactions may therefore push the twin neutrons to accumulate in a shell somewhere in the mirror star, possibly near or on its surface. 
Depending on subsequent details of mirror star evolution, this  could result in very low-mass twin neutron stars far below the minimum mass implied by the scaling relations in, e.g., Refs.~\cite{Hippert:2021fch, Hippert:2022snq}. 
On the other hand, it may be the case that a large majority of the mirror star mass ends up undergoing twin electron capture, in which case even low-mass mirror stars may convert into twin neutron stars. High mass mirror stars above the twin Chandrasekhar mass may convert a much larger fraction of their mass to black holes than SM stars. 
Dominant production of twin multi-neutron bound states could also significantly
impact mirror star evolution and signatures.
These possibilities are a highly motivated subject for future dedicated study of this naDM scenario, since the detection of degenerate dark stellar relics is a promising discovery channel.

Mirror white dwarfs~\cite{Ryan:2022hku}, representing the evolutionary end-state of some fraction of mirror stars, have a mass that scales as $m_{N,B}^{-2}$, i.e. $\sim (0.05 - 0.1) M_\odot$ in naDM, and mirror white dwarf mergers across almost the entire range of motivated dark electron masses we have been considering might be detectable by future gravitational wave experiments like DECIGO~\cite{Kawamura:2008zza,Ryan:2022hku}, though such estimates are subject to large background uncertainties. 
Mirror neutron stars can have similar if slightly higher masses~\cite{Hippert:2021fch, Hippert:2022snq}, and their mergers would then be even more detectable due to their higher compactness, with earlier parts of the inspiral falling into the Advanced LIGO~\cite{LIGOScientific:2017vwq,LIGOScientific:2018cki,LIGOScientific:2018hze} sensitivity band, though existing searches for sub-solar-mass binaries~\cite{Andres-Carcasona:2022prl,
Bentara:2026crj,
Kacanja:2026byy}, mostly formulated to search for primordial black holes (PBHs), impose lower bounds on the mass which may exclude such candidates. 
On the other hand, black holes born from the collapse of more massive mirror stars may fall into the sensitivity window of these PBH searches, though current constraints on the dark matter fraction in such black holes are still too weak to meaningfully constrain naDM models~\cite{Kacanja:2026byy}.

Future prospects for the detection of stellar-scale dark compact objects in naDM are very exciting. Mirror stars can be looked for in existing stellar catalogues~\cite{Howe:2021neq, Armstrong:2023cis, Cabral:2026tjb}, and the naDM parameter space favoured to solve the DM coincidence problem should give rise to detectable gravitational wave signatures from sub-solar mirror white dwarf, neutron star or black hole mergers.
Finally, the likely mass range of these compact objects lies in the maximum sensitivity range of microlensing searches utilizing the LSST  survey by the Rubin observatory~\cite{Winch:2020cju}, meaning the next decade could realistically see a detection or much stronger constraints on naDM models.


\subsection{Cosmological Phase Transitions}
\label{subsec: phase transition}

The dark sector of naDM contains two possible sources of detectable first-order phase transitions: the dark electroweak phase transition, and the dark QCD confinement phase transition. The latter case is unlikely to be detectable in our studied regions of naDM parameter space: recent lattice studies~\cite{Cuteri:2020yke, Pham:2021ftz,Kaiser:2022vjg} suggest that a strong first order phase transition in QCD with two ``light'' quark flavours requires $m_{\pi,B} \gtrsim 4$~GeV for SM-like $\LQCD$, which requires $m_{\pi,B}/m_{N,B} \sim 2/3$ in the IR parameterization of our model. 
This is significantly above the heaviest dark pion masses we study, though given the significant theoretical uncertainties in our current understanding of the confining phase transition, a strong dark QCD phase transition may be possible in extreme corners of our naDM parameter space.

On the other hand, the dark electroweak phase transition is entirely unconstrained, and may well be first-order. 
It would therefore be interesting if a twin version of electroweak baryogenesis~\cite{Fujikura:2018duw,Matsedonskyi:2020kuy,Badziak:2022ltm,vandeVis:2025efm,Badziak:2025fdp}
could generate the $\mathbb{Z}_2$ symmetric abundances in both sectors. 
In that case, a stochastic gravitational wave signal corresponding to a strong phase transition~\cite{Caprini:2019egz,Huang:2020crf,LISACosmologyWorkingGroup:2022jok} at $T \sim f \sim \mathcal{O}(\mathrm{TeV})$ would not only lend support to the existence of an extended scalar sector, but possibly specify, in the context of other signatures discussed in this work, the baryogenesis mechanism operating within the naDM scenario to address the matter coincidence problem.


\section{Terrestrial Observables}
\label{sec: terrestrial observables}

We now discuss terrestrial observables of naDM, most importantly direct detection, where the presence of multiple dark sector species can give rise to simultaneous multi-component direct detection signals that may allow for the twin-nucleonic nature of the dark sector to be determined.
We also briefly discuss the standard collider signatures of this type of Twin Higgs scenario.

For direct detection, we consider five possible twin nuclear candidates: atomic twin deuterium, atomic twin helium, ionized twin protons, ionized twin deuterons, and ionized twin helium nuclei. 
We also consider ionized twin electrons and signals from TNDM. 
We comment on how our results would extrapolate to heavier ions or bound states.  
The results of \cref{subsec: phase diagram,subsec: naDM abundances} can be used to determine the dominant naDM candidate for a particular point in the parameter space.


\subsection{naDM Direct Detection}\label{subsec: constraints direct detection}

Because the dark sector is coupled to the SM via, at the very least, the Higgs portal, it can generically scatter off of SM particles, making it a candidate for direct detection experiments. We compute the direct detection signals following the methods of Ref.~\cite{DelNobile:2021wmp}. 
Which naDM candidate is present will be determined by the microphysics parameters, as detailed in~\cref{sec: cosmo}. We will assume that the TNDM follows the standard halo model (SHM) velocity distribution, while the naDM velocity parameters may vary, see~\cref{subsec: cosmo ionization}. A review of the basic formalism for nuclear recoil and electron recoil direct detection can be found in Appendix \ref{app: direct detection BG}. In the main body, we restrict our attention to elements of the calculation that differ from a standard single-component WIMP candidate. 

While the TNDM component is assumed to follow the standard CDM distribution, to bracket different possibilities, we consider either that the naDM follows the standard CDM distribution or that the naDM collapses into a dark disk that co-rotates with the SM disk: see \cref{app: velocity distribution} for details. We then consider the possibility of a fully ionized or fully atomic population. In the case of a fully ionized naDM population, the escape velocity of the twin nuclei $v_{\rm esc, DM}$ would effectively go to infinity because the naDM population scatters strongly enough among itself to be collisionally trapped. We comment on the plausibility of the ionized scenario in \cref{subsec: cosmo ionization}. 

If the naDM halo is instead atomic, the mean free path can be estimated from the geometric cross section $\sigma \sim \pi a_{0,B}^2$ with $a_{0,B}$ the Bohr radius of the twin atom:
\begin{equation}
\begin{aligned}
    \ell_B &
    \sim \frac{1}{n\sigma} 
    \sim \left(\frac{m_{H,B}}{\fnaDM\rho_{DM}}\right)
    \left( \frac{\alpha_B^2 m_{e,B}^2}{\pi} \right)
    \\
    &\sim 
    10^{-4}\ \mathrm{kpc}
    \left(\frac{m_{H,B}}{\rm{GeV}}\right)
    \left(\frac{0.1}{\fnaDM}\right)
    \left(\frac{0.4 \ \rm{GeV/cm^3}}{\rho_{DM}}\right)
    \left(\frac{\alpha_B}{\alpha_A}\right)^2
    \left(\frac{m_{e,B}}{\rm{MeV}}\right)^2 \, .
\end{aligned}
\end{equation}
The Milky Way DM halo radius is $r_{\rm MW} \sim 10$~kpc, thus for a large part of our parameter space $\ell_B\ll r_{\rm MW}$. We therefore expect that the twin atomic halo will also be collisionally trapped, and we can take the same $v_{\rm esc, DM}\rightarrow \infty$ limit as in the ionized case. This can be important, especially for light naDM candidates, as the high velocity tail of the distribution may furnish the only particles with sufficient energy to cause any recoil in a detector whatsoever. For the heavier candidates, in which interactions are limited by both number density and the smaller Bohr radius, $v_{\rm esc, DM}$ should be reinstated at the standard galactic escape speed $v_{\rm esc} \sim 500\ {\rm km}/{\rm s}$, though we find that the high velocity tail in those mass ranges does not contribute substantially to the signal and therefore the results are insensitive to the choice of $v_{\rm esc, DM}$. 

There is another important modification to the velocity distributions in the ionized case. For species in thermal equilibrium, the velocity dispersion is modified for each species \cite{Chacko:2021vin}
\begin{equation}
    v_{0,i} = v_{0}\sqrt{\frac{\bar m}{m_{i}}}. 
    \label{eq:v0_eqb}
\end{equation}
Here $v_{0}$ is the standard DM velocity dispersion (given in \cref{tab: SHM params}), $\bar m$ is the average mass per mirror particle in the region and $v_{0,i}$ is the resulting velocity dispersion for species $i$. By charge neutrality, the number density of dark electrons will always be equal to the twin proton number density. In the limit that the electron is light, the average mass per DM particle is $\sim m_{D}/2$ for 
deuteron and $\sim m_{He}/3$ for helium nuclei, since there are two electrons per helium nucleus. This represents a small slow down in the twin nuclear velocity that can have a modest effect on the direct detection signal. However, the dark electron can be boosted to many times the naive expected velocity dispersion, resulting in significant enhancements in direct detection signals.

There are two potential portals relevant for direct detection:
\begin{itemize}
    \item \textbf{Photon kinetic mixing.}
    The Lagrangian in Eq.~\eqref{eq: twin Lagrangian} also motivates~\cite{Gherghetta:2019coi} adding the kinetic-mixing term~\cite{Holdom:1985ag}
    \begin{equation}
        \mathcal{L}
        \supset
        \frac{\kappa}{2}F_A^{\mu\nu}F_{B,\mu\nu}\, ,
        \label{eq:kinetic_mixing}
    \end{equation}
    which induces $\kappa$-suppressed interactions between the SM photon and twin-sector charged fields.
    For relics carrying twin electromagnetic charge, this portal gives rise to an effective nanocharge under SM electromagnetism.\footnote{The term \textit{millicharge} is frequently used in the literature; we prefer nanocharge as it more accurately reflects the size of the interaction.} 
    In the early universe, strong bounds on the mixing $\kappa$ can be obtained both from $\Delta N_{\rm eff}$ at the CMB epoch and from BBN \cite{Vogel:2013raa}, both of which broadly require $\kappa\lesssim 10^{-9}$. 
    For heavier $m_{e,B}$, which sets the bound due to being the lightest dark-charged state, the bound can be somewhat weaker, as discussed below.
    Neutral twin states can instead interact through higher-dimensional dark electromagnetic form factors and moments, including a charge radius, a magnetic dipole moment, and, if the twin $\bar{\theta}$ angle is nonzero, an electric dipole moment (EDM).\footnote{We note that the $\bar\theta_B$ term would also contribute to the twin nucleon mass \cite{Lee:2020tmi}. Since we have neglected it above, we will restrict our attention to models with $\bar\theta_B\lesssim0.1$, such that our prior results for the masses remain valid, though it is possible that a viable dark sector could have $\bar\theta_B\sim \OfOrder{1}$. }

    \item \textbf{Higgs portal.}
    The Higgs interaction in Eq.~\eqref{eq: twin Lagrangian} can also mediate interactions between the two sectors. 
    The coupling of a twin relic to this portal is proportional to the fraction of its mass generated through interactions with the twin Higgs, as well as to the mixing between the SM and twin Higgs bosons. 
    The latter is, in turn, related to the ratio of the Fermi constants in the two sectors.
    
\end{itemize}
In what follows, we study the direct-detection signals associated with both portals.

We also point out that if relatively small twin multi-neutron bound states are formed, the direct detection signals discussed below would be altered and potentially enhanced.  For Higgs portal direct detection, binding of multiple dark neutrons would raise the cross section (assuming coherent scattering) by $N_n^2$, while the number density is reduced by $1/N_n$, leading to enhanced direct detection bounds as $N_n$ increases.
If twin nucleosynthesis proceeds to the formation of very large ``dark nuggets,'' then the direct detection signals can change qualitatively \cite{Hardy:2015boa,Coskuner:2018are, Grabowska:2018lnd}.

\subsubsection*{Kinetic Mixing Portal}
In this subsection, we study direct detection mediated by the kinetic mixing interaction of Eq.~\eqref{eq:kinetic_mixing}.  The interactions we expect to be relevant are the nanocharge interaction for the charged species and, if $\theta_B$ is allowed to be large, twin neutron EDM for TNDM.\footnote{For values of the kinetic mixing that do not violate CMB bounds $\kappa\lesssim 10^{-9}$ \cite{Vogel:2013raa}, the twin neutrons do not have an appreciable signal from other electromagnetic moments such as charge radius or magnetic dipole moment interactions. 
The cross sections for these processes are given in Appendix \ref{app: direct detection BG}.}
The cross sections for these processes are
\begin{equation}\label{eq: direct detection NC and EDM cross section}
\begin{aligned}
    \left(\frac{d \sigma_T}{d E_R} \right)_{\rm NC} &= \frac{m_T}{2\pi}\frac{1}{v^2} \left(\frac{ e^2 \kappa Q_{DM }}{q^2} \right)^2  F^{(p,p)}_M(E_R),
     \\
     \left(\frac{d \sigma_T}{d E_R} \right)_{\rm EDM} & = \frac{3 m_T}{2\pi v^2} \left(\frac{ e \kappa d_{n,B}  }{q} \right)^2F^{(p,p)}_M(E_R)\, ,
\end{aligned}
\end{equation}
where $Q_{\rm DM}$ is the DM electric charge in units of the dark charge $e_D$, $q$ is the momentum exchanged, $d_{n,B}$ is the twin neutron EDM, $m_T$ is the mass of the target species, $v$ is the relative velocity of the DM-nuclei system, $E_R$ is the recoil energy and $F^{(p,p)}_M(E_R)$ is a nuclear form factor encoding the effective charge distribution of the SM nucleus. 
In writing this equation we implicitly assume that the DM can be treated as a point particle---an assumption that may break down in our parameter space and introduces new form factors in the calculation. When calculating the direct detection signal, it is necessary to calculate twin form factors; our general method will be to rescale SM form factors according to the relevant scales of our dark sector. 
Additionally, if the twin nuclei have bound electrons, there could be screening effects, which we will parameterize with a dark electronic form factor. 
We describe the two types of form factors below. 

We will focus on the sensitivity of the xenon-based experiments and use the results of Ref.~\cite{Fitzpatrick:2012ix} for the nuclear SM xenon form factors. 
On the twin side, when discussing different naDM candidates we use the results of Ref.~\cite{Catena:2015uha} to parametrize the nuclear form factors. 
These form factors are a function of the dimensionless parameter
\begin{equation}
    y = \left(\frac{bq}{2} \right)^2,
\end{equation}
where $b$ is the length parameter related to the extent of the nucleus and can be parameterized as 
\begin{equation}
    b_A(\Anum) = \sqrt{\frac{41.467}{45 \Anum^{-1/3}-25 \Anum^{-2/3}}} \ \ \text{fm}
\end{equation}
for single particle states, with $\Anum$ the atomic mass number of the nucleus. We note that these form factors are in good agreement with the empirical Helm form factor \cite{Helm:1956zz}. When we consider twin nuclei, we use the same functional form of the form factors, but the length parameter $b$ must be rescaled to account for the dark sector strong coupling:
\begin{equation}
    b_{B}(\Anum) = b_A(\Anum) \left(\frac{\Lambda_{A}}{\Lambda_{B}}\right).
\end{equation}
In general, these form factors parameterize the loss of coherence when $q\gtrsim b^{-1}$, that is, when the momentum is \textit{large }enough that it probes the internal structure of the nucleus. 
Since the SM target nuclei have a much larger atomic number than, e.g., the twin deuteron, their effective size is generally larger and the loss of coherence happens at a lower energy. For $m_{N,B}\lesssim0.7$ GeV, however, $\Lambda_B/\Lambda_A$ can be as small as 0.23, and the scattering becomes incoherent due to the extent of the twin helium before the xenon scattering becomes incoherent. For the rest of our parameter space, the effect of the dark nuclear form factor is negligible; we include it for the light nucleons where it is relevant, but the effects are minimal. 

For sufficiently \textit{small} momentum transfer, the interaction of a neutral twin atom is screened by the twin electrons bound to its nucleus. 
The characteristic atomic size and therefore the momenta at which screening effects become relevant is approximately set by the Bohr radius, $a_{0,B} = \left( m_{e,B} \alpha_B \right)^{-1}$. 
This effect can be encoded in an atomic form factor that suppresses the interaction for
$q\lesssim a_{0,B}^{-1}$.
For SM atoms, the small electron mass implies a large Bohr radius and hence a screening scale well below the momentum transfers probed by direct-detection experiments. 
Electronic screening can therefore usually be neglected in ordinary nuclear-recoil calculations.

Twin deuterium and helium atoms are only stable for sufficiently small electron masses, see Eqs.~\eqref{eq:stability_cond_D_EC},~\eqref{eq:stability_cond_He_EC}. For the range of binding energies allowed by Eqs.~\eqref{eq: binding energy deuteron} and~\eqref{eq: binding energy helium}, this implies that electrons bound in deuterium must have $m_{e,B}/m_{N,B}\lesssim 0.04$, and electrons bound in helium must have $m_{e,B}/m_{N,B}\lesssim 0.06$. This corresponds to $m^{\max}_{e,B}\simeq 400$ MeV and 600 MeV respectively, for $m_{N,B}=10$ GeV. Above these masses, these light twin atoms are unstable to electron capture in all regions of our parameter space (see~\cref{subsec: phase diagram}). We now estimate if electrons at the upper end of this mass range can affect direct detection observables of light twin atoms. 

The screening scale in the twin sector is approximately
\begin{equation}
    q_{\mathrm{scr},B}\sim \frac{1}{a_{0,B}}\sim \frac{1}{a_{0,A}}\frac{m_{e,B}}{m_{e,A}},
\end{equation}
where we have assumed $\alpha_{EM,B}=\alpha_{EM,A}$.
The characteristic SM screening scale $a_{0,A}^{-1}\sim$ keV. For the heaviest twin electrons considered in our parameter space, this gives $q_{\mathrm{scr},B}\sim$ MeV.
For the SM atom, this corresponds to a recoil-energy scale
\begin{equation}
    E_{r,\rm scr,A}
    \sim
    \frac{q_{\rm scr,A}^2}{2m_{N,A}}
    \sim
    \frac{\mathrm{keV}^2}{\mathrm{GeV}}
    \sim
    \mathrm{meV}\, ,
    \label{eq:ERmin_SM}
\end{equation}
which lies far below typical detector thresholds. 
For twin atoms this becomes
\begin{equation}
    E_{r,\rm scr,B}
    \sim
    \frac{q_{\rm scr,B}^2}{2m_{N,B}}
    \sim
    \frac{\mathrm{MeV}^2}{\mathrm{GeV}}
    \sim
    \mathrm{keV}\, ,
    \label{eq:ERmin_twin}
\end{equation}
which is approaching experimental recoil thresholds. However, this effect is minor for the light twin atoms considered here ($\sim2\%$ reduction in the event rate for the heaviest twin electrons in the helium analysis). While we include atomic screening in our calculations, it cannot be visually seen in the plots.
For hypothetical heavier atoms which are stable to heavier electron masses, these effects could affect the event rate. Details on how we implemented the electronic screening can be found in Appendix~\ref{app:electronic_screening}.

\begin{figure}
    \centering
    \includegraphics[width = 0.97\linewidth]{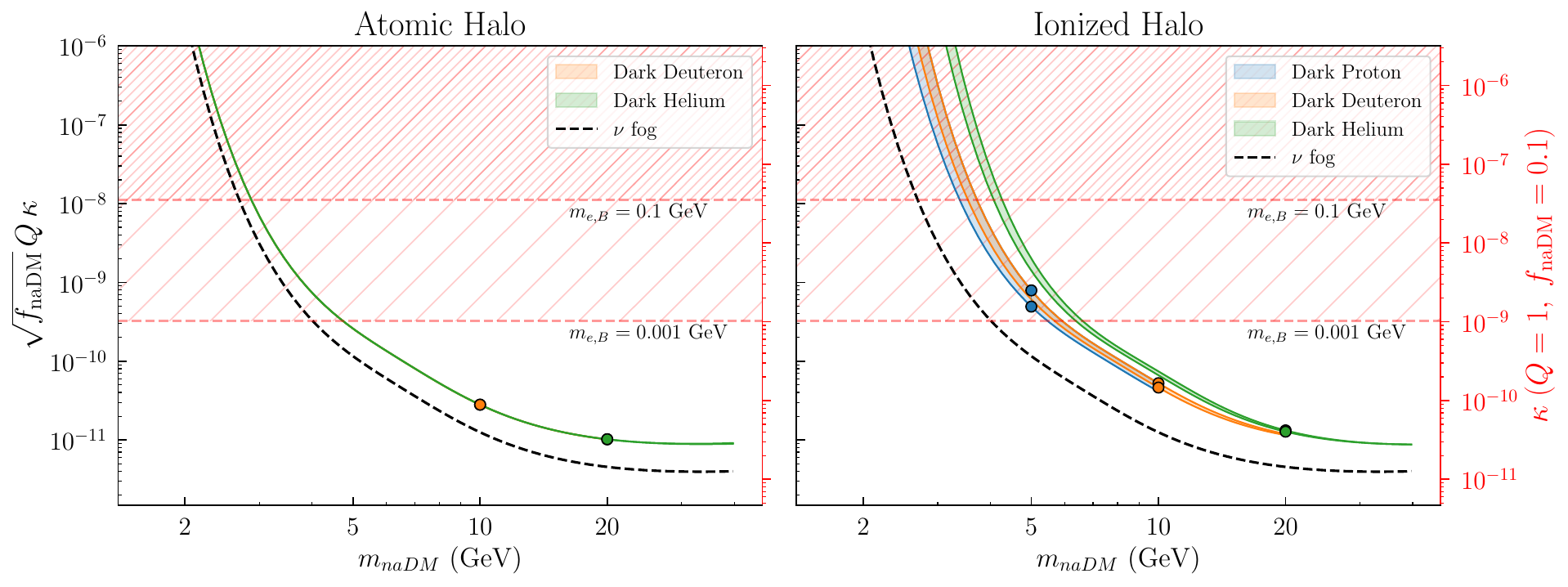}
   
    \caption{
    LUX-ZEPLIN (LZ) limits~\cite{LZ:2024zvo} on kinetic mixing for an atomic naDM component (\textbf{left}) and an ionized naDM component (\textbf{right}) in the halo distribution. 
See the discussion at the beginning of this section for the effect of ionization on the relic velocity distribution. 
The exclusion limits are obtained by requiring fewer than three signal events.
The spread in the limits for each relic species arises from varying the twin electron mass, as detailed in the text. 
The \textcolor{DarkBlue}{blue}, \textcolor{D_He_naDM}{orange}, and \textcolor{N_SIDM}{green} points highlight the maximum and minimum LZ limits for twin proton, deuterium, and helium respectively at the benchmark point $m_{N,B} = 5$ GeV.
The left vertical axis shows the limit on $\sqrt{\fnaDM}\,Q\kappa$, while the right vertical axis gives the corresponding limit on $\kappa$ for the benchmark value $\fnaDM=0.1$. The cosmological $\kappa$-bounds~\cite{Vogel:2013raa} for light ($m_{e,B} = 0.001~\mathrm{GeV}$) and heavy ($m_{e,B} = 0.1~\mathrm{GeV}$) twin electrons are then indicated with horizontal \textcolor{red}{red} dashed lines and are insensitive to $f_\mathrm{naDM}$.
The neutrino floor curve (dashed black) is calculated assuming the SHM velocity distribution and should therefore be regarded as representative. 
Direct-detection searches can surpass existing cosmological constraints, particularly for heavier twin relics.
    }
    \label{fig: DD results Kin Mixing}
\end{figure}

Our nuclear-recoil direct-detection results are shown in \cref{fig: DD results Kin Mixing}. 
We use the recoil threshold and acceptance function from Ref.~\cite{LZ:2024zvo} to calculate the expected number of events at LUX-ZEPLIN (LZ), and constrain $\kappa$ by requiring fewer than three signal events. 
We consider the two limiting cases in which the naDM component is either fully atomic or fully ionized. 
The widths of the coloured regions reflect variation of the twin electron mass over the range in \cref{tab: IR params} for ionized candidates and up to the maximum allowed by the atomic stability conditions Eqs.~\eqref{eq:stability_cond_D_EC},~\eqref{eq:stability_cond_He_EC} for atomic candidates.
For atomic naDM, this only changes the screening of the nuclear charge (which is irrelevant for light species);
for ionized naDM, it slightly changes the velocity dispersion of the twin ions, as discussed around Eq.~\eqref{eq:v0_eqb}.

The cosmological bounds on $\kappa$, shown as shaded \textcolor{red}{red} regions for different twin electron masses, are independent of the naDM fraction, whereas the direct-detection limits scale as $\fnaDM^{-1/2}$. 
The results demonstrate that nuclear-recoil searches can probe regions of parameter space beyond existing cosmological constraints.

We estimate the neutrino floor by finding the value of $\kappa$ for which the expected DM and neutrino event rates are equal; further details are given in Appendix~\ref{app: neutrino floor}. 
Because the mediator is massless, the DM recoil spectrum is strongly peaked toward low energies and can differ appreciably from the neutrino spectrum, particularly for heavier naDM masses. 
The resulting \textit{floor} is therefore not a hard limit: with sufficient statistics, the two spectra may remain distinguishable. 
A full neutrino-\textit{fog} analysis, along the lines of Ref.~\cite{OHare:2021utq}, would provide a more accurate estimate, but our treatment should capture the approximate transition from the systematics-limited to the background-limited regime.

\begin{figure}
    \centering
    \includegraphics[width=0.97\linewidth]{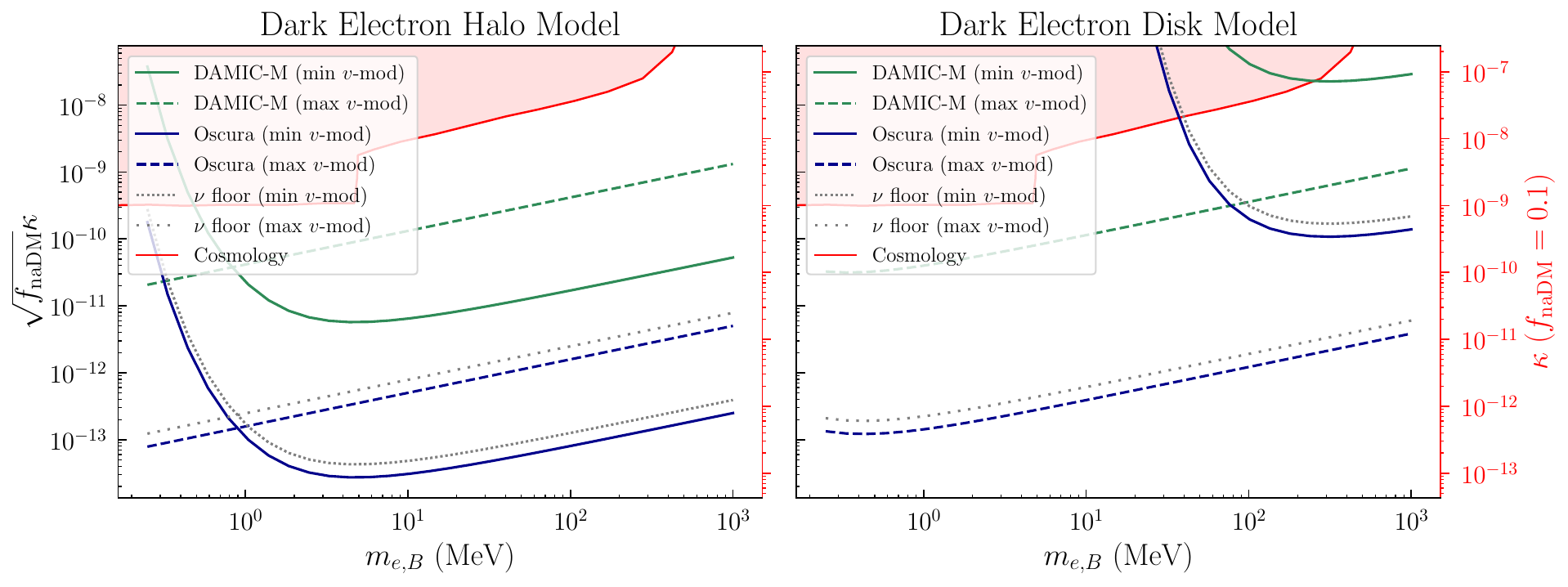}
    \caption{
    Electron-recoil direct-detection limits on the kinetic-mixing parameter $\kappa$ from scattering off twin electrons, computed using \texttt{darkELF} \cite{Knapen:2021bwg}. 
The \textbf{left} panel assumes the twin electrons follow a halo distribution, while the \textbf{right} panel assumes they have collapsed into a disk. 
The left vertical axis shows the limit on $\sqrt{\fnaDM}\,\kappa$, while the right vertical axis gives the corresponding bound on $\kappa$ for the benchmark choice $\fnaDM=0.1$. In this case, cosmological mixing constraints~\cite{Vogel:2013raa} are shown in \textcolor{red}{red}, which are determined by $\kappa$ alone. 
Current constraints from DAMIC-M \cite{DAMIC-M:2025luv} are shown in \textcolor{N_SIDM}{green}, while the projected reach of Oscura \cite{Oscura:2022vmi}, assuming zero background, is shown in \textcolor{DarkBlue}{blue}. 
For DAMIC-M, we impose $N_{\rm events}<15$, chosen to reproduce the published limits for WIMP candidate and appropriate for the $\sim 144$ observed background events, given that signal discrimination scales roughly as $\sqrt{N_{\rm BG}}$. 
For Oscura, we assume the zero-background criterion $N_{\rm events}<3$. 
The dotted curves show the neutrino floor for each velocity distribution.}
    \label{fig: electron recoil DD}
\end{figure}

Direct detection of twin electrons can probe the model even more effectively. 
Their scattering with SM electrons is conventionally parameterized by the reference cross section
\begin{equation}
    \bar{\sigma}_e
    =
    \frac{\mu_{\chi e}^2 g_e^2 g_\chi^2}
    {\pi\left(\alpha_{A}^2m_{e,A}^2+m_\phi^2\right)^2}
    =
    \frac{16\pi\kappa^2\mu_{\chi e}^2}
    {\alpha_{A}^2m_{e,A}^4}\, ,
\end{equation}
where the second equality applies to a massless mediator, $m_\phi=0$, with $g_e=e$ and $g_\chi=\kappa e$. 
The reference momentum $\alpha_{A}m_{e,A}$ corresponds to the characteristic momentum of a valence electron in a material. 
The material response is encoded in the energy-loss function
$\mathrm{Im}\!\left[-1/\epsilon(\omega,\vec{k})\right]$, where $\epsilon$ is the dielectric function. 
We use \texttt{darkELF} \cite{Knapen:2021bwg} to derive current constraints from DAMIC-M \cite{DAMIC-M:2025luv} and projected sensitivities for Oscura \cite{Oscura:2022vmi}. 
The results are shown in \cref{fig: electron recoil DD}.

In deriving the electron-scattering limits, we assume that naDM is fully ionized, so that the twin electrons can be treated as free particles, as required by the \texttt{darkELF} calculation. 
Scattering from twin electrons bound in neutral atoms requires a separate treatment and is left for future work.

Charge neutrality requires that the twin electron number density track that of the primary naDM component of mass $m_{\rm naDM}$ and twin proton number $\Znum$, with which it is in thermal contact.
The twin electron number density is therefore suppressed by $\sim \Znum m_{e,B}/ m_{naDM}$, relative to a single component DM candidate of mass $m_{e,B}$. 
This competes with an enhanced electron velocity dispersion from thermal contact with the heavy species. 
For $m_{e,B}/m_{\rm naDM}\sim 10^{-4}$ (dashed lines in~\cref{fig: electron recoil DD}), the density suppression is large, but the velocity enhancement is also substantial, allowing twin electrons with  $m_{e,B}\lesssim 1-10\, \mathrm{MeV}$ to produce signals where SHM particles would fall below threshold. 
For  $m_{e,B}/m_{\rm naDM}\sim 10^{-1}$ (solid lines in~\cref{fig: electron recoil DD}), both effects are modest, and the signal resembles that of a single-component candidate, reaching slightly lighter masses than in the SHM.

We also present results for the case in which naDM collapses into a dark disk; details of the disk model are given in Appendix~\ref{app: velocity distribution}. 
Although the slowly moving twin nuclei produce negligible nuclear-recoil signals for either ionized or atomic case in the disk model, in the ionized case the twin electrons can remain sufficiently energetic to yield appreciable signals.
Modifying the velocity dispersion can have two different effects.
When $v_0\sim v_{\min}$, with $v_{\min}$ the smallest velocity that can generate a detectable event, increasing $v_0$ enhances the signal by increasing the fraction of electrons energetic enough to produce a measurable recoil. 
By contrast, the cross sections in Eq.~\eqref{eq: direct detection NC and EDM cross section} scale as $1/v^2$, so that for $v_0\gg v_{\min}$ the total rate scales approximately as $1/v_0$ and decreases as the velocity dispersion grows.

This competition is visible for very small $m_{e,B}/m_{\rm naDM}\sim 10^{-4}$ (dashed lines in~\cref{fig: electron recoil DD}). 
The disk constraints can be slightly \textit{stronger} than those obtained for the halo distribution because the twin electron velocity dispersion is near the optimal regime between the lower threshold $v_{\min}$ and the asymptotic $1/v_0$ suppression. 
In the halo case, by contrast, the velocity dispersion is larger and is in the regime where the $1/v_0$ suppression becomes important.
For $m_{e,B}/m_{\rm naDM}\sim 10^{-1}$, the twin electron disk velocity dispersion is close to that of the slow moving twin nuclei and is closer to $v_{\min}$, thus the signal is suppressed compared to the halo case.  

The low-mass cutoff is controlled by the minimum twin electron momentum that can yield a signal, and thus by $v_{\min}$. 
The corresponding turnover is therefore at larger masses for the disk distribution, which has a smaller characteristic velocity than the halo distribution. 
For $m_{e,B}/m_{\rm naDM}\sim 10^{-4}$ the twin electron velocity is strongly enhanced and the turnover is pushed to much lower masses compared to $m_{e,B}/m_{\rm naDM}\sim 10^{-1}$.

Let us also briefly comment on the prospect of direct detection signal from a twin neutron DM relic.
As the twin neutron is neutral, it will not interact with SM as a nanocharge relic, but rather through its twin electromagnetic moments, in particular EDM (see the second line of Eq.~\eqref{eq: direct detection NC and EDM cross section}).
For a detailed discussion on CP-violating dark baryons as DM, see Ref.~\cite{Figueroa:2026bmi}. 
Here we estimate the neutron EDM signal at nuclear recoil experiments. 
The neutron EDM derived from the chiral $\mathrm{SU}(2)_L\times \mathrm{SU}(2)_R$ of up and down quarks is approximately \cite{Crewther:1979pi, Srednicki:2007qs}
\begin{equation}\label{eq: neutron EDM}
    d_n \simeq e\frac{e \bar \theta g_a m_um_d (m_n-m_p)}{8 \pi^2 f_\pi^2(m_d+m_u)(m_d-m_u)}\ln{\left(\frac{m_N}{m_\pi}\right)^2}.
\end{equation}
This expression makes a number of approximations~\cite{Crewther:1979pi}, but gives similar results to lattice QCD calculations \cite{Blum:2026hul}. Fortunately, these approximations are secondary to the dominant effect of modifying $\bar\theta_B$.

Since at best we can know this at the order of magnitude level, we simplify this expression by assuming the light quark reduced mass is approximately the twin down quark mass, and the mass splitting of the twin neutron and proton is the same as the light quark mass splitting. 
Based on this estimate, the largest twin neutron EDM we can achieve in our parameter space is when $\Lambda_B/\Lambda_A$ is minimized and the down quark mass is maximized. It is approximately given by
\begin{equation}\label{eq: neutron EDM max value}
    d_{n,B}^{\max} \sim 0.35 \frac{\bar \theta_B}{m_{N,B}}.
\end{equation}

\begin{figure}
    \centering
    \includegraphics[width=0.6\linewidth]{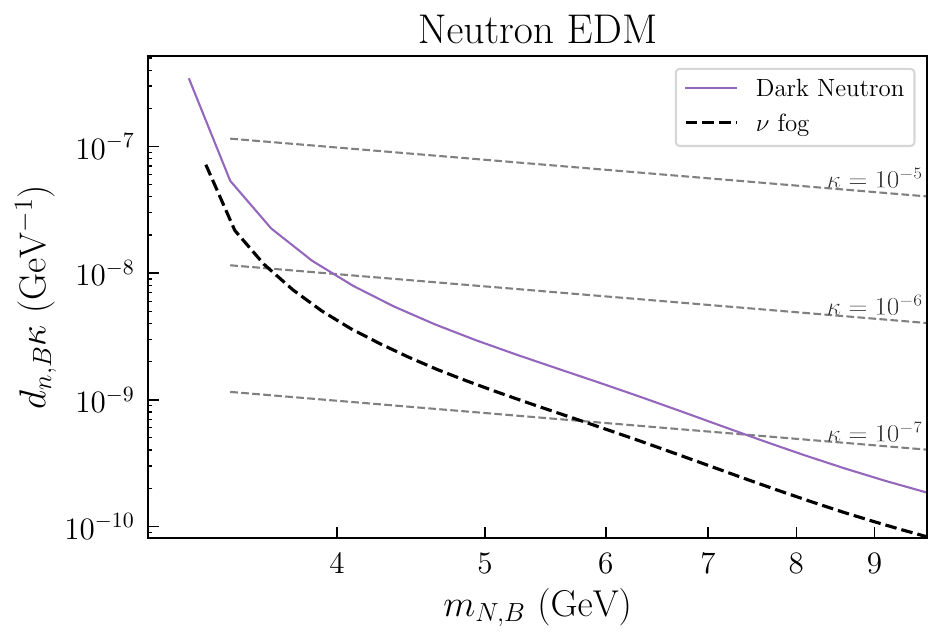}
    \caption{
    Constraints on the kinetic mixing suppressed twin neutron EDM (\textcolor{violet}{purple}) based on LZ (2025) \cite{LZ:2024zvo}. The black dashed line gives the approximate location of the EDM neutrino floor. Grey dashed lines show the value of the kinetic mixing assuming that $d_n$ saturates Eq. \eqref{eq: neutron EDM max value} with $\theta =0.1$. Unless the twin electron is heavy, this parameter space is mostly excluded by cosmological bounds.}
    \label{fig: DD results neutron EDM}
\end{figure}

The twin neutron EDM signals are shown in \cref{fig: DD results neutron EDM}. Unsurprisingly,
the neutron would be difficult to observe even for the maximal value of $\bar \theta_B$ allowed. The interaction arises at dim-5 in the effective Lagrangian compared to the dim-4 nanocharge interaction, and is always greatly suppressed compared to the charged species. However, in the case that the twin protons are completely depleted, which is the majority of the viable parameter space, $\kappa$ can be as large as $10^{-8}$ if the lightest dark fermion mass is $\OfOrder{0.1 \ \rm{GeV}}$ \cite{Vogel:2013raa}. If the mass of the lightest dark fermion is $\gtrsim 0.1 $ GeV, constraints weaken dramatically as CMB and BBN bounds cease to be effective and collider searches are the primary probe. 
Constraints in this scenario can be as weak as $\kappa \lesssim  10^{-2}-10^{-3}$ \cite{milliQan:2021lne}, with proposed/future experiments probing up to $\kappa \sim 10^{-4}$ \cite{Bailloeul:2025fde}. 
In this case, it may be possible to detect twin neutrons via their twin EDM interaction. 
If naDM candidates were also present, this would yield much stronger constraints on $\fnaDM$ due to the huge direct detection signal a kinetic mixing this large would cause.

To summarize, if naDM forms and is present in the local halo, photon kinetic mixing can allow upcoming nuclear- and electron-recoil direct detection experiments to probe naDM parameter space above the neutrino floor. It may be possible to detect several naDM components simultaneously and determine their integer mass relation~\cite{Chacko:2021vin}. In particular, dark matter masses in multiples of 5 GeV would lend credence to a Twin solution of the matter coincidence problem.


\subsubsection*{Higgs Portal}\label{sec: DD Higgs portal}

The second possible portal is through Higgs mixing between the A and B sectors. In some parts of our parameter space a Higgs portal interaction can give rise to a detectable direct detection signal for the twin neutron. While much of the generic prediction for this signal lies beneath the current neutrino floor, future directional direct detection experiments could probe this regime.

As discussed above, the prospect for twin neutron direct detection scattering via electromagnetic moments is dismal, even for very large kinetic mixing $\kappa$ allowed by CMB measurements. 
However, the underlying Twin Higgs nature of the naDM setup gives rise to a Higgs portal that induces interactions between the fermions of the two sectors suppressed by the mixing angle $\theta \approx v/f$ for $v/f \ll 1$, related to the Fermi constant via
$G_{F,A}/G_{F,B} = f^2/v^2$.
The interactions between the physical Higgs $h$ and the quark fields are given by
\begin{equation}
    \mathcal{L} =  \sum_q \frac{y_{B}}{\sqrt{2}}\left( \frac{v}{f} \right)\bar q_B q_B h+ \sum_q  \frac{y_{A}}{\sqrt{2}}\left( 1-\frac{v^2}{2f^2}\right) \bar q_A q_A h.
\end{equation}
The factor of $v/f$ suppressing the B sector interactions can be seen, e.g., from the analysis in Ref.~\cite{Curtin:2021alk}. 
Under these assumptions, it is clear that if the mixing angle is too large, the physical Higgs would decay approximately equally to both sectors. Requiring the Higgs to invisible decay branching ratio and measured Higgs couplings to satisfy current constraints~\cite{Craig:2015pha, ParticleDataGroup:2024cfk} requires $v/f \lesssim 1/3$ and therefore restricts $G_{F,B}\lesssim G_{F,A}/9$. 

For those models which can satisfy the above criteria, we can calculate the interactions via the Higgs portal. 
The couplings of A or B nucleons to the Higgs are related to the quark couplings via
\begin{equation}
    c_{N,X} \equiv \sum_{q} \frac{m_{N,X}}{m_{q,X}}c_{q,X} f_{q,X}^{(N)}\, . 
\end{equation}
Here $c_{q,X}$ is the coupling of the quarks to the Higgs, which in our case will be proportional to the Yukawas. The nucleon scalar form factor $f_{q,X}^{(N)}$ encodes the quark mass contributions to the nucleon mass. In the non-relativistic limit, these are

\begin{equation}\label{eq: scalar form factors and def of sigmaN}
    f_{q,X}^{(N)} = \frac{\bra{N_X}m_{q,X} \bar q_X q_X \ket{N_X}}{ m_{N,X}} =  \frac{\sigma_{q,X}^{(N)}}{m_{N,X}} = \frac{m_{q,X}}{m_{N,X}}\frac{\partial M_{N,X}}{\partial m_{q,X}}.
\end{equation}
In general, the momentum scale of DM-nucleon scattering is much less than the Higgs mass, so we can approximate the nuclear recoil DM-nucleon total cross-section for a single point-like nucleon as

\begin{equation}\label{eq: Higgs cross section}
   \sigma_N \approx \frac{\mu_N^2  }{\pi m_{h,A}^4  }
        c_{N,B}^2 
        c_{N,A}^2,
\end{equation}
with $\mu_N$ the reduced mass of the DM-nucleon system and the effective A and B sector nucleon Yukawas are
\begin{equation}
    \begin{aligned}
        c_{N,A} &\equiv \frac{m_{N,A}}{v}\sum_q  f_{q,A}^{(N)}, 
        \\
        c_{N,B} & \equiv \left(\frac{m_{N,B}}{f} \right) \left(\frac{v}{f}\right) \sum_q  f_{q,B}^{(N)}(m_{\pi,B}) 
        \label{eq:cN_Higgs_portal}
    \end{aligned}
\end{equation}
These sums can be reduced to a sum over light quarks only via the methods outlined in e.g. Ref.~\cite{Shifman:1978zn}. The calculation is sketched in Appendix~\ref{app: effective baryon yukawas}. The result is 
\begin{equation}
    \sum_q  f_{q}^{(N)} = \frac{2}{9} + \frac{7}{9}\sum_{q=u,d,s} f_{q}^{(N)}.
\end{equation}
The values of these quantities for the SM fields as derived from the 2024 FLAG report \cite{FlavourLatticeAveragingGroupFLAG:2024oxs} are given in \cref{tab: FLAG best fit values} under the assumption of degenerate up and down quark masses. The SM nucleon effective Yukawa is thus
\begin{equation}
    c_{N,A} \simeq 1.2 \times 10^{-3}.
\end{equation}
For the twin nucleon scalar form factor we use the lattice results of Ref.~\cite{DeGrand:2019vbx}. See Appendix~\ref{app: lattice fits} for the exact fit expression used. Note that these data assumed that the masses of the up and down quark were degenerate, an assumption that is fairly accurate for the SM but could be significantly violated in the twin sector. However, since we don't consider mass splittings larger than $\sim 10\%$ of the overall nucleon mass, we expect this estimate to be reasonable. For the strange quark we do not know either the explicit mass or the pion dependence of the form factor and therefore use the SM value as a rough estimate. 

\begin{table}
\renewcommand{\arraystretch}{1.4} 
    \centering
    \begin{tabular}{c|c|c}
     Quantity    & Value & Reference  \\
     \hline
      $f_u^{(N)} + f_d^{(N)}$   &  0.065(7) & \cite{Alexandrou:2014sha,Gupta:2021ahb} 
      \\
     $ f_s^{(N)}$ & 0.04(1) & \cite{Freeman:2012ry}
    \end{tabular}
    \caption{Values of the scalar form factor derived from lattice simulations with number of flavours $N_f=2+1+1$. }
    \label{tab: FLAG best fit values}
\end{table}

Note that in the heavy pion limit, for $\mu_N\sim 1$ GeV and $m_{\pi,B}/m_{N,B}\simeq0.56$, and for an optimistic $v/f\sim 1/3$, $c_B$ is approximately $10^{-3}$, the cross section Eq.~\eqref{eq: Higgs cross section} will be approximately 
\begin{equation}
    \sigma_N \sim \left(10^{-3}\right)^4\frac{\left(1 \ \textrm{GeV}\right)^2}{\left(100 \ \textrm{GeV}\right)^4}\sim 10^{-48} \ \textrm{cm}^2.
\end{equation}
This will be below the neutrino floor for direct detection at $m_{N,B}=5$ GeV. However, future experiments with directional sensitivity would be capable of filtering out solar neutrinos, which would reduce the background in that region to primarily cosmogenic neutrinos \cite{Grothaus:2014hja, OHare:2015utx, Vahsen:2020pzb, Sassi:2021umf}. Our findings highlight a motivated target for these prospective experiments.

\begin{figure}
    \centering
    \includegraphics[width=0.7\linewidth]{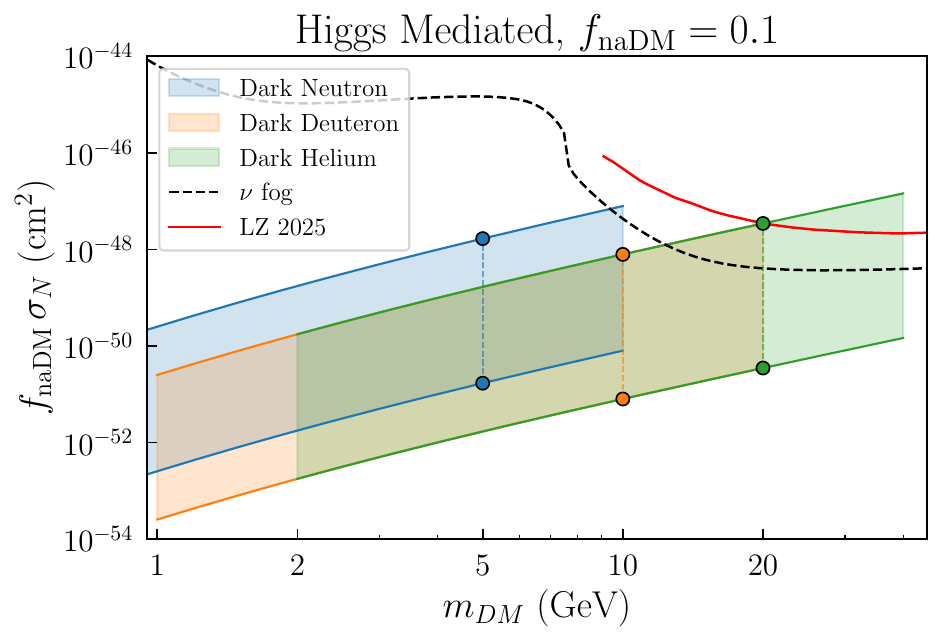}
    \caption{Predicted cross sections for Higgs mediated direct detection of twin neutrons, deuterons, and helium, under the assumption that twin neutrons and twin nuclei are in a halo distribution. 
    Variations in the signal from each relic arise from scanning a range of values for $G_{F,B}$ and $m_{\pi,B}$ (see \cref{tab: IR params} for the ranges used). The latter controls the mass of twin quarks, which in turn determine their coupling to the twin Higgs, see Eq.~\eqref{eq:cN_Higgs_portal}.
    Note that only the naDM candidates have the $\fnaDM$ suppression, while the twin neutron is assumed to constitute the majority of DM. Constraints shown are from LZ (2025) \cite{LZ:2024zvo} and the neutrino fog is taken from Ref.~\cite{OHare:2021utq} assuming the SHM for WIMP DM.}
    \label{fig:Higgs_cross_section}
\end{figure}

For dark deuterium or helium, we expect the per-nucleon cross section to be well approximated by 
\begin{equation}\label{eq: Higgs cross section nuclei}
   \sigma_N \approx \frac{\hat \mu_{N}^2  }{\pi m_{h,A}^4  }
        \Anum_{B}^2 c_{N,B}^2 
        c_{N,A}^2,
\end{equation}
where $\hat \mu_N$ represents the reduced mass of the SM nucleon with the twin atom and $\Anum_B$ is the atomic mass number of the naDM candidate. For our models $\hat \mu_{N}\sim  m_{N,A}$ since the nuclei will be heavier than the SM nucleons. On dimensional grounds, we expect that since $f^N_{q}$ is dimensionless, it does not depend on the magnitudes of $m_{\pi,B},\ m_{N,B}$ but only the dimensionless combination $m_{\pi,B}/m_{N,B}$ (this can also be seen explicitly from the form of the fits in Appendix~\ref{app: lattice fits}). Therefore, for fixed $m_{\pi,B}/m_{N,B}$ and $G_{F,B}$, the effective baryon Yukawa $c_{N,B}\propto m_{N,B}$, and  
\begin{equation}
    \sigma_{N}\propto \Anum_B^2 m_{N,B}^2 = m_{naDM}^2.  
\end{equation}
The observed $\sigma_{N,obs}= \fnaDM\sigma_N $ for the naDM candidates is reduced compared to the twin neutron by the lower number density of the atomic state. These results are shown in \cref{fig:Higgs_cross_section}. The width of the bands is set by the variation of the pion mass within our allowed bands and $G_{F,A}/100<G_{F,B}<G_{F,A}/10$.

Our results indicate that, over the parameter range considered in this work (see \cref{tab: IR params}), the model can produce observable signals in upcoming direct-detection experiments.
For the  Higgs portal, the situation is particularly promising for twin deuterium and twin helium (either ionized or atomic), which can produce signals above the neutrino floor for $m_{N,B} = 5~\mathrm{GeV}$. Detecting multi-component dark matter with masses in multiples of 5 GeV would already be highly suggestive of a connection to the matter coincidence problem.
The twin neutron itself is currently inaccessible, but future direct detection experiments that could screen out solar neutrinos might, with sufficiently large exposure, detect TNDM through its irreducible Higgs portal interaction. Confirming its mass to be 5 GeV would constitute a rather spectacular resolution of the matter coincidence problem, and this scenario should serve as a sensitivity goal for future generations of directional direct detection experiments.


\subsubsection*{Heavy Twin Elements}

A further comment concerns the direct detection of heavier twin nuclei. As discussed above, in some regions of parameter space the naDM component may contain nuclei heavier than helium and/or neutron-rich isotopes of these nuclei. Extrapolating the scaling of Higgs-mediated scattering in \cref{fig:Higgs_cross_section}, these states would extend the corresponding bands in the figure to larger dark-matter masses and higher scattering cross sections.
For $m_{N,B}\simeq 5\,\mathrm{GeV}$, as motivated by the matter coincidence problem when the visible- and twin-sector baryon asymmetries are equal, portions of this parameter space may already be excluded by LZ, while others can lie below the current LZ bound but above the neutrino fog and may therefore be imminently accessible to future searches. 
Furthermore, the rates shown in \cref{fig:Higgs_cross_section} assume $\fnaDM=0.1$ and scale linearly with the naDM abundance. Smaller abundances weaken the constraints and can leave otherwise excluded parameter points viable. If the TNDM were bound into multi-neutron states the Higgs detection cross section would also trivially scale along the same bands as the twin neutron signal, strongly constraining these scenarios. The possible presence of heavy naDM elements therefore provides additional motivation for future direct-detection searches.

If sufficiently heavy twin atoms aggregate into larger bound structures, the naDM component could form twin dust or even larger macroscopic composite objects~\cite{Jacobs:2014yca}. 
Depending on their masses, sizes, and effective interactions with SM matter, the passage of such objects through terrestrial detectors or the atmosphere could produce distinctive non-pointlike signatures. 
These include multiple-scatter events in large-volume detectors and delayed time-of-flight coincidences between spatially separated scintillators \cite{DEAPCollaboration:2021raj,Cappiello:2020lbk}, persistent damage tracks in ancient-mineral paleo-detectors \cite{Baum:2018tfw} or large-area etched-plastic detectors \cite{Acevedo:2021tbl,Bhoonah:2020fys}, or unconventional bright $\gamma$-ray bursts \cite{Kaplan:2024dsn}. 
Energy deposition by rare, massive objects could also overheat cold interstellar gas clouds \cite{Bhoonah:2020dzs} or, for sufficiently large geometric cross sections, produce localized macroscopic damage or injury \cite{SinghSidhu:2019loh}. 
Establishing whether such aggregates form in our scenario, and mapping their mass, size, abundance, and stopping power onto these searches, requires a dedicated analysis that we leave to future work.

\subsection{Collider Signatures}

As in other Twin Higgs models, our realization can be probed at future lepton colliders through precision measurements of Higgs-coupling shifts of order $v^2/f^2$~\cite{Craig:2013xia,Blondel:2021ema,Batell:2022tif,McKeen:2023uzo}. 
Evidence for this characteristic pattern would, however, neither determine the cosmological history or low-energy particle content of the twin sector nor uniquely identify our naDM scenario. 
An important ingredient in our scenario is asymmetric reheating, which suppresses 
the twin-sector contribution to $\DNeff$ while leaving a potentially observable residual signal~\cite{Chacko:2016hvu,Craig:2016lyx}. 
Alternative solutions instead eliminate the light twin degrees of freedom responsible for dark radiation, such as fraternal Twin Higgs, instead predict long-lived-particle signatures~\cite{Craig:2015pha}. 
A future measurement of $\DNeff\gtrsim0.01$ without such long-lived particle signatures and with Higgs coupling shifts (and invisible Higgs decays, see below) could therefore point toward an asymmetrically reheated twin sector with light twin states. 
Determining whether the twin baryons form naDM would then rely on its cosmological and astrophysical signatures, making these compelling targets for future searches.

Other collider signals, such as invisible Higgs decays, are more model dependent. 
Particular to naDM is the substantially modified twin quark spectrum due to the increased effective twin Yukawas. 
This will generically modify the precise $\mathrm{Br}(h \to \mathrm{invisible})$ prediction~\cite{Burdman:2014zta}, and its correlation with measurable Higgs coupling shifts, compared to the standard MTH scenario.
Our IR parameterization of naDM predicts a \textit{minimum} exotic Higgs decay branching fraction to invisible twin up- and down-type quarks and twin electrons, which in the majority of our parameter space is less than $\sim 10\%$, consistent with current bounds~\cite{ATLAS:2023tkt,CMS:2023sdw}. 
The \textit{total} Higgs to invisible branching fraction depends on the full twin fermion spectrum, and is therefore model dependent. 
We give specific predictions for the Froggatt--Nielsen UV completions we study in~\cref{sec: uv realization}.
It may even be possible to map out the heavier twin quark masses using off-shell Higgs production by scanning in $\sqrt{s}$ at future lepton colliders. 
This would provide the most direct collider evidence for the naDM scenario, and is highly motivated for future study.


\section{Comments on Possible UV Realizations}\label{sec: uv realization}

In this section, we investigate whether the IR parameter ranges that support naDM can arise naturally in a motivated ultraviolet framework. 
We focus on a supersymmetric Twin Higgs model in which fermion hierarchies are generated by a Froggatt--Nielsen (FN) mechanism and the twin symmetry is spontaneously broken by unequal flavon expectation values.

Our aim is not to construct a complete UV model, but to establish two concrete, but limited, results. 
First, for a broad set of FN charge assignments that naturally reproduce visible-sector flavour, an order-one difference between the two FN expansion parameters can reverse the twin up--down quark hierarchy and \textit{generically} yield a twin-baryon mass in the few-GeV range. 
The resulting pion, nucleon-splitting, and electron masses overlap substantially with the regions in which naDM arises. 
Second, we show that the same Yukawa asymmetry can radiatively generate a Higgs-sector $\mathbb Z_2$-breaking mass of the order required to obtain $f/v>1$ in a supersymmetric Twin Higgs completion.

If baryogenesis produces equal surviving baryon asymmetries in the two sectors, a twin-baryon mass near 5~GeV reproduces the observed ratio of dark- to visible-matter densities. 
The coincidence problem is therefore reduced to explaining why spontaneous flavour breaking places the twin confinement and quark-mass scales in the few-GeV regime. Our scan indicates that this behaviour is robust across well-performing FN textures (although a complete treatment of the flavon vacuum, supersymmetry breaking, baryogenesis, and cosmological evolution is left for future work).
Thus, this construction provides a proof of principle that the IR parameter region motivated by the matter coincidence problem can arise from spontaneous \Ztwo\ breaking in a FN construction. 

\subsection{Spontaneous $\mathbb{Z}_2$ breaking in a Twin Froggatt--Nielsen model}

We adopt the Froggatt--Nielsen mechanism~\cite{Froggatt:1978nt}, which has been extensively studied in both the quark and lepton sectors~\cite{Leurer:1992wg,Leurer:1993gy,Dudas:1996fe,Irges:1998ax,Sato:2000ff, Sato:2000kj,Suematsu:2001sp, Dreiner:2003hw,Nir:2004my,Kamikado:2008jx,Plentinger:2008up,Buchmuller:2011tm,Krippendorf:2015kta,Bauer:2016rxs,Ema:2016ops,Calibbi:2016hwq,Nishiwaki:2016xyp,Vien:2019zhs,Feruglio:2019ybq,Berger:2019rfk,Bordone:2019uzc,Smolkovic:2019jow,Fedele:2020fvh,Nishimura:2020nre,Aloni:2021wzk,Ringe:2022rjx, Asadi:2023ucx,Qiu:2023igq,Cornella:2023zme,Nishimura:2024apb,Ibe:2024cvi,Cornella:2024jaw}, see e.g.\ Ref.~\cite{Altmannshofer:2022aml} for a modern review,
 as the underlying framework to explain the fermion mass hierarchies.
Fermions are charged under a new global $\mathrm{U}(1)_F$ horizontal flavour symmetry, generally forbidding the Yukawa interactions in the UV, except for the top quark. However, this U(1)$_F$ is broken spontaneously by the vev of a flavon $\phi$ (where without loss of generality we set its flavour charge to  $X_\phi=1$), allowing the generation of the Yukawa couplings as non-renormalizable operators with hierarchical $\phi/\Lambda_F$ suppressions determined by the quark flavour charge assignments:
\begin{equation}\label{eq: FN lagrangian}
    \mathcal{L}\supset -c^u_{ij}\bar{Q}_i\Tilde{H}u_j \left(\frac{\phi^{\left(*\right)}}{\Lambda_F}\right)^{|X_{Q_i}-X_{u_j}|}
    -c^d_{ij}\bar{Q}_i H d_j \left(\frac{\phi^{\left(*\right)}}{\Lambda_F}\right)^{|X_{Q_i}-X_{d_j}|} + \mathrm{h.c.} \, .
\end{equation}
Here, $c^u_{ij}$ is an \Oone\ complex coupling matrix, $\Lambda_F$ is the scale of some UV physics that actually generates these operators, and $i,j \in \{1,2,3\}$ flavour indices. Operators for which the charge difference $X_{Q_i} - X_{f_j}$ is positive (negative) couple to the $\phi$ ($\phi^*$). 
The simplest UV completion of the FN mechanism is a set of vector-like FN-messenger-fermions $\Psi$ with masses $\sim \Lambda_F$ and couplings of the form $\phi \bar \Psi Q$.
The Yukawa matrices are then given by 
\begin{equation}
    Y^{u,d}_{ij} = c^{u,d}_{ij} \epsilon^{|X_{Q_i} - X_{{u,d}_j}|} \ ,
\end{equation}
where $\epsilon \equiv \langle \phi \rangle/\Lambda_F \ll 1$, and suitable choices of quark flavour charges can then naturally generate the observed mass and mixing hierarchies, usually (though not always) with $\epsilon \sim 10^{-1}$, directly generating the Cabibbo angle. Typically, the large top Yukawa is the only Yukawa coupling that is permitted in the UV ($X_{Q_3} = X_{u_3}$ if $X_H = 0$), explaining $Y_t \approx 1$, while other Yukawas are suppressed by powers of $\epsilon$.

A straightforward generalization of the FN mechanism to the Twin Higgs framework suggests the existence of two flavon fields, $\phi_A$ and $\phi_B$, each charged under separate U(1)$_{F,A}$, U(1)$_{F,B}$ flavour symmetries. 
The discrete $\mathbb{Z}_2$ twin symmetry enforces that flavons, Higgs bosons and fermions in both sectors have the same flavour charges. 
However, it has been observed ~\cite{Barbieri:2017opf} that if the two flavons have different vevs, $\langle \phi_A \rangle \neq \langle \phi_B \rangle$, the different dependence of the up and down Yukawa matrices on their respective $\epsilon_{A,B} \equiv \langle \phi_{A,B} \rangle / \Lambda_F^{A,B}$ --- $Y^{A,B}_u \sim \epsilon_{A,B}^{|X_{Q_1} - X_{u_1}|}$ and $Y_d^{A,B} \sim \epsilon_{A,B}^{|X_{Q_1} - X_{d_1}|}$ --- can lead to the up-down quark mass ordering being flipped in the twin sector relative to the SM, resulting in stable twin neutrons which could act as a DM candidate. 

Recent work on the general signatures and features of FN models that fit the SM~\cite{Cornella:2023zme,Cornella:2024jaw} indicates that this is actually a very common outcome.
For instance, $|X_{Q_1}-X_{u_1}| = |X_{Q_1}-X_{d_1}|+1$ is realized by half of the top twenty natural charge assignments from Ref.~\cite{Cornella:2023zme}; then once the coefficients are fitted to give rise to the SM fermion textures, one typically finds that $c_{11}^d/c_{11}^u \gtrsim \epsilon_A$.  In this case, $\epsilon_B \gtrsim c_{11}^d / c_{11}^u$ will flip the the up-down mass ordering in the twin sector for modest values of $\epsilon_B / \epsilon_A \sim \mathcal{O}(1)$.
Importantly, since the top Yukawa is typically the only Yukawa coupling which is independent of $\epsilon$, this spontaneous $\mathbb{Z}_2$ breaking in the flavon sector leaves the important $y_t^A = y_t^B$ symmetry which underlies the twin mechanism untouched.

To explore this possibility more systematically in the context of the matter coincidence problem, we examine the most natural FN charge assignments, as determined by~\cite{Cornella:2023zme,Cornella:2024jaw}.
We limit our attention to charge assignments that are compatible with supersymmetry, justified in the next section, which forbids $\phi^*$ in the operators that generate Yukawas. All of the top twenty quark textures in Ref.~\cite{Cornella:2023zme}  and a subset of the top twenty lepton textures in Ref.~\cite{Cornella:2024jaw} have this feature.

\begin{figure}
    \centering
    \includegraphics[width=0.65\linewidth]{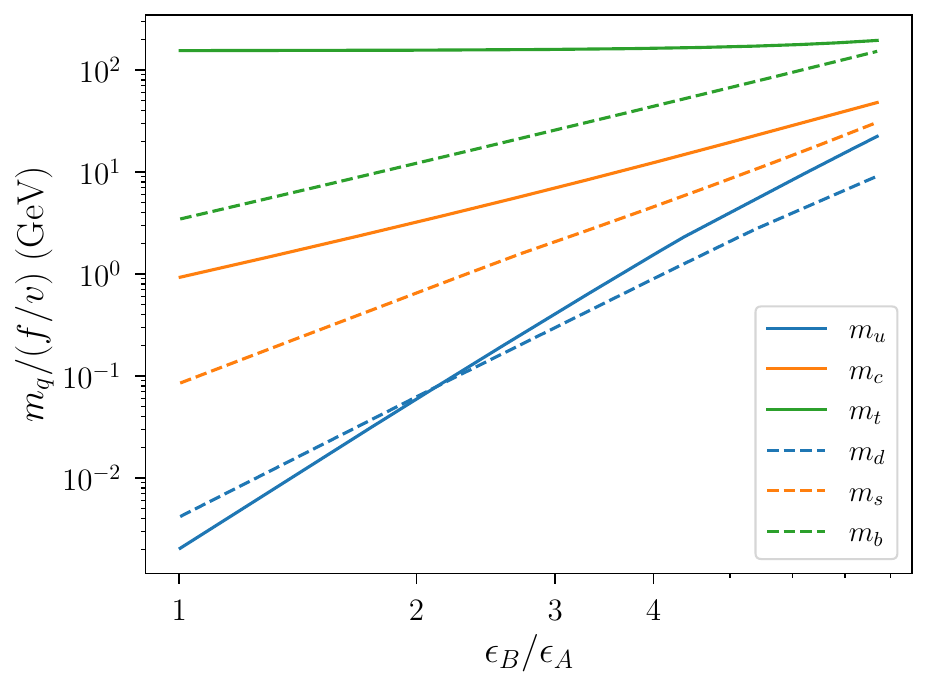}
    \caption{Quark masses as a function of $\epsilon_B/\epsilon_A$ for texture $X_Q=\{ 2,1,0\}$, $X_u =\{-3,-1,0\}$, $X_d =\{-2,-2,-2\}$ with $\epsilon_A=0.06$. At the SM point an exact coefficient fit for the couplings $c_{ij}^{u,d}$ was found to match the measured quark masses and mixings. The point $\epsilon_B/\epsilon_A=1$ corresponds to the SM. The up-down mass hierarchy flips at $\epsilon_B/\epsilon_A\approx 2$. These values will vary slightly for different coefficient fits. We restrict $\epsilon_B<0.5$. }
    \label{fig: quark masses for RG}
\end{figure}

\cref{fig: quark masses for RG} shows the different quark masses in the $B$-sector as a function of $\epsilon_B/\epsilon_A$ assuming one of the top quark flavour charge assignments from Ref.~\cite{Cornella:2023zme}. 
At the SM point ($\epsilon_B/\epsilon_A=1$), the masses are evaluated for a particular set of $\OfOrder{1}$ $c_{ij}$ coefficients that reproduces the SM quark mass matrix in the visible A-sector at the scale $\mu=1$ TeV. We then varied $\epsilon_B/\epsilon_A$, holding $c_{ij}$ fixed, and used the boundary condition $\alpha_{s,B}(m_{\max}) =\alpha_{s,A}(m_{\max}) $ to run the masses and strong coupling to low energies using the one-loop renormalization group. Here $m_{\max}$ is the heaviest field in either sector (effectively always the twin top quark), neglecting the possibility of UV coloured states. This allows us to identify the effective scale of twin QCD.\footnote{Because the exact definition of $\LQCD$ is ambiguous, we run both SM and twin sectors down to their respective poles and take the ratio between the two, and do not work with the scales directly.} 
The PDG conventions evaluate light quark masses at $\mu_{\rm{ref},A}= 2$ GeV and heavy quark masses at $m(\mu)= \mu$. We matched this convention with the substitution $\mu_{\rm{ref},B}=(\Lambda_B/\Lambda_A)\mu_{\rm{ref},A}$. 

The resulting up-down quark mass ordering inverts at $\epsilon_B/\epsilon_A \approx 2$, when their respective Yukawas are $\mathcal{O}(100)$ times larger than the SM value, as can be seen in \cref{fig: quark masses for RG}. 
Note that the twin confinement scale is increased modestly, due to all twin quark masses being higher due to both the increased twin Yukawa coupling and the usual $f/v$ Twin Higgs vev ratio. 

\begin{figure}
    \centering
    \includegraphics[width=0.98\linewidth]{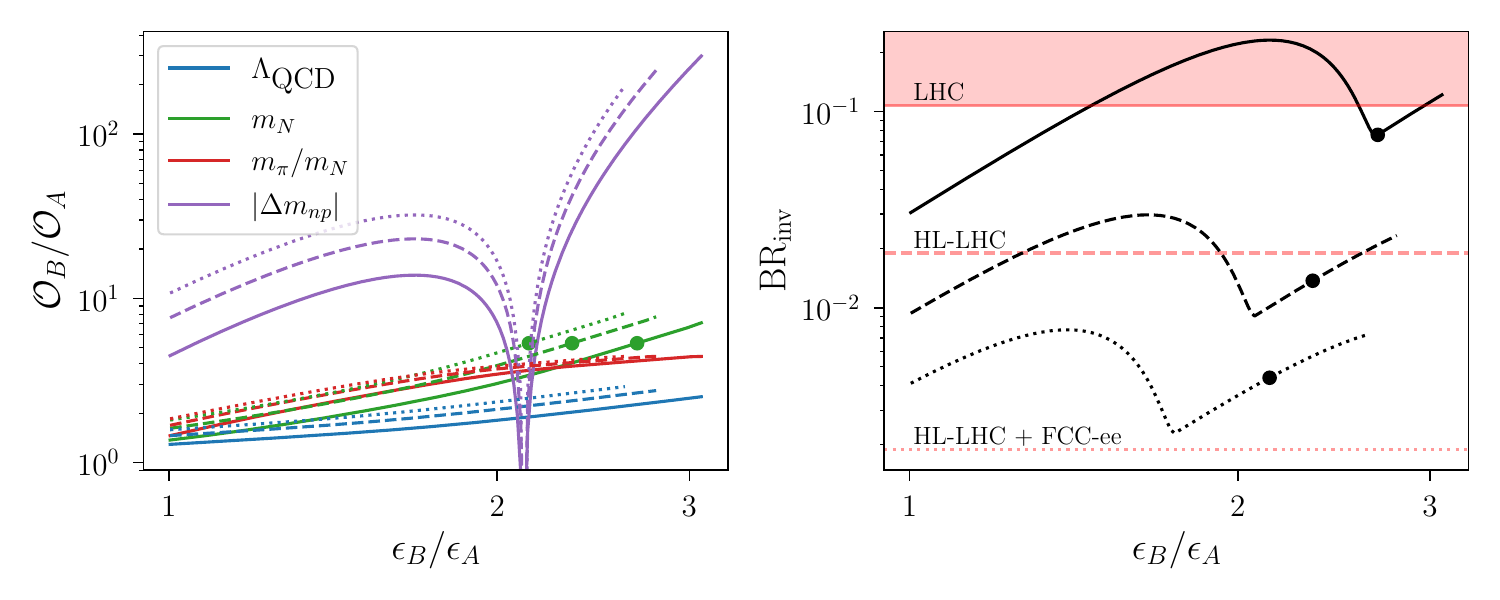}
    \caption{\textbf{Left:} Derived hadronic quantities as a function of $\epsilon_B/\epsilon_A$ for the texture in \cref{fig: quark masses for RG}. The solid-dashed-dotted lines give $f/v=\{3,5,7 \}$ respectively. 
    Here $\Lambda_{\mathrm{ QCD},B}/\Lambda_{\mathrm{ QCD},A}$ denotes the ratio of Landau poles in the strong couplings. The \textcolor{DarkGreen}{green} dots show the point at which $m_{N,B}=5$ GeV. \textbf{Right:} Higgs branching fraction to twin quarks. Horizontal lines indicate the current LHC bounds (\textcolor{red}{red}) \cite{ParticleDataGroup:2022pth}, projected HL-LHC bounds (\textcolor{red}{red dashed}), and combined HL-LHC and FCC-ee (\textcolor{red}{red dotted}) \cite{deBlas:2019rxi} . Note that in these plots, we have cut off $\epsilon_B$ at the maximum value that yields light quark parameters for which there is lattice data.}
    \label{fig: FN hadron masses}
\end{figure}

We show various quantities in the twin QCD sector as a function of $\epsilon_B/\epsilon_A$ in \cref{fig: FN hadron masses}. As expected from the previous figure, at $\epsilon_B/\epsilon_A \simeq 2$ the twin neutron--proton mass hierarchy flips.
Interestingly, for $\epsilon_B/\epsilon_A \sim 2-3$ we find $m_{N,B} \sim 5~\mathrm{GeV}$, $m_{\pi,B}/m_{N,B} \sim 0.5$, and $\DeltMnp$ spans the entire range explored in our phenomenological analysis of the earlier sections. 
As mentioned before, this behaviour is fairly generic amongst the top-20 supersymmetry-compatible quark flavour charge assignments identified by Ref.~\cite{Cornella:2023zme}.

The right panel of \cref{fig: FN hadron masses} shows the Higgs invisible branching fraction for this benchmark FN charge assignment and several values of $f/v$. 
The kink in each contour occurs because, as $\epsilon_B/\epsilon_A$ increases, the twin fermions become heavier and a dominant Higgs decay channel becomes kinematically inaccessible, sharply reducing the invisible width. 
The figure demonstrates that (HL-)LHC and future FCC-ee measurements can probe portions of the viable parameter space, see Ref.~\cite{McKeen:2023uzo} and references therein. This includes the region supporting naDM and underscores the importance of Higgs invisible-decay searches for testing this framework.

There is a discrete model-building assumption to be made regarding flavour in the lepton sector. One option is that the lepton sector either does not have its Yukawa couplings generated via an FN mechanism, or that this FN mechanism does not exhibit discrete $\mathbb{Z}_2$-breaking. In that case, the twin lepton masses are purely determined by $m_{e,B} = (f/v) m_{e,A}$, giving $m_{e,B} \sim 10^{-3}~\mathrm{GeV}$.
On the other hand, if the lepton sector Yukawas are generated by the FN mechanism and couple to the same flavon(s) as the quark sector, then 
the twin electron mass will be increased by a steep factor of $(\epsilon_B/\epsilon_A)^{|X_{L_1} - X_{e_1}|}$. For the top SUSY-compatible lepton charge assignments identified in Ref.~\cite{Cornella:2024jaw}, this predicts $m_{e,B} \sim 10^{-1}~\mathrm{GeV}$.  
Finally, it is of course possible that the lepton FN sector has its own pattern of $\mathbb{Z}_2$ breaking independent of the quark sector, in which case $m_{e,B}$ could be between or even above these two extremes.

\begin{table}
    \centering
    \begin{tabular}{|l|l|}
    \hline
    $m_{n,B}$ & $\sim 5~\mathrm{GeV}$
    \\ \hline
    $m_{\pi,B}/m_{n,B}$ & $\sim 0.5$
    \\ \hline
    $|\DeltMnp|/m_{n,B}$ & $\sim (10^{-3},~10^{-1})$ 
    \\ \hline
    $G_{F,B}$ & $\sim (10^{-2} - 10^{-1}) \times G_{F,A}$
    \\ \hline
     $m_{e,B}$
     & \hspace*{-3.5mm}
     \begin{tabular}{ll}
     $\ \sim$ MeV & ($\mathbb{Z}_2$ symmetric $Y_e$)
     \\ 
     $\ \sim$ 0.1 GeV & (shared $\cancel{\mathbb{Z}_2}$ with quark FN sector)
     \end{tabular}
     \\ \hline
    \end{tabular}
    \caption{
    Motivated naDM parameters in the Twin FN framework, for $f/v \sim 3-9$ and assuming the coincidence problem is solved by an unbroken high-scale $\mathbb{Z}_2$ dictating equal $A$- and $B$-sector baryon number production, while spontaneous $\mathbb{Z}_2$ breaking in the quark flavon vevs gives flipped twin up-down mass ordering with $m_{n,B} \approx 5~\mathrm{GeV}$ due to the combination of increased twin quark Yukawas, larger twin Higgs vev and modified twin QCD running.
    }
    \label{t.FNtwinmotivatedparameters}
\end{table}

In \cref{t.FNtwinmotivatedparameters}, we summarize the ranges of IR parameters obtained from this UV completion for the preferred FN charge assignments of Ref.~\cite{Cornella:2023zme}. 
We assume a \(\mathbb Z_2\)-symmetric baryogenesis mechanism produces equal surviving baryon asymmetries in the two sectors.
Explaining the observed ratio of DM to SM abundances therefore requires $m_{N,B} \sim 5\,\mathrm{GeV}$.
Comparing the parameter ranges in \cref{t.FNtwinmotivatedparameters} with the phase diagrams in \cref{subsec: phase diagram} and \cref{subsec: naDM abundances}, we find that this UV completion can naturally yield parameters consistent with the existence of naDM. In particular, the prediction $m_{\pi,B}/m_{N,B}\sim 0.5$ suggests that this UV completion gives rise to the \Ions\ or the \DnaDM\ phase; see \cref{fig: phase diagram mpi}.

\subsection{Supersymmetric Twin Froggatt--Nielsen UV Completion}

Since the FN mechanism operates at scales $\gtrsim 10^{5}~\mathrm{TeV}$, much above the $\sim 2-10~\mathrm{TeV}$ scale where the Twin Higgs framework must be UV-completed to protect the Higgs mass at higher scales, we must necessarily discuss how the above Twin FN setup could be UV-completed. 
Rather than build a full UV complete model, we sketch out one set of plausible ingredients and how they could generate the spontaneous $\mathbb{Z}_2$ breaking in the flavour sector, ideally with the required soft $\mathbb{Z}_2$ breaking in the Twin Higgs sector being a radiative consequence of the former. This can serve as a roadmap for future model-building studies.

\subsubsection*{Supersymmetric Twin Higgs}

Here we elaborate on a potential UV-completion for our FN Twin mechanism. 
We must first choose whether the full hierarchy problem at high scales is solved via supersymmetry~\cite{Martin:1997ns}, or via compositeness/extra dimensions~\cite{Csaki:2018muy}. We choose a supersymmetric UV completion, mainly due to our choice of the FN mechanism to address the flavour and coincidence problems.\footnote{In a composite Higgs setup, other mechanisms to resolve the flavour problem, like partial compositeness \cite{Kaplan:1991dc,Arkani-Hamed:1999ylh,Grossman:1999ra,Gherghetta:2000qt,Huber:2000ie}, are more natural, and it would be interesting to investigate how a twin version of those mechanisms could solve the coincidence problem in the future.}
Several examples of supersymmetric UV completions of the Twin Higgs have been constructed~\cite{Katz:2016wtw, Craig:2013fga, Chang:2006ra, Falkowski:2006qq,Badziak:2017syq, Badziak:2017kjk, Badziak:2017wxn, Asadi:2018abu}, typically featuring stop masses and SUSY breaking scales around $\sim 2~\mathrm{TeV}$. When adding a FN flavour sector to such a model, we must take care to not re-introduce a severe tree-level Higgs mass tuning problem due to the large hierarchy $\langle \phi \rangle = v_\phi \gg f \sim v_B >  v_A$ via biquartics like $\lambda_{H\phi} |H_{u,d}|^2 |\phi|^2$ (which can arise from a $\phi H_u H_d$ term in the superpotential).
This is related to the $\mu$-problem of the MSSM, and one possible solution is to use `holomorphic zeros'~\cite{Leurer:1992wg, Leurer:1993gy, Nir:1995bu}, where a negative choice of total flavour charge for the $H_u H_d$ superpotential operator prevents any $\mu$-term in the UV. After SUSY- and flavour-breaking, K\"{a}hler terms generate a $\mu$-term at a lower scale. However, this mechanism is unavailable to us, since the twin FN setup with $\epsilon_B \neq \epsilon_A$ would result in very different Higgs mass terms in the two sectors, destroying the Twin mechanism.

We therefore adopt an NMSSM-style~\cite{Ellwanger:2009dp} UV completion~\cite{Katz:2016wtw, Craig:2013fga, Chang:2006ra, Falkowski:2006qq} of the Twin Higgs. In such a setup with an additional gauge-singlet chiral superfield $S$, we can introduce a $\mathbb{Z}_3$ discrete symmetry which must commute with the twin $\mathbb{Z}_2$, and under which the flavons $\phi$ are neutral and all other fields (SM and twin fermions, Higgs fields, and $S$) have equal unit charge.
This forbids the $\mu$ terms, as well as superpotential operators like $\phi H_u H_d$ which would generate large Higgs mass shifts after the flavon gets its vev.
SUSY breaking soft terms then generate the singlet and Higgs masses/vevs.\footnote{The $\mathbb{Z}_3$ NMSSM also comes with some well-known cosmological challenges, like the domain-wall and tadpole problems~\cite{Abel:1995wk, Larsson:1996sp, Mazumdar:2015dwd, Bagger:1993ji, Bagger:1995ay, Abel:1996cr, Panagiotakopoulos:1998yw,Ellwanger:2009dp}, but these issues are orthogonal to our flavour- and Twin Higgs oriented discussion. For instance the domain wall problem can be averted by a small explicit bias.}
Crucially, this still allows the effective operators in Eq.~\eqref{eq: FN lagrangian} which generate the Yukawa couplings in the FN mechanism. In such a twin FN setup, it is also required that $X_\mu = X_{H_u} + X_{H_d} = 0$ and $X_S = 0$.  Otherwise, the SU(4) symmetric part of the Twin Higgs superpotential:\footnote{Note that the $\mathbb{Z}_3$ allows the existence of a $S^3$ term in the potential, which can give rise to a common $\mu$ term on the SM and the twin side (as $S$ is neutral under \Ztwo ). Such a shared $\mu$-term is not dangerous and can exist without spoiling any of our discussions.}
\begin{equation}
    W_{\mathrm{SU(4)}} \supset \lambda_S S (H_u^A H_d^A + H_u^B H_d^B)
\end{equation}
could only be generated with flavon insertions, which in our FN setup would generate SU(4)-violating masses and quartics in the Higgs sector---generically producing an uncontrolled \Ztwo-breaking contribution to the Higgs mass parameters that could exceed the amount compatible with a natural Twin-Higgs vacuum. 
We therefore set $X_S = X_{H_u} = X_{H_d} = 0$, which means that we can adopt all conclusions about the flavour sector studied in the previous section verbatim.\footnote{Requiring the singlet interaction to arise without flavon insertions imposes \(X_S+X_{H_u}+X_{H_d}=0\). We adopt the stronger simplifying choice \(X_S=X_{H_u}=X_{H_d}=0\), which leaves the FN textures studied above unchanged.}

All in all, this shows that there exists a plausible supersymmetric field-content and symmetry assignment in which the Higgs-sector interactions are independent of the asymmetric FN order parameters, allowing the latter to modify fermion Yukawas without automatically generating unacceptably large Higgs masses.

\subsubsection*{Flavon Sector}

We now discuss how the flavon sector of such a Twin NMSSM FN setup could spontaneously break the twin $\mathbb{Z}_2$. Based on some earlier models of spontaneous twin $\mathbb{Z}_2$ breaking~\cite{Batell:2019ptb}, one might think to implement a `flip flop' flavon potential, where a positive bi-quartic,
\begin{equation}
    V_\phi \, \supset \, \lambda_{AB} |\phi_A|^2 |\phi_B|^2 \, \quad (\lambda_{AB} > 0) \, ,
    \label{eq:V_flip_flop}
\end{equation}
would drive the vacuum towards $\phi_A = 0, \phi_B = v_\phi$ (or vice versa).\footnote{This potential can give rise to two disconnected vacua $\phi_A = 0, \phi_B = v_\phi$ or $\phi_A = v_\phi, \phi_B = 0$, thus it has a domain-wall problem. 
We can, however, simply assume this spontaneous breaking takes place before inflation and the symmetry is never restored, thereby solving this domain wall problem while still respecting the \Ztwo\ symmetry at the Lagrangian level.} 
To ensure a nonzero flavour breaking order parameter in both sectors, one should at least add a second pair of flavons, $\varphi_A$ and $\varphi_B$, with a $\mathbb{Z}_2$-symmetric vev, resulting in
$\epsilon_B/\epsilon_A \sim \mathcal{O}(1)$.

We can exploit flat directions of the flavon potential and introduce a new superfield to generate this quartic. 
Let us gauge the U(1)$_{X,A}$ $\times$ U(1)$_{X,B}$ flavour symmetries,\footnote{If the flavour symmetry is gauged, its anomalies may be cancelled through the Green--Schwarz mechanism~\cite{Ibanez:1994ig,Dudas:1995yu,Rathsman:2019wyk}, or through a suitable choice of charged matter, including anomaly-free light-field assignments, additional chiral fermions, or heavy FN messenger sectors~\cite{Rathsman:2019wyk,Bonnefoy:2019lsn,Smolkovic:2019jow}.} 
and introduce \textit{four} flavons on each side ($\phi_{A,B},~\bar \phi_{A,B}$, $\varphi_{A,B},~\bar \varphi_{A,B}$) with charges ($i=A,B$)
\begin{equation}
\begin{array}{c|cccc}
 & \phi_i & \bar{\phi}_i & \varphi_i & \bar{\varphi}_i \\ \hline
U(1)_{X,i} & +1 & -1 & +1 & -1
\label{eq:charges_flavons}
\end{array}
\end{equation}
and the superpotential 
\begin{equation}
    W\supset \lambda_Z Z(\phi_A \bar \phi_A + \phi_B \bar \phi_B -  v_\phi^2) \, + \, \lambda_Y Y(\varphi_A \bar \varphi_A + \varphi_B \bar \varphi_B - 2 v_\varphi^2) \ ,
\end{equation}
where $Z$ and $Y$ are neutral chiral superfields. 
$D$-flatness of each U(1)$_{X,i}$ demands:\footnote{These $D$-flatness conditions are the reason we need to introduce at least four flavons on each side in order to obtain the desired asymmetric vevs. If $\varphi_i$ and $\bar{\varphi}_i$ fields were not present in the spectrum, the $D$-flatness conditions enforce $|\phi_i| = |\bar{\phi}_i|$, thus in the desired asymmetric vev vacuum no flavons on the $A$ side would have a non-trivial vev to generate a mass for the SM fermions. }
\begin{align}
    D_{\mathrm{U(1)}_{X,i}}: |\phi_i|^2 - |\bar{\phi}_i|^2 + |\varphi_i|^2 - |\bar{\varphi}_i|^2 = 0 \,, \qquad i=A,B\, .
\end{align}
The $F$-term potential for the flavon fields will be given by
\begin{align}
    \label{eq:F_term_V}
    V_\phi &= |\lambda_Z|^2 |\phi_A \bar{\phi}_A + \phi_B\bar{\phi}_B - v_\phi^2|^2 + |\lambda_Y|^2 |\varphi_A \bar{\varphi}_A + \varphi_B\bar{\varphi}_B - 2v_\varphi^2|^2 \\
    &+  |\lambda_Z|^2 \, |Z|^2 \, \sum_i (|\phi_i|^2 + |\bar{\phi}_i|^2 ) +  |\lambda_Y|^2 \, |Y|^2 \, \sum_i (|\varphi_i|^2 + |\bar{\varphi}_i|^2 ) \, , \nonumber
\end{align}
which sets the sum in quadrature of various $A$- and $B$-sector vevs but leaves the allocation of flavon vev to each sector undetermined.  
From $F$-flatness conditions we will find $\langle Z \rangle = 0$ and $\langle Y \rangle = 0$ in the supersymmetric limit and the second line in \cref{eq:F_term_V} vanishes.

After SUSY breaking, a variety of effects will lift these flat directions, which depending on details could lead to spontaneous $\mathbb{Z}_2$ breaking in the flavon sector. 
For example, the Coleman-Weinberg potential will give 
\begin{equation}
    V = V_\phi + V_{CW} \supset V_\phi + m_{\phi_A}^2(\mu = \phi_A) |\phi_A|^2 + m_{\phi_B}^2(\mu = \phi_B) |\phi_B|^2 
\end{equation}
where $m_{\phi_A}(\mu) = m_{\phi_B}(\mu) \equiv m_{\phi}(\mu)$ are the identically running $\phi_{A,B}$ masses, which are evaluated at their respective vevs. 
Here we display only the $\phi_i$ contributions: assuming approximately identical running within each conjugate pair, $D$-flatness makes $\bar\phi_i$ track $\phi_i$, while the $\varphi_i,\bar\varphi_i$ contributions remain \Ztwo\ symmetric, so the additional flavons modify only the quantitative coefficients and not the vacuum-alignment argument.
Importantly, with the right sign in the RGE of $m_\phi^2$ ($m_\phi^2 \propto - \log (\phi/\mu)$)
the flavons will be driven towards the asymmetric vev $\langle \phi_A \rangle \ll \langle \phi_B \rangle \approx \sqrt{2} v_\phi$ (or vice versa). 
This will be the case  if flavour-gaugino contributions to the running mass dominate over Yukawa contributions from FN messengers, see SUSY renormalization group equations (RGEs) in Ref.~\cite{Martin:1997ns}.
Together with a symmetric pair of flavon vevs, this can therefore give the $\epsilon_A \neq \epsilon_B,\, \epsilon_B/\epsilon_A \gtrsim 1$ phenomenology our setup requires.

This demonstrates that a manifestly \Ztwo-symmetric supersymmetric flavon sector can possess flat directions that are lifted by radiative or SUSY-breaking effects in favour of one of two \Ztwo-related asymmetric vacua.

\subsubsection*{Higgs Sector}

We now show how the spontaneous $\mathbb{Z}_2$ breaking in the flavon sector may provide the common origin of the flavour- and Higgs-sector \Ztwo-breaking.\footnote{Although the spontaneous \Ztwo-breaking induced by the flavon vevs could in principle source an  asymmetric reheating---required to suppress otherwise excessive contributions to $\Delta N_{\mathrm{eff}}$---we remain agnostic about the origin of this asymmetry and leave a concrete realization, including its possible connection to the flavon sector, to future work. } 
To streamline the presentation, let us define 
\begin{equation}
    H_X = \sin \beta_X H_{u,X} + i \sigma_2 \cos \beta_X H^*_{d,X}\,,
    \label{eq:H_X}
\end{equation}
where $\sigma_2$ is a Pauli matrix. $H_X$ is the lighter Higgs doublet direction in each sector after the heavy one is integrated out. 
Here $\tan\beta_X\equiv v_{u,X}/v_{d,X}$, and we assume the approximately aligned limit \(\beta_A\simeq\beta_B\equiv\beta\). We focus on \(\tan\beta\simeq1\), which maximizes the SU(4)-symmetric singlet \(F\)-term quartic, proportional to \(\sin^22\beta\), while suppressing the SU(4)-violating MSSM \(D\)-term quartic, proportional to \(\cos^22\beta\) \cite{Katz:2016wtw}. 
This limit also helps keep the FN-enhanced twin-bottom Yukawa perturbative.

With this definition, the standard Twin Higgs potential \cite{Chacko:2005pe, Craig:2015pha} can be written as: 
\begin{eqnarray}
    V = &  m_H^2&(|H_A|^2 + |H_B|^2) + \lambda_H(|H_A|^2 + |H_B|^2)^2  \nonumber \\ &+&  \delta (|H_A|^4 + |H_B|^4) + \Delta m_H^2 |H_A|^2 \, ,
    \label{eq:pot_H}
\end{eqnarray}
where $m_H$ and $\lambda_H$  are the SU(4) invariant mass and quartic, $\delta$ is the SU(4)-breaking but $\mathbb{Z}_2$ preserving quartic.
In this equation, the effective quadratic coefficient $m_H^2 =  m_{B,\mathrm{eff}}^2$ is given by
\begin{equation}
    m_{X,\mathrm{eff}}^2 = (m_{H_u,X}^2 + |\mu|^2) \sin^2 \beta_X + (m_{H_d,X}^2 + |\mu|^2) \cos^2 \beta_X - b_X \sin 2\beta_X \, ,
    \label{eq:m_eff_def}
\end{equation}
where $m_{H_{u/d},X}$ are soft SUSY-breaking parameters, we have assumed a \Ztwo-symmetric $\mu$-term (motivated by the discussion in the previous section), and $b_X$ is the usual soft SUSY-breaking mass term on each side.
$\Delta m_H^2=m_{A,\mathrm{eff}}^2 - m_{B,\mathrm{eff}}^2$ in \cref{eq:pot_H} is the soft $\mathbb{Z}_2$-breaking Higgs mass required for $f/v > 1$, where $f^2 = 2(\langle H_A^2 \rangle + \langle H_B^2 \rangle) \sim v_B^2$ and $v^2=v_A^2 = 2 \langle H_A^2 \rangle$. 

At tree-level, the mass of the light Higgs boson, identified with our 125 GeV SM-like Higgs, is $m_h^2 \approx 4 \delta v^2$, setting $\delta \approx 0.065$. 
We can also relate the required size of the $\mathbb{Z}_2$ breaking soft mass to the desired vev ratio $f/v$:
\begin{equation}
    \Delta m_H^2 = \delta v_B^2\left( 1-\frac{2 v_A^2}{v_B^2}\right) \approx \left( \frac{f}{v} \  (60~\mathrm{GeV})\right)^2 \, .
\end{equation}
For $f/v \sim 3 - 7$, we therefore need ${\Delta m_H^2} \sim (180~\mathrm{GeV})^2 - (420~\mathrm{GeV})^2$.

Now consider $\mathbb{Z}_2$ asymmetric contributions to the running Higgs mass in our SUSY UV completion. As is evident from the representative example in \cref{fig: quark masses for RG}, the biggest difference between the A and B sector is introduced by the modified bottom Yukawa $y_{b,B}\sim \mathcal{O}(10)\times y_{b,A}$. This drags the $m_{H_d, B}$ mass down further in the IR than $m_{H_d, A}$, generating the needed positive contribution to ${\Delta m_H^2}$. We can estimate the magnitude of the generated radiative correction~\cite{Martin:1997ns}. Assuming $\tan \beta \approx 1$ and defining $\Delta m^2_{H_{u,d}} \equiv m^2_{H_{u,d}^A}-m^2_{H_{u,d}^B}$, $\Delta b \equiv b_A - b_B$, 
\begin{eqnarray} 
\Delta m_H^2
&=& \sin^2\beta \Delta m_{H_u}^2 + \cos^2\beta \Delta m_{H_d}^2 - \sin 2 \beta \Delta b 
\\ \nonumber 
&\approx & \frac{1}{2}\left( \Delta m_{H_u}^2 + \Delta m_{H_d}^2 \right) \sim \frac{1}{2} \Delta m_{H_d}^2
\\ \nonumber 
&\approx&\frac{3}{32 \pi^2} (\mathcal{X}_{b,B}  - \mathcal{X}_{b,A}) \log \frac{v_\phi}{m_\mathrm{soft}}
\\ \nonumber 
    &\sim& \frac{3}{16 \pi^2}   (y_{b,B})^2(3 m_\mathrm{soft}^2) \log \frac{v_\phi}{m_\mathrm{soft}}
 \\  \nonumber 
    &\approx& (230~\mathrm{GeV})^2\left(\frac{y_{b,B}}{10 y_{b,A}}\right)^2 \left( \frac{m_\mathrm{soft}}{1.5 \mathrm{TeV}}\right)^2 \left(\frac{\log \left(\frac{v_\phi}{m_\mathrm{soft}} \right)}{10}\right)
\end{eqnarray}
where $\mathcal{X}_{b,i}$ is the standard quantity appearing in MSSM RGEs \cite{Martin:1997ns}
\begin{equation}
    \mathcal{X}_{b,X} = 2 |y_{b,X}|^2 \left(  m^2_{H_d,X} + m^2_{Q_3,X} + m^2_{d_3,X} + |A_{b,X}|^2  \right)\, ,
    \label{eq:def_RGE_X}
\end{equation}
and we assumed all soft supersymmetry-breaking terms in $\mathcal{X}_{b,X}$ are $\mathcal{O} (m_\mathrm{soft}^2)$.
The $\Delta b$ contribution can be comparable to that from \(\Delta m_{H_d}^2\). Since our purpose here is only to establish the parametric size of the induced Higgs-sector \Ztwo-breaking, we neglect $\Delta b$; a complete calculation must include the coupled running of all Higgs-sector parameters.\footnote{The contribution from different $A$ vs $B$ sector FN messengers is much smaller due to the at most very modest $\mathcal{O}(1)$ mass difference between vector-like fermions in the two sectors of the FN UV completion. The charm Yukawa is also increased in the $B$ sector, but in absolute terms its effect is subdominant compared to the above twin-bottom quark contribution.}
However, our estimate clearly demonstrates that it is plausible for the spontaneous $\mathbb{Z}_2$ breaking in the flavon sector to radiatively generate the $\mathbb{Z}_2$-breaking Higgs mass needed for a natural MTH realization with $f/v \sim 3 - 9$.

\medskip

All in all, the ranges summarized in \cref{t.FNtwinmotivatedparameters} show that a Twin FN construction can populate the parts of our IR parameter space that support naDM. In particular, among the well-performing FN textures considered in Ref.~\cite{Cornella:2023zme}, the values of $\epsilon_B/\epsilon_A$ that reverse the twin up--down hierarchy also tend to enhance the twin confinement scale and produce a twin-nucleon mass in the few-GeV range, see \cref{fig: FN hadron masses}. 
Using our hadronic interpolation, these points characteristically give $m_{\pi,B}/m_{n,B}\sim0.5$, while the twin neutron--proton splitting varies substantially across the texture and coefficient choices. Depending on this splitting and on the realization of the lepton sector, the resulting cosmology can lie in different naDM regions of the phase diagrams from \cref{subsec: phase diagram}. 

If a symmetric baryogenesis mechanism generates equal surviving baryon asymmetries in the two sectors, the observed abundance ratio further selects $m_{n,B}\simeq 5\,\mathrm{GeV}$. We therefore find a nontrivial alignment between the mass scale suggested by the matter coincidence problem and the region obtained from spontaneous flavour-sector $\mathbb{Z}_2$ breaking. This conclusion is robust at the level of the FN texture scan. 
A complete UV realization must additionally demonstrate the asymmetric flavon vacuum, specify the ordering of the flavour-, $\mathbb{Z}_2$-breaking, and SUSY-mediation scales, and verify that baryogenesis and subsequent thermal evolution preserve the required relation between the two sector asymmetries. 
The fate of any light flavon degrees of freedom, or twin lightest-supersymmetric-partner~\cite{Badziak:2019zys, Badziak:2022eag}, should also be investigated---we leave all these studies to future works.


\section{Discussion and Conclusions}\label{sec: conclusion}

We have studied the cosmology and phenomenology of twin atomic dark matter in the Mirror Twin Higgs---with a careful treatment of nuances arising from twin nuclear physics---applying this developed theoretical framework to the coincidence problem and fleshing out the arguments of this paper's short companion letter~\cite{Asadi:2026oae}.  
The twin $\mathbb{Z}_2$ symmetry is used to correlate both the dark matter mass and number density with their SM baryonic counterparts.
We assume that twin electromagnetism is unbroken (to allow for the formation of twin atoms and minimize sources of $\mathbb{Z}_2$ breaking); asymmetric reheating leaves the twin sector colder than the visible sector (to satisfy $\DNeff$ bounds); and twin baryons make up $100\%$ of dark matter, with some $\mathbb{Z}_2$-respecting baryogenesis mechanism generating equal or at least similar baryon and twin-baryon asymmetries.
Motivated by ultraviolet constructions based on the Froggatt--Nielsen mechanism, we assume that the twin neutron--proton mass ordering is inverted, allowing for stable $\sim 5~\mathrm{GeV}$ twin neutron DM (TNDM) to make up the majority of cold dark matter to solve the coincidence problem. A small nuclear atomic dark matter (naDM)  sub-component is comprised of surviving twin nuclei and electrons.
We highlight the regions of parameter space in which a $0.01 - 20$\% fraction of the dark matter survives as (possibly astrophysically relevant) naDM and carefully examine its phenomenology.

We characterized the naDM scenario in the IR in terms of twin-hadronic effective parameters, including the twin nucleon and pion masses, the twin neutron--proton mass splitting, the twin electron mass, the twin Fermi constant, and the relative temperature of the two sectors.
We mapped these parameters onto the underlying twin quark masses and confinement scale, identified the regions in which different twin nuclei are stable, and determined the total surviving naDM fraction. 
We find that the phenomenology is most sensitive to the twin electron mass, the twin neutron--proton mass splitting, and the twin Fermi constant.

We also examined the effects of recombination and subsequent electron capture by nuclei synthesized during twin BBN, which can destabilize twin atoms and alter the late-time composition of naDM. 
This gives rise to a rich phase structure for the dark sector (see \cref{tab: phase diagram regions}) and motivates more detailed future studies.

An appreciable but subdominant naDM component generally requires a fractional twin neutron--proton mass splitting near the percent level, with the precise viable range extending approximately from $|\DeltMnp|/m_{N,B}\sim0.1\%$ to $10\%$.
To determine this abundance, we generalized the standard treatment of nucleon freeze-out and BBN to the twin sector. 
We treated the significant non-perturbative uncertainties as conservatively as possible. Where microscopic quantities could not be reliably calculated throughout our parameter space, we bracketed their effects using alternative lattice-motivated fits, heavy-quark estimates, and limiting cases defined by extreme parameter choices. In particular, we considered both the possibility that the twin deuteron remains bound and the possibility, suggested by recent lattice calculations, that it is unbound away from the SM point. 
These alternatives can qualitatively change the nature of the dominant naDM.

We have also shown that viable Twin Higgs scenarios can produce substantial primordial abundances of twin nuclei heavier than helium. For each model, we estimate the maximum atomic number, $\Znum_{\max}$, that can be synthesized in appreciable abundance during twin BBN. 
Since this estimate depends on uncertain twin-nuclear binding energies and reaction pathways, it should be interpreted as an approximate upper limit. 
Such twin metals would recombine and evolve very differently from the light atoms usually considered in aDM, potentially crystallizing into twin-solids or twin-dust grains and opening qualitatively new regimes of self-interaction, structure formation and direct detection, which deserve future study.

The resulting mixed dark sector gives rise to a broad range of observable phenomena. The most striking cosmological signature is a unique two-scale DAO signal which lies at the boundary of current UVLF constraints. Astrophysically, the 5 GeV twin neutron naDM benchmark supplies a highly motivated target for detailed simulation studies of naDM galactic dynamics, and the formation and signatures of mirror stars with densities (sizes) much larger (smaller)  than that of SM stars. 
While naDM satisfies current Bullet Cluster self-interaction bounds, future cluster measurements might probe TNDM self-interaction cross sections. Stochastic gravitational waves from the twin electroweak phase transition may also be motivated, in particular if twin electroweak baryogenesis could generate $\mathbb{Z}_2$ symmetric (twin) baryon abundances. 

Terrestrial observables include relatively standard collider probes of the Twin Higgs, including Higgs coupling deviations and invisible Higgs decays, though the latter are modified  from the standard Mirror Twin Higgs due to the modified twin Yukawa couplings that give rise to a $\sim 5~\mathrm{GeV}$ stable twin neutron. Future lepton colliders may be able to detect this difference through off-shell Higgs production modes of twin states, another motivated subject for future investigation.

The direct detection phenomenology of naDM is very rich. 
Direct detection can proceed through both the irreducible Higgs portal of the Twin Higgs framework and the photon kinetic-mixing portal. 
The Higgs portal can mediate nuclear recoils even when the dark matter is almost entirely composed of twin neutrons, while kinetic mixing can produce both nuclear and electron recoils from the naDM component. The predicted rates depend strongly on the ionization fraction and galactic distribution of naDM, so we bracketed the possibilities by considering fully atomic and fully ionized populations with both halo- and disk-like distributions; for light twin electrons, thermal coupling within an ionized component can significantly enhance the electron-recoil signal. For a benchmark naDM fraction of $10\%$, nuclear and electron recoils can probe kinetic mixing down to approximately $\kappa\sim10^{-11}$, depending on the ionization state and galactic distribution. 
In the event of a signal, detailed measurements of recoil spectra may reveal the multi-component nature of the dark sector; if these measurements furthermore reveal masses in multiples of $\sim 5~\mathrm{GeV}$, then this would provide tantalizing evidence of a twin solution to the coincidence problem.
The coincidence-motivated twin nucleon mass range also places much of the irreducible Higgs-portal signal near the solar-neutrino background, making this framework a particularly motivated target for future directional detectors.

Two main sources of twin hadronic uncertainty stand out from our investigation. First, the precise character of the twin nucleons (D, He and/or beyond) requires convergence on lattice calculations with pion masses higher than the SM value, to determine whether the twin deuteron and possible multi-neutron states are bound, and to more precisely determine the corresponding deuterium, helium, and heavier-nucleus binding energies.
Second, more precise determinations of nucleon scattering lengths (again with heavier pion masses) would allow for DAO constraints to be sharpened to possibly exclude part of naDM parameter space, or pinpoint DAO signatures at galactic or sub-galactic scales that could be discovered in the future.

Our analysis has been primarily bottom-up: we identified the regions of IR parameter space that produce viable and phenomenologically rich naDM without committing to a particular ultraviolet realization, though we expect naDM to be a generic consequence of many classes of theories that solve the coincidence problem~\cite{Foot:2003jt,Lonsdale:2018xwd,Beauchesne:2020mih,Bodas:2024idn}. 
We supply a proof of principle that a supersymmetric Twin Higgs solution to the hierarchy problem, together with a Froggatt--Nielsen solution to the flavour problem, can naturally generate this naDM scenario.  We showed that spontaneous $\mathbb{Z}_2$ breaking in the flavon sector can reverse the twin up- and down-quark hierarchy for generic choices of flavour charges; preserve the equality of the top Yukawa couplings that underlies the Twin Higgs mechanism; and radiatively generate the soft $\mathbb{Z}_2$-breaking in the Higgs sector needed to realize natural Twin Higgs models with $f/v \sim 3-10$. Fully developing this UV completion would be an interesting future model-building investigation.

The matter coincidence remains the strongest empirical indication that the visible and dark sectors may be connected non-gravitationally, and twin asymmetric dark matter provides arguably the most fully-developed solution to this problem to date. Our work constitutes the first detailed, systematic exploration of the combined nuclear, cosmological, and observational consequences of this class of models. It demonstrates that a viable twin-neutron dark sector need not be phenomenologically sterile: a small, cosmologically selected nuclear and atomic component can provide observable traces of the underlying connection between flavour, naturalness, and dark matter.

\section*{Acknowledgements}
We thank Wolfgang Altmannshofer, Stefania Gori, Michael Luke, Jared Barron, and Nick Rodd for helpful discussions. 
The work of PA is supported in part by the U.S. Department of Energy grant number DE-SC0010107.
ZC is supported in part by the National Science Foundation
under Grant Number PHY-2514660 and by the Maryland Center for Fundamental Physics.
The work of DC is in part supported by Discovery Grants from the Natural Sciences and Engineering Research Council of Canada, the Canada Research Chair program, the Ontario Early Researcher Award, and the University of Toronto McLean Award.
The work of MM is also supported in part by the NSERC Postgraduate Scholarship-Doctoral.
The work of ETN is supported in part by the U.S. Department of Energy under grant number DE-SC0010005.
In our analysis and preparation of this manuscript, Claude and ChatGPT were used to assist with literature searches, resolve stubborn \LaTeX{} errors, refine the language, and for iterative discussions and clarifications of selected physics questions. 
The authors take full responsibility for the content of the manuscript.


\appendix

\section{Auxiliary Material}
\label{app: auxiliary}

In the auxiliary material we include several additional documents for reference. These files contain plots of the $\fnaDM$ fraction and values of $\Znum_{\max}$ as in~\cref{fig: nadm abundance benchmark} for the full parameter space defined by~\cref{tab: IR params}. These are named according to the convention \texttt{bound\_mn\_<mn>.pdf}, where \texttt{<mn>} is the numerical value of the twin neutron mass in GeV. The unbound case has \texttt{unbound} leading and matches the conventions of~\cref{fig: nadm abundances unbound deuteron benchmark}.

\section{Lattice Fits}\label{app: lattice fits}

We follow the conventions of Ref.~\cite{DeGrand:2019vbx} and parameterize the lattice derived quantities in terms of $(m_{\pi,B}/m_{\rho,B})^2$, which itself can be found as a function of $m_{\pi,B}$. 
We fit these data to polynomial functions, emphasizing that these are not meant to be interpreted as physical models.
For a quantity $\mathcal{O}$ with mass dimension $M$, we use the ansatz 
\begin{equation}
    \mathcal{O} = \sum_{n=0}^N \left(A_n x^n\right) \left(\frac{\Lambda_B}{\Lambda_A} \right)^{M-qn}, 
\end{equation}
where $q$ is the mass dimension of the variable $x$ and $A_n$ are fit coefficients to be determined from lattice data. The order of the fit (largely $N=3$ in this work) is chosen to get good agreement with the data for the range in which lattice data exists and is not used to extrapolate beyond this region. Note that this implies that the fit coefficients $A_n$ have the dimensions of $M-qn$.

We include the physical point in the data, since most lattice results are computed at $m_{\pi,B}\geq 200$ MeV. The resulting fit coefficients can be found in \cref{tab: lattice fits}.

\begin{table}[]
    \centering
    \renewcommand{\arraystretch}{1.4} 
    \begin{tabular}{|c|c|c|c|c|c|}
        \hline
        Observable  & Variable& $A_0$ & $A_1$  & $A_2$ & $A_3$ 
        \\
        \hline
        $ \left(m_{\pi,B}/m_{\rho,B} \right)^2$ & $m_{\pi,B}$ (GeV) &0.00132 & -0.0157 & 1.738 & -1.466
        \\
       $m_{N,B}$ (GeV) & $(m_{\pi,B}/m_{\rho,B})^2$ & 0.877 & 2.376 & -3.609 & -3.851
       \\
       $m_{\Delta,B}$ (GeV) & $(m_{\pi,B}/m_{\rho,B})^2$ & 1.155 & 3.473 & -8.543 & 8.808
       \\
       $f_{(u,d),B}^{N}$ 
       &
       $(m_{\pi,B}/m_{\rho,B})^2$
       & 0.02 & 0.52 & --- & --- 
       \\
        \hline
    \end{tabular}
    \caption{Fit coefficients used for different lattice quantities, found by digitizing the data in Ref.~\cite{DeGrand:2019vbx}.}
    \label{tab: lattice fits}
\end{table}


\section{Cosmological Evolution Background}\label{app: cosmo BG}

In this appendix we review basics of cosmological evolution of the two sectors, their decoupling, and milestone epochs. The naDM abundances presented in the main text are computed from twin neutron--proton freeze-out on top of this background evolution. While the late-time temperature ratio of the twin and SM sector is determined when we fix $\DNeff$, the timing of each decoupling can vary depending on the model content and thus the temperature ratio during twin neutron--proton freeze-out may be different. There are three milestone temperatures at which the evolution is modified:

\begin{itemize}
    \item Decoupling of twin and SM sector and subsequent asymmetric reheating
    \item Decoupling of twin neutrinos and photons
    \item Twin QCD phase transitions
\end{itemize}

Fixing $\DNeff$ corresponds to setting a late time boundary condition on the temperature ratio of the two sectors. For one massless photon and three approximately massless neutrinos at a single temperature $T_B$, the instant decoupling approximation yields~\cite{Bansal:2021dfh}

\begin{equation}\label{eq: Dneff in terms of temp ratio}
    \DNeff = \frac{29}{7}\left( \frac{11}{4}\right)^{4/3} \left(\frac{T_B}{T_A}\right)^{4}\simeq 15.96\left(\frac{T_B}{T_A}\right)^{4} ,
\end{equation}
where the temperature ratio is evaluated at CMB times. We use $T_A=T_B$ as our early time boundary condition, however, we note that our results are largely insensitive to the temperature evolution above $T\sim $ GeV, and we expect the results to be essentially unchanged even under the assumption that the two sectors are decoupled at scales much higher than the naive twin-SM decoupling of our numerical computation. 

We use conservation of entropy to evolve the scale factor above the temperature of the SM-twin decoupling. We use the full thermal integrals for the energy density $\rho(T)$ and pressure $P(T)$ to compute the entropy at each temperature. 
When the sectors decouple, the entropy in each bath is conserved separately. We take a hard decoupling approximation such that above temperature $T_{dec,A/B}$ the two baths are in equilibrium, and below the two baths are completely secluded. We can then use the common value of the scale factor $a$ to determine the twin sector temperature as a function of the SM temperature. We perform a similar hard decoupling at the neutrino decoupling of each sector. Therefore our primary aim in this section is to determine the temperatures of these decouplings. 

The first decoupling is between the SM and the twin sector.  
The decoupling occurs when the Higgs portal interaction and the kinetic mixing portal go out of equilibrium. However, as mentioned above, we find numerically that so long as the decoupling temperature is above the mass splitting of the nucleons, the final naDM abundance is insensitive to the exact temperature of the SM-twin decoupling. Therefore we use a simplified approach to select our decoupling temperature based on the Twin Higgs scaling. 
We estimate this rate using a similar method to Refs.~\cite{ Chacko:2016hvu, Curtin:2021alk}. These works assumed a mirror Twin Higgs scaling. The Higgs interaction can be written as

\begin{equation}
    \angl{\sigma v} \simeq \left(\frac{2 m_A^i m_B^j}{v_{A}^2}\right)^2 \left(\frac{G_{F,B}}{G_{F,A}}\right)^2 \frac{T^2}{m_{h,A}^4}.
\end{equation}
The masses $m_X^i$ correspond to the heaviest fermion of each sector that is in equilibrium with the bath, which will dominate the Higgs interaction, and $T$ is the common temperature $T=T_A=T_B$. We have used $G_F \propto m_W^{-2} \propto m_h^{-2}$ to write this in terms of the ratios of the Fermi constants and the SM Higgs vev $v_A$. 
Decoupling occurs when $n\angl{\sigma v} \sim H$, with $n$ the fermion number density.
For $G_{F,B}\lesssim G_{F,A}$, this condition is satisfied around $T\gtrsim$ few GeV. The exact spectrum of twin quark masses does not substantially change this temperature, as $\gstarS$ only varies by a percent-level amount. We find our results largely insensitive to the exact temperature of the twin decoupling and therefore use the curve found in Ref.~\cite{Curtin:2021alk} with the substitution $f/v \rightarrow \left(G_{F,A}/G_{F,B} \right)^{1/2}$. 

After the two sectors decouple, asymmetric reheating must occur to ensure that the dark sector temperature is sufficiently suppressed not to run afoul of $\DNeff$ bounds \cite{Chacko:2016hvu, Craig:2016lyx}. 
Different methods have been proposed to achieve this \cite{Adshead:2016xxj, Craig:2016lyx, Chacko:2016hvu,Alonso-Alvarez:2023bat}, and we do not specify which in this work. We set the dilution temperature at $T_{\gamma,B}\sim 0.9\ T_{dec,A/B}$, just below the twin-SM decoupling temperature.

After the two sectors decouple, the neutrinos will eventually also decouple from their respective baths, which can modify the background twin photon temperature during twin neutron--proton freeze-out. 
The neutrinos in both sectors are taken to be approximately massless and will decouple from their respective baths when their weak interactions with fermions (primarily electrons) go out of equilibrium. This is usually estimated using the weak rate $\Gamma \simeq G^2_F T^5\sim H$. 

For the SM neutrinos this decoupling proceeds virtually unchanged from the SM scenario except for $H^2 \propto \gstarS \sim 2 g_{\star S, SM}$; that is, there are roughly twice as many degrees of freedom in the bath as in standard cosmology. This shifts the decoupling temperature by only $\sim 2^{1/6}$. In practice, the shift is usually even smaller, since the dark sector is cool relative to the SM sector and $\gstarS$ is less than twice the standard expectation. For completeness, we compute the SM neutrino decoupling temperature for each model consistently, according to the prescription below. 

For the twin neutrinos, we generally expect similar behaviour, however, the interaction rate has a strong dependence on the twin electron density since it is the lightest charged field that the neutrinos interact with. In the SM, neutrino decoupling occurs before electron-positron annihilation occurs, but in the twin sector this may no longer be true. If the twin electrons are sufficiently heavy, twin neutrino decoupling will be triggered early when the twin electron density drops. The cross section for neutrino interactions with electrons in either sector is approximately given by Ref.~\cite{Lesgourgues:2013sjj}

\begin{equation}
    \sigma_{e\nu,X} \simeq \frac{4 G_{F,X}^2  E_\nu^2}{\pi}.
\end{equation}
Here $E_\nu$ is the energy of the neutrinos. 
Following the general prescription of Ref.~\cite{Mukhanov:2003xs}, treating Pauli blocking as insignificant at our level of precision, we get:

\begin{equation}
    \Gamma_{\nu,B} = n_{e,B} \sigma_{e\nu,B} v_{\nu,B} \simeq  \frac{2 G_{F,B}^2 m_{e,B}^5}{\pi^3} \int_{1}^\infty \frac{ x^4}{1+ e^{m_{e,B} x/T_B}}\left(1- \frac{1}{x^2}\right)^{3/2}  dx, 
\end{equation}
where $x= E_e/m_{e,B}$ with $E_e$ the electron energy, which is related to $E_\nu$ by $E_\nu^2 = E_e^2-m_{e,B}^2$ in the centre of mass frame, with the neutrino treated as massless such that the average over the velocity distribution is exactly $\angl{v_\nu}=1$. Note that here $T_B$ is the temperature of the twin electrons, thus the temperature of the twin photons. Twin neutrino decoupling occurs when $H \sim \Gamma_{\nu,B}$. In the SM-like regime, this is determined largely by $G_{F,B}$, but in the case that the electron is much more massive the decoupling occurs shortly after $T_{\gamma,B}\sim m_{e,B}$. 

Finally, we need to account for the twin QCD phase transition (QCDPT). This impacts our analysis in two ways. Firstly, the number of degrees of freedom in the bath is modified at this epoch.  Secondly, and perhaps more importantly, our computation assumes that the degrees of freedom during freeze-out are baryons. However, if the twin QCDPT temperature is comparable to $|\DeltMnp|$, then it is possible that the weak interactions will freeze out before the QCDPT occurs. In this case, the DOF which undergo freeze-out would be quarks, not baryons, and the final baryon abundance of the phase transition would be dependent on the quark content of the bath at the time of the QCDPT. While this scenario is interesting, further modelling would be required to determine the exact output at these points, so we exclude it for our present investigations.

Lattice computations have found that the temperature of the SM QCDPT is approximately 157 MeV \cite{HotQCD:2018pds, Borsanyi:2020fev,Guenther:2020jwe,Aarts:2023vsf}. For QCD with a different number of active flavours at the time of the phase transition, we use the results of Refs.~\cite{Lombardo:2014fea, Laermann:1996xn, Karsch:2000kv, Engels:1996ag}. These works computed the ratio $T_\mathrm{QCD}/\sigma^{1/2}$ for different numbers of active light twin flavours $N_f$, where the string tension $\sigma^{1/2}\sim \LQCD$. Since there is no data at $N_f=1$, we use a linear interpolation between the zero- and two-flavour cases.  For a given $N_f$, it is then straightforward to rescale the SM temperature to find the twin $T_\mathrm{QCD}$:
\begin{equation}
    T_{\mathrm{QCD},B}(N_f) =\left( \frac{\Lambda_B}{\Lambda_A} \right)\frac{C_{N_f}}{C_3}T_{\mathrm{QCD},A}.
\end{equation}
Here $C_{N_f}$ are the lattice values of $T_{\mathrm{QCD},{N_f}}/\sigma^{1/2}$ for $N_f$. Because $T_{\rm QCD}$ increases when the number of flavours decreases, it is possible that a quark field decoupling will trigger a spontaneous phase transition. We find the cosmological output largely insensitive to $T_{\rm QCD}$, so we take a hard decoupling approximation at $T=m_q$ and compute the phase transition temperature above and below threshold to determine the approximate transition temperature. 

To determine the number of quarks in the bath at a given temperature, we need to know the mass spectrum of quarks in the theory. However, the results of \cref{subsec: dark hadron masses} only determine the masses of the two lightest quarks. Further assumptions would have to be included to specify the masses of the heavy quarks. Fortunately, the phase transition temperature varies slowly with $N_f>3$, so the approximation $N_f \leq 2$ is reasonable, and we only use the twin up and down quarks to determine $T_{\mathrm{QCD},B}$ in our analysis.

Additionally, since the heavy quarks participate in none of the freeze-out interactions to follow, the effect of their masses is confined to a weak dependence in the Hubble rate via $\gstarS$. We explicitly verified by computing the cosmological evolution for the mass spectra of representative models from \cref{sec: uv realization} that the effect on the final freeze-out fraction was negligible. Therefore we set all twin quark masses other than up and down to be $\gg $ GeV and neglect them in our freeze-out analysis. If one of the other quarks was in fact light enough to remain in the bath until the phase transition, it would cause the phase transition temperature to be $\OfOrder{5\%}$ lower, and have a negligible effect on the freeze-out. 

To account for the effect of both the visible and twin sector phase transitions on $\gstarS$, we make the approximation that at $T_{\rm QCD}$, all quark and gluon degrees of freedom (DOF) are instantaneously removed from the bath and hadronic DOF enter.


\section{Dark Neutron-Proton Freeze-out}\label{app: np freeze-out}

In this appendix we present the interaction rates that dictate twin neutron--proton freeze-out presented in \cref{sec: cosmo}, specifically Eq.~\eqref{eq: np boltzmann evolution}. There are two relevant rates to compute. The first are the two-to-two scattering processes that maintain equilibrium between twin neutrons and protons, and the second is the twin proton decay rate. 
The dominant processes maintaining thermal equilibrium between twin neutrons and protons are 
\begin{equation}\label{eq: freezeout processes appendix}
    p_B^+ + \bar{\nu}_B \leftrightarrow n_B + e_B^+, \qquad p_B^+ + e_B^- \leftrightarrow n_B + \nu_B.
\end{equation}
These processes can be computed in the Fermi effective theory; for what follows, we follow the treatment of Ref.~\cite{Mukhanov:2003xs}. The effective 4-Fermi operator is 
\begin{equation}
    \mathcal{L}\supset \frac{ G_{F,B}}{\sqrt{2}} (\bar \nu_{e,B} \gamma_\mu(1- \gamma_5) e_B) (\bar n_B \gamma^\mu(g_{v,B}- g_{a,B}\gamma_5)p_B).
\end{equation}
Here $g_{v,B}$ is the effective vector coupling of the nucleons, $g_{a,B}$ is the effective axial coupling, and $e_B,\ \nu_{e,B},\ p_B,\ n_B$ are the twin electron, electron neutrino, proton, and neutron spinors, respectively. We assume the conserved vector current hypothesis such that $g_v=1$. We also make use of the simplifying assumptions that $m_{N,B} \gg |\DeltMnp|$, $m_{N,B}\gg m_{e,B}$, and that twin neutrinos are effectively massless. The relative concentration of a species $X_i$ is given by Eq.~\eqref{eq: p/n relative number density} and evolves according to
\begin{equation}
    \frac{dX_i}{dt} = \sum_{j \neq i}\left(\Gamma_{j\rightarrow i}X_j - \Gamma_{i\rightarrow j}X_i \right)
\end{equation}
with $\Gamma_{i\rightarrow j}$ the rate for $i$ to convert into $j$, subject to the constraint $\sum_j X_j = 1$. Because the initial equilibrium abundance of nuclei is negligible, we will restrict ourselves to $\{i,j\}=\{p_B, n_B\}$. For the first process in Eq.~\eqref{eq: freezeout processes appendix} one finds
\begin{equation}\label{eq: rate p nu to e n}
   \Gamma_{p\nu \rightarrow n e,B} = \frac{1+ 3 g_{a,B}^2}{2\pi^3} G_{F,B}^2|\DeltMnp|^5 J\left[ \max{\left(1, \frac{m_{e,B}}{|\DeltMnp|}\right)},\infty\right] \, .
\end{equation}

The integral $J[a,b]$ is defined as
\begin{equation}
    J[a,b] = \int_a^b \frac{dq\ q^2  (q-1)^2 }{1 + e^{-q|\DeltMnp|/T_{\gamma,B}}}\frac{1}{1+ e^{(q-1)|\DeltMnp|/T_{\nu,B}}} \sqrt{1-\left( \frac{m_{e,B}}{q\DeltMnp}\right)^2},
\end{equation}
where $q \equiv E_e/|\DeltMnp|$ and $E_e$ the electron energy. This treatment is identical to the SM version except for the lower limit of integration in the $J$ function in Eq. \eqref{eq: rate p nu to e n}; in the case where $m_{e,B} > |\DeltMnp|$, the minimum electron energy is $m_{e,B}$ rather than $|\DeltMnp|$, and the initial state neutrino cannot be arbitrarily soft.

The second process can also be written in terms of $J[a,b]$:
\begin{equation}\label{eq: rate p e to n nu}
    \Gamma_{p e \rightarrow n \nu} = \frac{1+ 3 g_{a,B}^2}{2\pi^3}G_{F,B}^2 |\DeltMnp|^5 J\left[-\infty, - \frac{m_{e,B}}{|\DeltMnp|} \right].
\end{equation}
The rates of the inverse reactions are related to the rate of the direct reactions (at $T_{\nu,B}=T_{\gamma,B}=T_B$)
\begin{equation}
    \Gamma_{cd\rightarrow ab,B} =\Gamma_{ab\rightarrow cd,B} e^{-|\DeltMnp|/T_B}.
\end{equation}
Since the neutrinos decouple at a similar temperature as the baryons, for most of the evolution this will hold. If the dark electrons annihilate after freeze-out, the temperatures are the same throughout the entire freeze-out process. If electron annihilation triggers freeze-out, the photons and neutrinos will remain in equilibrium until the abundance of electrons is exponentially suppressed, and the entropy dump to photons will be even less than in the SM: $T_\nu \approx T_\gamma$ will hold at late times as well.

In models where $m_{e,B}< |\DeltMnp|$, the twin proton can decay. If the timescale of this decay is very rapid, then the twin protons can be depleted prior to BBN. This scenario has the advantage of being safe for self-interaction bounds, since the twin neutron does not have any long-range interactions, but is difficult to detect and phenomenologically somewhat sterile. In other parts of our parameter space, $m_{e,B}>| \DeltMnp|$ and the twin proton is stable, and a relic population can exist irrespective of whether or not heavier nuclei are bound. We wish to estimate the decay process
\begin{equation}
    p_B^+ \rightarrow n_B + e_B^+ + \nu_{e,B}.
\end{equation}
A first order estimate can be found using the V-A theory of neutron beta decay from the SM \cite{Griffiths:2008zz}
\begin{equation}\label{eq: proton decay}
    \Gamma_{p,B} =f^R \left(\frac{E_{max}}{m_{e,B}}\right)\frac{  m_{e,B}^5}{2\pi^3}G^2_{F,B} |V_{ud}|^2  (1+ 3 g_{a,B}^2).
\end{equation}
We assume that $|V_{ud}|\approx 1$ as it is in the SM. The maximum electron energy $E_{max}$ is given by
\begin{equation}
    E_{max} = \frac{m_{p,B}^2 -m_{n,B}^2 + m_{e,B}^2 }{2 m_{p,B}}.
\end{equation}
The phase space factor $f^R(x)$ can be approximated:
\begin{equation}
    f^R(x) = \frac{1}{60}\left(2 x^4 - 9 x^2 -8 \right)\left(x^2 -1\right)^{1/2} + \frac{1}{4}x \ln{\left( x + \left(x^2 -1\right)^{1/2} \right)}.
\end{equation}
This calculation (with the appropriate replacements $n \leftrightarrow p$) gives $\OfOrder{10\%}$ agreement for SM values with the measured decay rate of the neutron. 

Finally, we use conservation of entropy to change variables to temperature:
\begin{equation}
     \frac{dT_A}{dt} = \frac{-H(T_A) T_A}{1+ \frac{T_A}{3g_{\star S}(T_A) }\frac{dg_{\star S}(T_A)}{dT_A}}.
\end{equation}
We include explicitly the variation of $\gstarS$ with temperature, since the usual approximation that $\gstarS$ is constant during freeze-out may not hold for some of our scenarios.  Regardless, the denominator is usually only an $\OfOrder{1}$ correction. 

Using the constraint $X_{n,B} + X_{p,B} =1$, we can write the total rate of depletion of the twin proton
\begin{equation}
    \frac{dX_{p,B}}{dT_A} = -\left(\frac{dT_A}{dt}\right)^{-1}
    \left[
    \left(X_{p,B}- 
    \left( 1- X_{p,B}\right)e^{-|\DeltMnp|/T_B} 
    \right)
    \Gamma_{p\rightarrow n,B}(T_B)
    + X_{p,B}\Gamma_{p,B}
    \right]. 
\end{equation}
Here $\Gamma_{p\rightarrow n,B}$ is the total rate for conversion of twin protons to neutrons, taken to be the sum of the rates in Eqs.~\eqref{eq: rate p nu to e n} and \eqref{eq: rate p e to n nu}, with the decay rate $\Gamma_{p,B}$ given in Eq.~\eqref{eq: proton decay}. The computed cosmological evolution from Appendix \ref{app: cosmo BG} allows us to write $T_B(T_A)$ and thus $\Gamma(T_A)$. This equation must be integrated numerically for each parameter point. Note, however, that the rate is explicitly dependent on $m_{e,B}$ and $\DeltMnp$, with only weak dependence on the masses of the pion and the nucleons via the Hubble evolution. We therefore compute the freeze-out curves $X_{p,B}(T)$ for each $m_{N,B}=1$, 5, and 10 GeV for the central value $m_{\pi,B}/m_{N,B}\simeq 0.35$, and obtain results for different pion masses by evaluating $X_{p,B}$ at the corresponding deuterium formation temperature $T_{D}$. We have verified that this approximation agrees with the full calculation to within 1--2\% wherever $X_{p,B}(T_D)\gtrsim 10^{-5}$. This allows us to capture the full behaviour of the model with a relatively sparse numerical scan.

\section{Direct Detection Background}\label{app: direct detection BG}

Below we review the standard treatment for direct detection experiments. \cref{app: velocity distribution} gives the DM velocity distribution, \cref{app:electronic_screening} gives the form of the electronic screening form factors, \cref{app: nuclear recoil} details the nuclear recoil set up, \cref{app: electron recoil} details the electron recoil set up, and \cref{app: neutrino floor} describes our approximation for the neutrino floor for nanocharge candidates.

\subsection{DM Velocity Distribution}\label{app: velocity distribution}

To estimate the distribution of the DM candidates we use the standard halo model (SHM) as defined in Ref.~\cite{DelNobile:2021wmp}. The details can be found therein, and the modifications we make to this picture are discussed in the main text in \cref{sec: observables}. Below we sketch the SHM parametrically. We write $v_{esc}$ as the galactic escape speed and $v_E$ as the speed of the Earth with respect to the galactic reference frame. We approximate the velocity distribution with a non-rotating, isotropic distribution following a Maxwell-Boltzmann distribution in the galactic rest frame $f_{SHM} \propto e^{-\vec \omega^2/v_0^2}$. For the SHM, we will assume that $v_0$ is equal to the ambient speed of the local standard of rest (LSR) with respect to the galactic centre. $\omega$ is the DM speed in the galactic reference frame and is related to the DM speed in Earth's reference frame by $\vec v = \vec \omega - \vec v_E$
\begin{equation}
    f_{SHM}(\vec v) \propto \frac{e^{-(\vec v + \vec v_E)^2/v_0^2}}{v_0^3 }\Theta(v_{esc}-  |\vec v + \vec v_E|) \, .
\end{equation}
Here the quantities without vectors indicate the magnitudes of the corresponding velocities. Assuming the distribution is isotropic, the angular variables can be integrated analytically to yield a one dimensional velocity distribution. 

\begin{table}
    \centering
    \begin{tabular}{|c|c|c|}
    \hline
        Parameter & Definition & Value \\ \hline
        $v_{\rm esc}$ & Galactic Escape Velocity & 544 km/s \\
        $v_\oplus$ & Speed of Earth wrt Sun & 30 km/s \\
        $v_S$ & Speed of Sun wrt GRF & 232 km/s \\
        $v_0$ & Speed of LSR wrt GRF & 220 km/s \\
        $\rho_{DM}$ & DM energy density & 0.4 GeV/cm$^3$\\ 
        \hline
        \end{tabular}
    \caption{Numerical values of parameters used in the computation of direct detection results, see Refs.~\cite{Lewin:1995rx, Baxter:2021pqo,DelNobile:2021wmp} and references therein. LSR is the local standard of rest and GRF is the galactic rest frame}
    \label{tab: SHM params}
\end{table}

We consider two benchmark models: the halo distribution and the \lq\lq dark disk". For the halo model, we take the Earth's velocity in the galactic reference frame $v_E$ to be equal to the time average of:
\begin{equation}
    v_E(t) = \sqrt{v_S^2 + v_\oplus^2 + v_S v_\oplus \cos[\omega (t-t_0)]},
\end{equation}
with $v_S$ the speed of the Sun with respect to the galactic reference frame, $v_\oplus$ the speed of the Earth with respect to the Sun and $\omega$ the angular frequency of rotation of the Earth around the sun. For the values in \cref{tab: SHM params}, this gives $v_E \approx 234$ km/s, which is approximately the speed of the Sun. Additionally, if the halo is optically thick to the naDM, there is no effective $v_{esc}$, as fast particles will be scattered within the halo and will not escape directly. 

The interacting sub-component of naDM may not follow the halo distribution. To account for the possibility that the naDM collapses to a disk, we follow Ref.~\cite{Chacko:2021vin}. The distribution described above clearly does not describe a co-rotating disk, however, we can approximate such a distribution by taking the velocity of the DM to be approximately the same as that of the Sun; in essence, assuming the DM is co-moving with the Sun such that the entire velocity dependence is from the Earth moving around the Sun. For benchmark parameters, we take $v_E= v_\oplus = 30$ km/s and $v_0 = 20$ km/s based on the disk results of Refs.~\cite{Dehnen:1997cq, Lopez-Santiago:2006kjq}. 

\subsection{Electronic Screening}\label{app:electronic_screening}

In this section we describe how the twin electronic form factors can be approximated. These form factors characterize the screening effect of electrons bound around a nucleus. Since the atomic form factor is the Fourier transform of the electron density, we use the SM form factors compiled in Ref.~\cite{brownIntensityDiffractedIntensities2006} as a reference. They can be approximated by the numerical fit
\begin{equation}
    F(q)
    \simeq
    \sum_{i=1}^{4}
    \mathfrak{a}_i
    \exp\left[
        -\mathfrak{b}_i
        \left(\frac{q}{4\pi}\right)^2
    \right]
    +\mathfrak{c}\, ,
\end{equation}
where the $\mathfrak{a}_i$ are $\mathcal{O}(1)$ coefficients and
the $\sqrt{\mathfrak{b}_i}\sim \text{\AA}^2\sim \mathrm{keV}^{-2}$ characterize variations in the electron density on atomic length scales. 
The constant $\mathfrak{c}$ ensures the correct long-wavelength normalization. 
In particular, for a neutral atom whose nucleus has charge $Q$, the effective charge
\begin{equation}
    Q_{\rm eff}(q)=Q-F(q)
    \label{eq:Qeff}
\end{equation}
vanishes as $q\to0$.

The characteristic atomic size, and therefore the coefficients $\mathfrak{b}_i$, are set by the Bohr radius, $a_{0,B} = \left( m_{e,B} \alpha_B \right)^{-1}$.
Taking $\alpha_B=\alpha_A$, the form-factor coefficients scale as
\begin{equation}
    \mathfrak{b}_{i,B}
    =
    \mathfrak{b}_{i,A}
    \left(\frac{m_{e,A}}{m_{e,B}}\right)^2.
\end{equation}
For the heaviest electrons in our analysis, this results in very minor reductions in the event rate, however, for heavier twin atoms with heavier bound electrons, the effect could be substantial.

\subsection{Nuclear Recoil}\label{app: nuclear recoil}

In the following, we review the basic aspects of nuclear recoil direct detection experiments. We follow largely the treatment of Ref.~\cite{DelNobile:2021wmp}. At expected halo velocities $v\sim 200$ km/s, we expect a GeV-scale DM candidate to be highly non-relativistic. The rate of events per unit detector mass at a given energy is expressed by:
\begin{equation}\label{eq: differential event rate recoil energy}
    \frac{dR}{dE_R}(E_R) = \sum_T\frac{\rho}{m}\frac{\zeta_T}{m_T} \int_{v\geq v_{\rm min}(E_R)}d^3v f_E(\vec{v})v \frac{d\sigma_T}{d E_R}(E_R,\vec{v}).
\end{equation}
The subscript $T$ is used to identify target-specific quantities for detectors made up of multiple types of nuclei. Here $E_R$ is the recoil energy, $\rho$ is the DM density, $m,\ m_T$ are the DM and target masses, respectively, $\zeta_T$ is the target nuclei mass fraction in the detector, $v_{\rm min}$ is the minimum DM velocity required to get a scattering event with the specified recoil energy, $f_E(\vec{v})$ is the DM velocity distribution at Earth's location in the detector's rest frame, and $\sigma_T(v)$ is the velocity dependent cross-section of the DM with the target nuclei. We neglect any time dependence arising from the annual modulation in Earth's velocity.

At sufficiently high recoil energies, scattering becomes inelastic and the DM particle probes the internal structure of the SM nucleus. We model this effect with form factors $F_{X,Y}^{(N,N^\prime)}$, which parameterize the effects of loss of coherence due to the finite size of the nucleus. The $X,Y$ range over six different nucleon-specific responses including number density, spin, and angular momentum related quantities, while $N,N^\prime$ denote neutron or proton. Where $X=Y$, we use the simplified notation $F_X$. We draw our nuclear form factors from Ref.~\cite{Fitzpatrick:2012ix} for xenon and Ref.~\cite{Catena:2015uha} for helium. For simplicity, for neutrinos, we use the Helm form factor \cite{Helm:1956zz}
\begin{equation}
    F(q) = \frac{3 j_1(qR)}{qR}e^{-\left(qs \right)^2/2},
\end{equation}
where $R$ is the nuclear radius $R\simeq 1.2 \Anum^{1/3}$ fm, with $\Anum$ the atomic mass number, and the skin thickness $s$ set to 0.9 fm \cite{Lewin:1995rx}. Note that the form factors $F_{X,Y}^{(N,N^\prime)}$ are approximately proportional to $|F(q)|^2$. 

The total rate of events at the detector is given by 
\begin{equation}
    R =  \int_0^\infty\epsilon(E_R)\frac{dR}{dE_R}(E_R)  dE_R.
\end{equation}
Here $ \epsilon(E_R) $ is the detector's efficiency and cut acceptance, which enforces the experimental thresholds, and we have treated quenching as negligible. For LUX-ZEPLIN (LZ), we digitize the acceptance function given in Ref.~\cite{LZ:2024zvo}. Constraints are determined by requiring that the number of events in a 4.2 tonne-year exposure is $N<3$, with the threshold of 3 events chosen to reproduce the constraints for a single-component WIMP candidate from Ref.~\cite{LZ:2024zvo}.

LZ has no overlap with the disk model in the mass range we consider. However, it is very notable that many components of the event rates are enhanced at small speeds by $1/v$. Because of the extremely low speeds involved, these interactions can be enhanced in the disk model and may be probed at detectors with lower thresholds. 

\subsubsection{Other Twin Neutron Interactions}

The cross sections for the nanocharge and electric dipole moment interaction are given in the main text Eq. \eqref{eq: direct detection NC and EDM cross section}. Here we show the cross sections for the magnetic dipole moment and charge radius interactions. While these are always sub-leading compared to nanocharge interactions, if the twin proton is completely depleted these interactions could in principle yield direct detection signals for the twin neutron. We find the constraints to be very weak for our parameter space. The cross sections are given by 
\begin{equation}
    \begin{aligned}
            \left(\frac{d \sigma_T}{d E_R}\right)_{MDM} & =  \frac{1}{8 \pi}\frac{m_T}{m_{N,A}^2}\frac{1}{v^2}e^2 (\kappa \mu_\chi)^2 
    \left[\left(\frac{v^2}{E_R} - \frac{m+ 2m_T}{2m m_T}
    \right) 2\frac{m_{N,A}^2}{m_T} F_M^{(p,p)}(q^2) + 4 F_\Delta^{(p,p)}(q^2) \right. 
    \\
    & \qquad \qquad\left.-2 \sum_N g_N F^{(N,p)}_{\Sigma^\prime \Delta}(q^2) + \frac{1}{4} \sum_{N,N^\prime} g_N g_N^\prime F_{\Sigma^\prime}^{(N,N^\prime)}(q^2)
    \right],
    \\
    \left(\frac{d \sigma_T}{d E_R} \right)_{CR} & = \frac{m_T }{72\pi}\frac{1}{v^2}  \left( e^2 \kappa \angl{r^2_{\rm DM}}_E  \right)^2  F^{(p,p)}_M(E_R), \\
    \end{aligned}
\end{equation}
In these expressions $\angl{r^2_{\rm DM}}_E$ is the mean-squared charge radius of the DM, and $\mu_\chi$ is the magnetic dipole moment. The two interactions can also interfere, though we find the effect is small. 

To complete our estimate of the event rate, we need an estimate of the twin nucleon magnetic dipole moment $\mu_\chi$ and the mean-squared charge radius $\angl{r^2_{\rm DM}}_E$. We use the lattice for these estimates. The magnetic moment can be expressed as
\begin{equation}
   \mu_\chi = \kappa_{N,B} \mu_{N,B} = \frac{\kappa_{N,B} e }{2 m_{N,B}}, 
\end{equation}
where $\mu_N$ is the nuclear magneton. The constant $\kappa_{N,B}$ for neutral baryons on the lattice was found to be approximately independent of the baryon mass and number of quarks in the theory up to a baryon mass of $\sim 2.2 m_{N,A}$ \cite{LatticeStrongDynamicsLSD:2013elk}. These results were computed at heavier than physical pion masses and were consistent with the experimental value $\kappa = 1.91$ \cite{ParticleDataGroup:2024cfk}, suggesting that this quantity is also independent (or slowly varying) with pion mass. We take for our estimate the central value $\kappa_{N,B} = 1.9$. 

The charge radius interaction, being at dimension 6, is expected to be subleading to the dimension 5 magnetic dipole moment interaction. The charge radius is expected to be sensitive to the twin pion mass, since the size of the pNGB cloud of the twin neutron is dependent on the mass of the pion. At the SM point, the neutron charge radius is $\angl{r^2_{n}}_E=-0.1155 \ \mathrm{fm}^2$ \cite{ParticleDataGroup:2024cfk}. For $m_{\pi,B}/m_{N,B}\sim 0.45$, Ref.~\cite{LatticeStrongDynamicsLSD:2013elk} found $\angl{r^2_{n}}_E\simeq -(0.009...0.025)\ \mathrm{fm}^2$. This is consistent with the expectation for length scales shrinking with increasing pion mass. 

Heavy baryon chiral perturbation theory \cite{Gockeler:2003ay, Djukanovic:2023beb} suggests that the functional form of the charge radius with pion mass should take the form $\angl{r^2_{n,B}}_E\simeq c_1 + c_2 \ln(m_{\pi,B}/\mu)$, with $\mu$ the regularization scale and $c_i$ dimensionful constants that would be rescaled by $\Lambda_B/\Lambda_A$. For our purposes it suffices to observe that, for constant $m_{\pi,B}/m_{N,B}$, we expect $\angl{r^2_{n,B}}_E \propto 1/m_{N,B}^2$. We thus take as our bracketing cases:
\begin{equation}
    \begin{aligned}
      \max  \angl{r^2_{n,B}}_{E} &= -\frac{2.99}{m_{N,B}^2}, \\
        \min \angl{r^2_{n,B}}_{E} &= -\frac{0.43}{m_{N,B}^2},
    \end{aligned}
\end{equation}
where the upper limit is consistent with the SM value $\angl{r^2_{n,A}}_{E}= -0.1155\ \textrm{fm}^2$ and the lower limit is the mean of the values found in Ref.~\cite{LatticeStrongDynamicsLSD:2013elk} for $m_{\pi,B}/m_{N,B}\approx 0.5$. 

We find that the constraints on the kinetic mixing that can be placed based on nuclear recoil direct detection are $\kappa \lesssim \OfOrder{10^{-5}}$ for $m_{N,B}\sim 10$ GeV, and weaker as the mass gets smaller. If the lightest fermion in the spectrum has a mass below 1 GeV, this is already ruled out by $\DNeff$ at CMB and BBN. However, if the lightest twin fermion is heavier than 1 GeV, the kinetic mixing is only constrained by collider considerations to be $\kappa \lesssim 10^{-3}$ \cite{milliQan:2021lne}. In this particular case, the charge radius and magnetic dipole moment constraints can be relevant.

\subsection{Electron Recoil}\label{app: electron recoil}

Below, we review the basic methods of computing electron recoil signatures, following mainly the treatment of Refs.~\cite{Essig:2015cda, Knapen:2021bwg}. Note that in this section, \textit{electron} will always refer to the SM electron unless specified otherwise. DM interactions with a material can excite electrons in several different ways, which are encoded in the energy loss function (ELF) $\mathrm{Im} \left(-1/\epsilon(\omega, \vec k) \right)$, with $\epsilon$ the dielectric function. 

We will be interested in the direct excitation of SM electrons by DM candidates in which an electron is excited above the bandgap of the material and creates an electron-hole pair. DM particles could also excite phonons below the bandgap, however, in the mass range we are interested in we expect direct excitation to provide the dominant signal. The energy scales of phonon excitations are set by the vibrational scale of the lattice and are typically $\OfOrder{\rm{meV}}$. For very light DM candidates with insufficient energy to excite an electron above the bandgap, this can be the dominant signal. However, for the MeV and above candidates we consider, excitation above the bandgap is possible and this process creates a charge signal which is experimentally easier to resolve than the meV-scale phonons. 

The DM-electron interaction is typically parameterized in terms of the reference cross section
\begin{equation}
    \bar\sigma_e = \frac{ \mu^2_{\chi e} g_e^2 g_\chi^2}{\pi \left(\alpha^2 m_{e}^2 + m_\phi^2\right)^2 } .
\end{equation}
Here $m_\phi$ is the mass of the mediator, $g_e$ and $g_\chi$ are the couplings of the SM electron and the DM candidate to the mediator, and the factor of $\alpha m_{e}$ in the denominator is the average momentum of a valence electron bound in a material. We consider the nanocharge interaction of Eq. \eqref{eq:kinetic_mixing}, thus $m_\phi=0$. The total rate is integrated over the momentum dependent ELF, the cross section, and the velocity distribution of the DM as described in Appendix \ref{app: velocity distribution}. 

Unlike in nuclear recoil detectors, where the target particle is well approximated as stationary, the bound SM electron does not have a definite momentum and large momentum transfers are possible due to fluctuations in the electron momentum. The momentum transfer to the DM $q$ is related to the change in energy of the electron via
\begin{equation}
    \Delta E = \vec q \cdot \vec v - \frac{q^2}{2 \mu_{\chi N}}.
\end{equation}
Here $\vec v$ is DM velocity and $\mu_{\chi N}$ is the reduced mass of the DM-nuclei system and accounts for the fact that the whole atom can recoil. The maximum allowed energy transfer is 
\begin{equation}
    \Delta E_{max} = \frac{1}{2} \mu_{\chi N} v^2,
\end{equation}
which is the limit where all the kinetic energy in the DM-atom collision is transferred to the electron. For materials of interest in most electron recoil experiments, the bandgap is $\OfOrder{\rm{eV}}$. Thus to achieve $\Delta E$ large enough to excite an electron to the conduction band where it may be detected, given $v\sim 200 $ km/s for a typical DM halo, DM masses of $\OfOrder{\rm{MeV}}$ can be probed. 

The likelihood of an event with momentum transfer $q$ (or energy transfer $\Delta E$) is encoded in the energy loss function. The ELF can be measured experimentally and calculated theoretically in various approximations. We make use of the GPAW computation of the dielectric of silicon~\cite{Enkovaara:2010, Knapen:2021run}. This computation makes use of time-dependent density functional theory methods to compute the response of electrons on a periodic lattice subject to an effective potential which encodes the structure of the material. We use the package \texttt{darkELF} \cite{Knapen:2021bwg} to compute the rates using the precomputed ELFs using a threshold of two electron excitations per event. The currently published results of DAMIC-M restrict the analysis to $<6e^-$ \cite{DAMIC-M:2025luv}, while Oscura intends to consider up to $10e^-$ \cite{Oscura:2022vmi}. 

The band-gap energy of silicon is 1.1 eV and the average energy required to create an electron-hole pair is 3.6 eV~\cite{Klein:1968, Streetman2005}. We take as the lower energy threshold of the rate integration the sum of the band-gap energy plus the energy to create one additional electron-hole pair, $E_{\rm min}= 4.7$ eV. For the upper thresholds we make the approximation that each additional 3.6 eV liberates exactly one additional electron, and thus have $E_{\rm max}^{\rm DAMIC-M}= 19.1$ eV and $E_{\rm max}^{\rm Oscura} =33.5$ eV. 

\subsection{Neutrino Floor}\label{app: neutrino floor}

Here we briefly discuss the neutrino floor in the case of nanocharge models. While a sophisticated calculation which accounts for the lineshape of the energy spectra based on e.g. the methods of Ref.~\cite{OHare:2021utq} is beyond the scope of this work, we sketch here a parametric estimate for the location of the neutrino floor for our models. 

The interaction cross-section of neutrinos with a nucleus is \cite{Freedman:1973yd, Freedman:1977xn, Drukier:1984vhf}
\begin{equation}
    \frac{d\sigma}{dE_R}(E_R, E_\nu) =  \frac{G_{F,A}^2}{4 \pi}Q^2_{W,A} m_{\Znum,A} \left(1- \frac{m_{\Znum,A} E_R}{2 E_\nu^2}\right)F^2(E_R) ,
\end{equation}
where $Q_W =  \Anum- 2\Znum (1-2 \sin^2 \theta_W)$ is the weak hypercharge of a SM nucleus with mass number $\Anum$, atomic number $ \Znum$, and mass $m_{\Znum,A}$. $F^2(E_R)$ is the nuclear form factor, discussed in the nuclear recoil section above, which we take to be the Helm form factor \cite{Helm:1956zz} in this section for simplicity. 

The recoil energy spectrum is found by integrating over the differential cross section multiplied by the neutrino flux
\begin{equation}
    \frac{dR_\nu}{dE_R} = \frac{1}{m_{\Znum,A}}\int_{E_\nu^{\rm min}} \frac{d\Phi}{dE_\nu}  \frac{d\sigma}{dE_R}(E_R, E_\nu) dE_\nu. 
\end{equation}
The minimum neutrino energy that can cause a recoil with energy $E_R$ is $E_\nu^{\rm min}= \sqrt{m_{\Znum,A} E_R/2}$ for nuclear recoil. We draw our flux $d\Phi/dE_\nu$ from Ref.~\cite{Vitagliano:2019yzm}. 

For the electron recoil experiments, direct ionization via elastic scattering of neutrinos with an electron \cite{Marciano:2003eq, Formaggio:2012cpf} is subleading to electron ionization resulting from coherent neutrino scattering with a nucleus \cite{Essig:2018tss}. The conversion between the recoil energy of the nuclei $E_R$ to the energy injected to an electron is parameterized by the quenching factor $Y(E_R)$ with $E_e = Y(E_R)E_{R}$. We take as our benchmark the high conversion efficiency estimate for silicon of Ref.~\cite{Essig:2018tss}
\begin{equation}
    Y = 0.18\left(1- e^{-(E_R -15)/71.3}\right),
\end{equation}
where $E_R$ is in units of eV. This expression is approximately valid between $15\ \textrm{eV}<E_R<250$ eV, which encompasses our entire range of interest. The rate with respect to electron energy is given by
\begin{equation}
    \frac{dR_\nu}{dE_e} = \frac{dR_\nu}{dE_R}\times \frac{1}{\left(Y(E_R) + E_R \frac{dY(E_R)}{dE_R}\right)}.
\end{equation}

To estimate the neutrino floor, we compute the value of $\kappa$ that yields the same number of events as the neutrino background, $N_{naDM}= N_{\nu}$. For the nuclear recoil floor, we take as reference the acceptance function for LZ \cite{LZ:2024zvo}. This means that the neutrino floor as presented in this work is tied to this experiment. Following this method for a standard WIMP DM candidate, we find a neutrino floor that is within a factor of a few of the neutrino fog of Ref.~\cite{OHare:2021utq}. 

For the electron recoil we use the 2 electron minimum and 10 electron maximum threshold of Oscura \cite{Oscura:2022vmi} to compute the floor. This yields a neutrino floor for Oscura which occurs at approximately an exposure of 5-10 $\textrm{kg}\cdot\textrm{yr}$, which agrees with the analysis of Ref.~\cite{Essig:2018tss} for the point at which the background becomes relevant. As demonstrated in that reference, the neutrino floor does not constitute an impenetrable wall and only represents a threshold at which the increasing backgrounds will reduce the effective reach of increasing exposure compared to the linear expectation.


\section{Effective Baryon Yukawas}\label{app: effective baryon yukawas}

In \cref{sec: DD Higgs portal} we estimated the coupling of the physical Higgs to both SM and twin baryons using only the couplings of the light quarks $u,d,s$. We justify this calculation below, following the methods of Refs.~\cite{Shifman:1978zn, DelNobile:2021wmp}. The couplings of A or B sector nucleons to the physical Higgs are given by
\begin{equation}
    \begin{aligned}
        c_{N,A} &\equiv \frac{m_{N,A}}{v}\sum_q  f_{q,A}^{(N)}, 
        \\
        c_{N,B} & \equiv \left(\frac{m_{N,B}}{f} \right) \left(\frac{v}{f}\right) \sum_q  f_{q,B}^{(N)}(m_{\pi,B}) .
    \end{aligned}
\end{equation}
Omitting the $A,B$ subscripts for clarity, $f_q^{(N)}$ is the effective contribution of the quark masses to the nucleon mass
\begin{equation}\label{eq: scalar form factor}
     m_{N}f_{q}^{(N)} = \bra{N}m_{q} \bar q q \ket{N}.
\end{equation}

Consider the sum over heavy quark contributions with  $h=c,b,t$. At lowest order in $\alpha_s$, the heavy quark expansion yields
\begin{equation}
    m_h \angl{ \bar h h } \simeq -\frac{\alpha_S}{12 \pi}\angl{ G^2 },
\end{equation}
where $G$ is the gluon field strength, $G^2 = G^{a\mu\nu}G^a_{\mu\nu}$, and we have used $\angl{\mathcal{O}}$ as a shorthand for $\bra{N^\prime} \mathcal{O} \ket N$. This equality arises because heavy quarks largely contribute to the nucleon mass via couplings to gluons. We can therefore write the sum over all quarks as
\begin{equation}\label{eq: sum over scalar FF heavy quark approx}
    \sum_q m_N f_{q}^{(N)} \simeq \sum_{u,d,s}m_N  f_{q}^{(N)} - \frac{\alpha_S}{4\pi}\angl{G^2}.
\end{equation}

In the nucleon picture, denoted by the subscript $N$, the energy-momentum tensor for a single nucleon at rest can be approximately written as
\begin{equation}\label{eq: energy momentum nucleon}
    \angl{\Theta^\mu_\mu}_N\simeq m_N.
\end{equation}

At the quark level, denoted by the subscript $q$, the trace of the energy momentum tensor is found to be
\begin{equation}\label{eq: energy momentum quark}
        \left(\Theta^\mu_\mu \right)_q
        = \sum_q m_q \bar q q  + \frac{\beta(\alpha_s)}{4 \alpha_s}G^2
       \simeq \sum_{q} m_q \bar q q - \frac{7 \alpha_s}{8 \pi}G^2.
\end{equation}
The gluon term arises from the trace anomaly and in the second equality we have used the lowest order QCD beta function. Evaluating the expectation value and equating $\angl{\Theta^\mu_\mu}_N = \angl{\Theta^\mu_\mu}_q$, we find
\begin{equation}\label{eq: energy momentum trace equality}
     \sum_{q}m_Nf_{q}^{(N)} \simeq \frac{7\alpha_s}{8 \pi}\langle G^2 \rangle + m_N .
\end{equation}
We then combine Eqs. \eqref{eq: sum over scalar FF heavy quark approx} and \eqref{eq: energy momentum trace equality} to eliminate $\angl{ G^2}$,
\begin{equation}
    \sum_q  f_{q}^{(N)} \simeq \frac{2}{9} + \frac{7}{9}\sum_{q=u,d,s} f_{q}^{(N)}.
\end{equation}
We therefore find that to lowest order in $\alpha_s$ and in the non-relativistic baryon limit, the scalar form factor can be written entirely as a function of the light quark contributions. It is worth noting that in the twin sector, it is possible there are a different number of effective light flavours, which would slightly modify this relation. Numerically, however, the difference is only percent-level. 

\end{spacing}

\bibliographystyle{apsrev4-2-fixed.bst}
\bibliography{Refs}
\end{document}